\documentclass[pre,amsmath,amssymb,onecolumn, showpacs, superscriptaddress,10pt]{revtex4-2}

\usepackage[english]{babel}
\usepackage[utf8]{inputenc}
\usepackage[T1]{fontenc}
\usepackage{amsmath}
\usepackage[colorlinks=true,citecolor=blue]{hyperref}
\usepackage{graphicx}
\usepackage{amsfonts}
\usepackage{amsthm}

\newcommand{\nn}{\nonumber}
\newcommand{\be}{\begin{equation}}
\newcommand{\ee}{\end{equation}}
\newcommand{\bea}{\begin{eqnarray}}
\newcommand{\eea}{\end{eqnarray}}
\newcommand{\moy}[1]{\ensuremath{\left\langle #1 \right\rangle}}

\usepackage{color}
\definecolor{Blue}{rgb}{0.00, 0.00, 1.00}
\definecolor{Red}{rgb}{1.00, 0.00, 0.00}
\definecolor{Green}{rgb}{0.00, 0.70, 0.00}

\begin{document}

\title{Ground state energy fluctuations of pinned elastic manifolds}

\author{Yan Fyodorov}
\affiliation{King's College London, Department of Mathematics, London  WC2R 2LS, United Kingdom}
\author{Bertrand Lacroix-A-Chez-Toine}
\affiliation{King's College London, Department of Mathematics, London  WC2R 2LS, United Kingdom}
\author{Pierre Le Doussal}
\affiliation{Laboratoire de Physique de l'Ecole Normale Sup\'erieure, CNRS, ENS and PSL Universit\'e, Sorbonne Universit\'e, Universit\'e Paris Cit\'e,
24 rue Lhomond, 75005 Paris, France}
\date{\today}

\begin{abstract}
    We describe the atypical fluctuations of the ground state energy of the random elastic manifold, a disordered model defined on a lattice of linear size $L$ with internal dimension $0\leq d<4$ embedded in a medium of dimension $N\gg 1$. The ground-state energy results from a competition between confinement, elasticity and disorder. We obtain an exact description of the large deviation rate function with speed $NL^d$ and its different phases, corresponding to different patterns of replica symmetry breaking (RSB). Our results show that the ground-state energy satisfies a central limit theorem and we obtain an explicit expression for the rescaled variance. In the (massless) limit of zero confinement, this variance vanishes for short-range disorder and the ground-state energy displays super-concentration. From our results on the large deviation function, we characterise explicitly the left tail of the distribution of the typical fluctuations of the ground state energy. It displays an exponential tail for a one step RSB pattern while for a full RSB pattern it decays super-exponentially with a non trivial exponent $\xi$ that we compute explicitly.
\end{abstract}

\maketitle

\section{Introduction}

The optimal configuration adopted at equilibrium by many physical systems results from the competition between elasticity, promoting flat, homogeneous configurations, and disorder, promoting rugged, heterogeneous configurations \cite{giamarchi1998statics}. The elastic manifold is a versatile and well-studied model to describe such physical systems. Let us define it on a $d$-dimensional lattice with linear size $L$
and a lattice spacing $a$. At each point ${\bf x}$ of the lattice we define a $N$-dimensional field ${\bf u}({\bf x})$. The Hamiltonian of the elastic manifold then reads
\be
{\cal H}[{\bf u}]=\frac{1}{2}\sum_{{\bf x},{\bf y}}{\bf u}({\bf x})(\mu \delta_{{\bf x},{\bf y}}-t \Delta_{{\bf x},{\bf y}}){\bf u}({\bf y})+\sum_{{\bf x}}\, V({\bf u}({\bf x}),{\bf x})\;,\label{Hamiltonian}
\ee
where $\delta_{{\bf x},{\bf y}}=1$ if ${\bf x}={\bf y}$ and zero otherwise, $\Delta$ is the Laplacian matrix, $\mu\geq 0$ is the stiffness of the harmonic potential (that we will call mass in the following) and $t\geq 0$ is the interaction strength. In the following we use units such that $t=1$. The random field $V({\bf u}({\bf x}),{\bf x})$ is Gaussian with zero average and covariance
\be
\moy{V({\bf u}_1,{\bf x}_1)V({\bf u}_2,{\bf x}_2)}=\delta_{{\bf x}_1,{\bf x}_2}\,N\,f\left(\frac{({\bf u}_1-{\bf u}_2)^2}{2N}\right)\;.\label{cor_func}
\ee
The first term in the Hamiltonian \eqref{Hamiltonian} is simply optimised by setting ${\bf u}({\bf x})={\bf 0}$ while the configuration optimising the Hamiltonian in presence of the second (disordered) term is non-trivial. It will be convenient to introduce the expression of the Hamiltonian \eqref{Hamiltonian} in Fourier space where the first term is diagonal 
\be
{\cal H}[{\bf u}]=\frac{1}{2}\sum_{{\bf k}}\nu({\bf k}){\bf u}({\bf k}){\bf u}(-{\bf k})+\sum_{{\bf x}}\, V({\bf u}({\bf x}),{\bf x})
\ee
where $k_j= 2 \pi n_j/L$ and the elastic modulus $\nu({\bf k})$ takes the form
\be
\nu({\bf k})\simeq {\bf k}^2 + \mu\;,\;\;|{\bf k}|a  \ll 1\;.\label{nu_k_small}
\ee 
In the following, we denote
$\int_{{\bf k}}=L^{-d} \sum_{{\bf k}} 
$ for finite $L$ and $a$. As we will be interested in the scaling of various observables with $L$ in the following, we will keep it finite while the limit $a\to 0$ (which does not yield any divergence in dimensions $0<d<4$) will often be considered explicitly. In the continuum limit $L\to \infty$ and $a\to 0$ one can replace
$\int_{{\bf k}}\to\int \frac{ d {\bf k} }{(2 \pi)^d}$.
The equilibrium static properties of this Hamiltonian are  studied at finite inverse temperature $\beta$ by computing the quenched free energy
\be
\moy{F_{N,L}(\beta)}=-\frac{1}{\beta}\moy{\ln Z_{N,L}(\beta)}\;,\;\;Z_{N,L}(\beta)=\int {\cal D}{\bf u}({\bf x})e^{-\beta{\cal H}[{\bf u}]}\;.
\ee
As the inverse temperature diverge, the latter converges towards the average ground-state energy
\be 
\moy{{\cal E}_{N,L}} = \moy{\min_{{\bf u}}  {\cal H}[{\bf u}]}=\lim_{\beta \to \infty} F_{N,L}(\beta)\;.
\ee 
Two main methods are available to study this model in the low temperature regime: (i) the functional renormalisation group, valid for any $N$  
but perturbative around $d=4$ \cite{balents1993large}, and (ii) the
replica saddle point method in the limit $N \to +\infty$
which is valid for any $d$ \cite{mezard1991replica}. The matching between these quite different approaches at large
but finite $N$ was investigated in \cite{le2002functional,le2003functional,le2004derivation,le2008cusps}. 
The focus of the present article is to compute the 
fluctuations of the ground state energy in the limit $N\to \infty$
using approach (ii). 
It is worth noting that the variance of the {\it difference} in ground state energy induced
by a small perturbing field was computed in that limit in \cite{le2008cusps}, with applications
to avalanches and shocks in the closely related problem of Burger's turbulence
\cite{bouchaud1995scaling,le2010avalanches,le2012equilibrium}. However, to our knowledge, neither the variance
of the ground state energy, nor its large deviation function, have been explicitly 
computed previously.


In that limit, the rescaled free-energy is a self-averaging random variables, i.e. 
\be
\lim_{N\to \infty}{\cal F}_{N,L}(\beta)=\lim_{N\to \infty}\frac{F_{N,L}(\beta)}{NL^d}\underset{\rm a.s.}{=}\lim_{N\to \infty}\moy{{\cal F}_{N,L}(\beta)}={\cal F}_{d}(\beta,L)\;,
\ee
and the limiting quenched free energy ${\cal F}_{d}(\beta,L)$ has been computed  explicitly in the physics literature using replica and saddle-point techniques, notably by Mezard and Parisi \cite{mezard1991replica} (see also \cite{balents1996large,engel1993replica,fyodorov2007classical}). These results have recently been re-examined using rigorous methods in a series of papers by Ben Arous and Kivimae \cite{arous2024larkin,arous2024free}.  The limiting free energy is then found to be represented via the "Parisi formula" as the solution of an optimisation problem over a space of probability measures, with the optimal probability measure describing the overlap distribution between different configurations optimising the free-energy. The self-averaging property holds similarly at zero temperature and a Parisi formula can similarly be derived for the limitng ground-state energy, which concentrates around the typical value $e_{L,\rm typ}(\mu)=\lim_{N\to \infty}{\cal E}_{N,L}/(NL^d)=\lim_{\beta\to\infty}{\cal F}_{d}(\beta,L)$. 

A particularly interesting outcome of the paper by Mezard and Parisi \cite{mezard1991replica} is the computation of the scaling behaviour of the variance of the free energy with the size $L$ of the lattice. Using dimensional analysis and renormalisation group arguments, they showed that in the massless limit $\mu\to 0$ a simple relation exists between the growth exponent of the free energy variance with the lattice size in the limit $L\to \infty$,
\be
{\rm Var}(F_{N,L}(\beta))\sim L^{2\chi}\;,\;\;{\rm where}\;\;\chi=2\zeta+d-2\;,
\ee
and the growth exponent of the transverse fluctuations of the manifold, i.e.
\be
C_{N,L}({\bf l},\beta)=\moy{\frac{1}{Z_{N,L}(\beta)}\int {\cal D}{\bf u}({\bf x})e^{-\beta{\cal H}[{\bf u}]}\left({\bf u}({\bf x}+{\bf l})-{\bf u}({\bf x})\right)^2}\sim |{\bf l}|^{2\zeta}\;.
\ee
They also computed explicitly the value of $\zeta$ in dimensions $0<d<4$ for a family of correlation functions $f(q)$ as defined in Eq. \eqref{cor_func}. The variance can be shown to diverge in the continuum limit $L/a\to \infty$ in any dimension $d\geq 4$. Interestingly, it was found that the exponent $\chi> 0$ is positive when the free-energy is given by its full replica-symmetry broken (FRSB) expression while it is negative or zero otherwise. For negative exponents, the free-energy $F_{N,L}(\beta)$ decreases with the lattice size $L$ and is thus a self-averaging variable in the limit $L\to \infty$. The rescaled free-energy ${\cal F}_{N,L}(\beta)$ is instead self-averaging in the limit $L\to\infty$ whenever the weaker criterion $\zeta< 1$ is satisfied.

The goal of this paper is to study the fluctuations of the rescaled ground-state energy defined as
\be
e_{\min}=\frac{{\cal E}_{N,L}}{NL^d}=\frac{1}{NL^d}\min_{\bf u({\bf x})}{\cal H}[{\bf u}]=\lim_{\beta\to \infty}{\cal F}_{N,L}(\beta)\;.
\ee 
In particular, we will obtain an explicit expression for the rescaled variance in the large $N$ limit, defined as
\be
{\cal V}_L(\mu)=\lim_{N\to \infty}NL^d {\rm Var}(e_{\min})=\lim_{N\to \infty}\frac{{\rm Var}({\cal E}_{N,L})}{N L^d}\;.
\ee
This rescaled variance depends non-trivially on the length $L$ and the mass $\mu$ and, in particular, in the massless limit $\mu\to 0$ the variance ${\cal V}_L(0)\sim L^{2\tilde \chi}$ is predicted to grow with the exponent $\tilde \chi=2\zeta+(d-4)/2$ \cite{mezard1991replica}. Thus, our result
not only reproduces the scaling exponent obtained by Mezard and Parisi \cite{mezard1991replica} 
 in an alternative way (since $\tilde \chi=\chi-d/2$), but provides an exact, quantitative expression much beyond.  In addition, we characterise the atypical fluctuations of $e_{\rm min}$ by computing the large deviation function with speed $NL^d$ in the limit $N\to \infty$,
\be
\Lambda_L(e,\mu)=-\lim_{N\to \infty}\frac{1}{NL^d}\ln \moy{\delta(e-e_{\min})}\;.\label{LD_emin}
\ee
We will also sometimes be interested in the continuum limit and define the following large deviation function in that limit
\be
{\cal L}(\epsilon,\mu)=\lim_{L\to \infty}\Lambda_L(e_{L,\rm typ}(\mu)+\epsilon,\mu)\;.
\ee



In the following, we need to distinguish between two classes of disordered potential (see \cite{fyodorov2018hessian} for a detailed physical discussion and \cite{schoenberg1938metric} for rigorous results):

(i) Short range disorder, such that the covariance function $f(q)$ satisfies $f(q)=\int_0^{\infty} dk\,e^{-q k^2}\,\tilde f(k)$ with positive coefficients $\tilde f(k)\geq 0$ for any $k\geq 0$. In this case the
fluctuations of the ground state energy $e_{\min}$ can be investigated directly.

(ii) Long range disorder, such that the disordered potential is
a random process with stationary increments, such that $f(q) - f(0) =\int_0^{\infty} dk\,e^{-q k^2}\,\tilde f(k)$ 
with positive $\tilde f(k)\geq 0$ for any $k\geq 0$. In this case, the fluctuations of the energy difference with respect to the flat reference configuration need to be considered
\be 
\Delta e_{\min}= \frac{1}{NL^d}\min_{{\bf u}({\bf x})} ( {\cal H}[{\bf u}]  - {\cal H}[{\bf 0}] ) = \frac{1}{NL^d}\left({\cal E}_{N,L} - \sum_{\bf x} V({\bf 0},{\bf x})\right)\;. 
\ee 
The random variable $\Delta e_{\min}$ is by definition a negative random variable and the probability that $\Delta e_{\min}>0$ is identically zero. As $V({\bf 0},{\bf x})$ is a zero average Gaussian random variable, it is simple to check that $e_{\min}$ and $\Delta e_{\min}$ have the same average value. Finally, let us mention that this average value  $\moy{e_{\min}}=\moy{\Delta e_{\min}}$ depends on the microscopic details, in particular the lattice sizes $a$ and $L$ and it can be shown to diverge for any internal dimension $d\geq 2$ in the continuum limit $L/a\to \infty$.

As has been shown both in the physics \cite{mezard1991replica} and mathematics \cite{arous2024larkin,arous2024free} literature, the typical value of the ground-state energy is determined by the replica symmetric expression for $\mu$ above the so-called Larkin mass $\mu_{c}$. The latter is defined as the unique solution $\mu$ of $f''(0)I_2(\mu=\mu_{c})=1$, with $I_2(\mu)$ defined by taking $p=2$ in the expression
\be
I_p(x)=\int_{\bf k}\frac{1}{\left(x-\mu+\nu({\bf k})\right)^p}\;.\label{I_p}
\ee
This Larkin mass has been recently shown not only to determine the onset of the replica-symmetry breaking transition but also the associated  topology trivialisation transition  from an energy function (Hamiltonian) with a single minimum (for $\mu\geq \mu_c$) to a Hamiltonian displaying a number of minima growing exponentially with the dimension $N$ (for $\mu<\mu_c$) \cite{fyodorov2020manifolds,BenArousBourgadeMcKenna24}. We will show in the following that the Larkin mass also plays an important role in the description of higher-order cumulants of $e_{\min}$.

In the following, we will focus our attention on the cumulants of $e_{\min}$ and its large deviation function $\Lambda_L(e,\mu)$ for short-range disorder and provide some results for the cumulants of $\Delta e_{\min}$ for long-range disorder. 

The paper is structured as follows. In section \ref{sec-summary} we summarise the main results of the paper, providing a Parisi formula for the cumulant generating function of $e_{\min}$ as well as explicit expressions for the rescaled variance ${\cal V}_L(\mu)$ and the large deviation function $\Lambda_L(e,\mu)$. In section \ref{sec-opt-prob} we provide the detailed derivation of the expression of the cumulant generating function as an optimisation problem. In section \ref{sec-analysis} we analyse the expression of the cumulant generating function and large deviation function at different levels of replica-symmetry breaking. In section \ref{sec-special-cases} we obtain explicit expressions for the large deviation function in the continuous limit in a few specific internal dimensions $d$ and for some choice of covariance function $f(q)$. Finally in section \ref{sec-conclusion} we conclude and provide some ideas for future directions. Some technical details are relegated to the appendices. In appendix \ref{app_typ_GS} we analyse the average of $e_{\min}$. In appendix \ref{app_comp} we compute the annealed complexity of minima at fixed energy which provides a bound for the large deviation function.




\section{Summary of the main results} \label{sec-summary}

Unless specified otherwise, the results below apply for the ground-state energy $e_{\min}$ for a short-range disorder such that the correlation function decays at infinite distance $\lim_{q\to \infty}f(q)=0$. The results for the ground-state energy difference $\Delta e_{\min}$ for a long-range disorder are summarised in the last subsection.

\subsection{Typical fluctuations}

For any value $\mu>0$, we obtain an explicit expression for the rescaled variance
\begin{align}
{\cal V}_{L}(\mu)=&\lim_{N \to \infty}\frac{{\rm Var}({\cal E}_{N,L})}{N L^d}=f\left(Q_{\rm typ}\right)-\frac{I_2(\mu) f'\left(Q_{\rm typ}\right)^2}{2}\;,\label{var_res_1}
\end{align}
where $Q_{\rm typ}$ is the difference between largest and smallest overlap values, i.e. the range of the support of "Parisi function" describing the typical distribution of overlaps. 
In the RS phase, the distribution of overlaps is supported on a single atom and $Q_{\rm typ}=0$ while in the replica symmetry broken phases (both 1RSB and FRSB) $Q_{\rm typ}>0$. The expression for $Q_{\rm typ}$ is RS, hence vanishes, for any mass larger than the Larkin mass $\mu>\mu_c$ and RSB for $\mu<\mu_c$ \cite{mezard1991replica,arous2024larkin}. The criterion for selecting the pattern of replica symmetry breaking is not simple but still explicit (see Eq. \eqref{sig_prime_typ} and the following paragraph).

In the FRSB phase, $Q_{\rm typ}$ is given by a simple expression
\be
Q_{\rm typ}={f''}^{-1}\left(\frac{1}{I_2(\mu)}\right)\;,
\ee
where we remind that $I_p(\mu)$ is defined from Eq. \eqref{I_p} and ${f''}^{-1}$ is the inverse function of $f''$. In the 1RSB phase, $Q_{\rm typ}$ is the positive solution of Eqs. (\ref{res_var_0}-\ref{res_var}).

This rescaled variance is strictly positive for any $\mu>0$ and as a direct consequence, the ground-state energy satisfies the central limit theorem 
\be
\lim_{NL^d\to \infty}{\rm Prob}\left[\frac{{\cal E}_{N,L}-\moy{{\cal E}_{N,L}}}{\sqrt{{\rm Var}({\cal E}_{N,L})}}\leq x\right]=\frac{1}{2}\left(1+{\rm erf}\left(\frac{x}{\sqrt{2}}\right)\right)\;.
\ee
The energy fluctuations are thus of order $\sim (NL^d)^{1/2}$.

As we show below, the value of $Q_{\rm typ}$ diverges in the limit $\mu\to 0$ in any internal dimension $0<d<4$ which implies that the rescaled variance in Eq. \eqref{var_res_1} vanishes.  This in turn implies that the ground-state energy density "super-concentrates", similarly to what is observed for spherical spin-models at zero magnetic field \cite{ChenSen2017}, and the central limit theorem breaks down in that limit. Our results indicate that in the case $\mu \to 0$ the limiting distribution 
 takes a different form for different patterns of replica symmetry breaking. Based on results derived below, we expect:
\begin{enumerate}
    \item For a 1RSB pattern, we conjecture the limiting distribution
\be
\lim_{NL^d\to \infty}{\rm Prob}\left[{\cal E}_{N,L}-\moy{{\cal E}_{N,L}}\leq x\right]={\cal P}_{d,\rm 1RSB}(x)\;.
\ee
The energy fluctuations are thus of order $O(1)$. This limiting distribution displays an exponential left tail 
\be
0<\Sigma_{\min}'(e_{\rm typ},0)=\lim_{x\to -\infty}\frac{1}{x}\ln{\cal P}_{d,\rm 1RSB}(x)<\infty\;,\label{exp-tail}
\ee
where $\Sigma_{\min}(e,\mu)$ is the annealed complexity of minima at energy density $e$ and for mass $\mu$ and $e_{\rm typ}=\lim_{NL^d\to \infty}{\cal E}_{N,L}/(NL^d)$.

\item In contrast, when the solution to Parisi optimization problem is controlled by FRSB pattern, we conjecture the following limiting distribution of typical fluctuations
\be
\lim_{NL^d\to \infty}{\rm Prob}\left[\frac{{\cal E}_{N,L}-\moy{{\cal E}_{N,L}}}{(NL^d)^{\chi/d}}\leq
x\right]={\cal P}_{d,\rm FRSB}(x)\;,
\ee
where $0\leq \chi<2$ is the exponent of the variance computed by Mezard and Parisi (which depends on the asymptotic behaviour of the covariance function $f(q)$). The energy fluctuations in this case are thus of order $\sim (NL^d)^{\chi/d}$ with $0<\chi/d<1/2$. Additionally, we conjecture that the limiting distribution displays the left tail behaviour
\be
-\ln{\cal P}_{d,\rm FRSB}(x)\approx r_d(-x)^{\xi}\;,\;\;x\to -\infty\;,\;\;{\rm with}\;\;\xi=\frac{d}{d-\chi}\;,\label{left-tail-FRSB}
\ee
and $r_d$ is a constant that can be computed explicitly. The exponent $\xi$ corresponds to that of the large deviation function in the vicinity of the typical value, i.e. 
\be
{\cal L}(\epsilon)=-\lim_{NL^d\to \infty}\frac{1}{NL^d}\ln{\rm Prob}\left[\frac{{\cal E}_{N,L}-\moy{{\cal E}_{N,L}}}{NL^d}\leq  \epsilon\right]=\lim_{L\to \infty}\Lambda_L(e_{L,\rm typ}(0)+\epsilon,0)\;,\label{LD-regime-res}
\ee
where the function ${\cal L}(\epsilon)\approx r_d(-\epsilon)^{\xi}$ as $\epsilon\to 0_-$. This exponent can be recovered by a simple matching argument between the tail of the regime of typical fluctuations described by Eq. \eqref{left-tail-FRSB} and the large deviation regime in Eq. \eqref{LD-regime-res}, expressing both the rescaled variables $x$ and $\epsilon$ in terms of ${\cal E}_{N,L}-\moy{{\cal E}_{N,L}}$ and matching the powers of $NL^d$ in both equations.
\end{enumerate}

The behaviour described above is reminiscent of what has been observed for the fluctuations of the spherical spin model in \cite{lacroix2024replica}. It was shown there that under the FRSB ansatz the ground-state fluctuations display a non-trivial exponent while such an exponent is always equal to  unity within the 1RSB ansatz. Using an explicit approach to derive the large deviation function ${\cal L}(\epsilon)$ in the following sections, we determine independently the FRSB value of $\xi$ in Eq. \eqref{left-tail-FRSB} and of the exponential tail (i.e.  an exponent $\xi=1$) in the 1RSB pattern in Eq. \eqref{exp-tail}. Let us now provide more details on our results for the cumulant generating function and large deviation function of the ground-state energy.

\subsection{General expression for the cumulant generating function}

The scaled cumulant generating function (CGF) of the ground-state energy $e_{\min}$ 
is expressed as
\be
\varphi_L(s,\mu)=\lim_{N\to \infty}\frac{1}{N L^d}\ln\moy{e^{-s N L^d e_{\min}}}=\begin{cases}
\displaystyle\max_{g({\bf k},z),G_0({\bf k}),v({\bf k}),w_M}{\cal S}_{0,L}(\mu)&\;,\;\;s<0\;,\\
0&\;,\;\;s=0\;,\\
\displaystyle\min_{g({\bf k},z),G_0({\bf k}),v({\bf k}),w_M}{\cal S}_{0,L}(\mu)&\;,\;\;s>0\;,
\end{cases}
\ee
where the functional ${\cal S}_{0,L}(\mu)$ reads
\begin{align}
    {\cal S}_{0,L}(\mu)=&-\frac{s}{2}\int_{\bf k}\nu({\bf k})\left[g({\bf k},s)+G_0({\bf k})\right]+\frac{s}{2}\left[-f'(0)\int_{\bf k}v({\bf k})+w_M\,f\left(0\right)-\int_{s}^{w_M}dz\,f\left(\int_{\bf k}g({\bf k},z)\right)\right]\label{S_action_res}\\
    &+\frac{1}{2}\int_{\bf k}\left[-\frac{s}{w_M}\ln v({\bf k})+\ln\left(\lambda({\bf k},s)+s G_0({\bf k})\right)-s\int_{s}^{w_M}\frac{dz}{z^2}\ln\lambda({\bf k},z)\right]\;,\nn\\
    {\rm with}&\;\;\lambda({\bf k},z)=v({\bf k})+z\,g({\bf k},z)+\int_{z}^{w_M}dw\,g({\bf k},w)\;.
\end{align}
Here the (breaking point) parameter $w_M$ satisfies $w_M\geq s$, the functions $v({\bf k})\geq 0$ and $G_0({\bf k})\geq 0$ are positive and the function $g({\bf k},z)$ is a positive non-increasing function defined over the interval $z\in[s,w_M]$. Optimising over $w_M$ yields the explicit boundary condition $g({\bf k},z\geq w_M)=0$. The expression above is valid for any pattern of replica-symmetry breaking (replica-symmetric ($w_M=s$), one-step replica symmetry breaking, full replica-symmetry breaking).

{\bf Remark:} For $s\neq 0$, the expression of the cumulant generating function results from the properties of optimal atypical configurations. The values of the different parameters $w_M,v({\bf k}),G_0({\bf k}),g({\bf k},z)$ in the optimisation problem above describe the statistical properties of these optimal atypical configurations (see section \ref{sec-inter} for a detailed discussion).

As we show in appendix \ref{app_comp}, using an exact result relying on the Kac-Rice approach, the cumulant generating function is bounded from above as
\be
\varphi_L(s,\mu)\leq \frac{s^2 f(0)}{2}+\frac{1}{2}\int_{\bf k}\ln\left(1-\frac{s f'(0)}{\nu({\bf k})}\right)+\Sigma_{\min}(\mu-s f'(0))\;,
\ee
where $\Sigma_{\min}(\mu)$ is the annealed complexity of energy functional minima for the mass parameter $\mu$. For $\mu-s f'(1)>\mu_c$ one finds $\Sigma_{\min}(\mu-s f'(1))=0$ and the term on the right-hand side is shown to coincide with the replica-symmetric (RS) expression of the cumulant generating function, corresponding to $w_M=s$. 

The value for the average/typical centered ground state energy can be obtained from the general expression for the CGF above as 
\be
e_{L,\rm typ}(\mu)=-\partial_s\varphi_L(0,\mu)=\displaystyle\max_{g({\bf k},z),G_0({\bf k}),v({\bf k}),w_M}{\cal S}_{L,\rm typ}(\mu)\label{e_typ_res}
\ee
where the expression of ${\cal S}_{\rm typ}$ is obtained as
\begin{align}
    {\cal S}_{L,\rm typ}(\mu)=-\lim_{s\to 0}\frac{{\cal S}_{0,L}(\mu)}{s}=&\frac{1}{2}\int_{\bf k}\nu({\bf k})g({\bf k},0)+\frac{1}{2}\left[f'(0)\int_{\bf k}v({\bf k})-w_M\,f\left(0\right)+\int_{0}^{w_M}dz\,f\left(\int_{\bf k}g({\bf k},z)\right)\right]\label{S_0}\\
    &+\frac{1}{2}\int_{\bf k}\left[\frac{1}{w_M}\ln \frac{v({\bf k})}{\lambda({\bf k},0)}+\int_{0}^{w_M}\frac{dz}{z^2}\ln\frac{\lambda({\bf k},z)}{\lambda({\bf k},0)}\right]+\frac{1}{2}\int_{\bf k}G_0({\bf k})\left[\nu({\bf k})-\frac{1}{\lambda({\bf k},0)}\right]\;.\nn
\end{align}
In the following, we use $g_{\rm typ}({\bf k},z)$ and $w_{\rm typ}$ as the values of $g({\bf k},z)$ and $w_{k}$ optimising Eq. \eqref{e_typ_res}. The expression obtained from Eq. \eqref{e_typ_res} corresponds to the limit of zero temperature of the free-energy derived using the replica method in \cite{mezard1991replica} and has recently been obtained rigorously in \cite{arous2024free} (although for the latter the equivalence between the formulae is much less obvious). 

The expression for the cumulant generating function obtained above allows to derive higher-order cumulants and, in particular, the variance of the rescaled ground state energy.
The latter is defined as 
\begin{align}
{\cal V}_{L}(\mu)=&\lim_{N \to \infty}\frac{{\rm Var}({\cal E}_{N,L})}{N L^d}=f\left(Q_{\rm typ}\right)-\frac{I_2(\mu) f'\left(Q_{\rm typ}\right)^2}{2}\;,\;\;
{\rm where}\;\;Q_{\rm typ}=\int_{\bf k}g_{\rm typ}({\bf k},0)\;.\label{var_res}
\end{align}
The integral $I_2(x)$ is defined in Eq. \eqref{I_p}. The expression in Eq. \eqref{var_res} is valid at any level of RSB while the expression of $Q_{\rm typ}$ depends instead on the level of replica symmetry breaking. In the simplest, replica-symmetric (RS) phase $Q_{\rm typ}=0$. In the full replica symmetry broken (FRSB) phase it reads instead $Q_{\rm typ}={f''}^{-1}\left(1/I_2(\mu)\right)$, i.e. it is expressed from the functional inverse of the second derivative of the covariance function. Finally, in the one-step replica symmetry broken (1RSB) phase it is obtained as the non-zero solution of the following set of self-consistent equations (eliminating the breaking-point parameter $w_{\rm typ}$ in the following equations)
\begin{align}
Q_{\rm typ}&=\int_{\bf k}\frac{1}{w_{\rm typ}}\left(\frac{1}{\nu({\bf k})}-\frac{1}{\nu({\bf k})+w_{\rm typ}(f'\left(Q_{\rm typ}\right)-f'(0))}\right)\;,\label{q-typ-1rsb1}\\
f\left(Q_{\rm typ}\right)-f(0)-Q_{\rm typ} f'\left(Q_{\rm typ}\right)&=\frac{1}{w_{\rm typ}^2}\int_{\bf k}\left[\frac{w_{\rm typ}(f'\left(Q_{\rm typ}\right)-f'(0))}{\nu({\bf k})+w_{\rm typ}(f'\left(Q_{\rm typ}\right)-f'(0))}-\ln\left(1+\frac{w_{\rm typ}(f'\left(Q_{\rm typ}\right)-f'(0))}{\nu({\bf k})}\right)\right]\;.\label{q-typ-1rsb2}
\end{align}
The variance of the ground state energy then can be computed as a function of the mass $\mu$ and we find that it undergoes a second order transition at the Larkin mass $\mu_{c}$ corresponding to the emergence of replica-symmetry breaking in the typical ground state energy (see Eqs. \eqref{var_trans_1rsb} and \eqref{var_trans_full}).


For short-range correlation functions, using that $f(q)=\int_0^{\infty} dk\,e^{-q k^2}\,\tilde f(k)\;,$
and $\tilde f(k)\geq 0$ for any $k\geq 0$, one can show explicitly that 
\begin{align}
&2f''(q)f(q)-f'(q)^2=\int_0^{\infty}\int_0^{\infty}dk_1 dk_2\,e^{-q(k_1^2+k_2^2)}\tilde f(k_1)\tilde f(k_2)\left(k_1^4+k_2^4-k_1^2 k_2^2\right)\\
&\geq\int_0^{\infty}\int_0^{\infty} dk_1 dk_2\,e^{-q(k_1^2+k_2^2)} \tilde f(k_1)\tilde f(k_2)\left(k_1^4+k_2^4-2k_1^2 k_2^2\right)=\int_0^{\infty}\int_0^{\infty} dk_1 dk_2 \,e^{-q(k_1^2+k_2^2)}\tilde f(k_1)\tilde f(k_2)\left(k_1^2-k_2^2\right)^2\geq 0\;.  \nn  
\end{align}
The variance is positive in the FRSB phase where ${\cal V}_L(\mu)=f(Q_{\rm typ})-f'(Q_{\rm typ})^2/(2f''(Q_{\rm typ}))$ and in the RS phase where ${\cal V}_L(\mu)=f(0)-I_2(\mu)f'(0)^2/2$ with $I_2(\mu)\leq 1/f''(0)$ as required from the de-Almeida-Thouless criteria on stability of the RS ansatz.

To compare our expression for the variance of the ground state energy with the results obtained by Mezard and Parisi, let us consider an explicit expression for asymptotic behaviour of the short-range correlation function
\be
f(q)\sim -\frac{g}{1-\gamma}q^{1-\gamma}\;,\;\;q\gg 1\;,\;\;\gamma\geq 1\;,\label{asymp_MP}
\ee
with $g>0$. We will be interested in the limit of large $L$ and small $\mu$ where both are finite while the limit $a\to 0$ can be taken explicitly. Using the behaviour in Eq. \eqref{nu_k_small}, we can replace $\nu({\bf k})=\mu+{\bf k}^2$ and introduce a finite circular lower cut-off $(2\pi)/L$ in the ${\bf k}$ integrals. Let us treat separately the case (i) $d<2$ and $\gamma>\gamma_c=2/(2-d)$ for which the value of $Q_{\rm typ}$ is obtained from the 1RSB pattern and the case (ii) $d\geq 2$ or $d<2$ and $\gamma < \gamma_c=2/(2-d)$ for which it is obtained from the FRSB pattern.

(i) In the 1RSB pattern, the expression of $Q_{\rm typ}$ is obtained from Eqs. (\ref{q-typ-1rsb1}-\ref{q-typ-1rsb2}). Analysing these equations in the limit $L\to \infty$ and $
\mu\to 0$ with a fixed dimensionless parameter $x=\mu L^2$, one obtains the scaling form
\be
Q_{\rm typ}\approx L^{2\zeta_{\rm 1RSB}}{\cal Q}_{\rm 1RSB}(\mu L^2)\;,\;\;\zeta_{\rm 1RSB}=\frac{2-d}{2}>0\;,\label{Q-typ-L-large}
\ee
where the scaling function
\be
{\cal Q}_{\rm 1RSB}(x)=\frac{1}{w_{\rm typ}}\frac{2\pi^{d/2}}{2(2\pi)^2\Gamma(d/2)}\left(-\frac{x}{4\pi^2}\right)^{\frac{d-2}{2}}B\left(-\frac{x}{(4\pi)^2};\frac{2-d}{2},0\right)\;,\;\;w_{\rm typ}=-\frac{1}{f'(0)}\left(\frac{2d\sin(\frac{d\pi}{2})}{\pi(2-d)}\frac{f(0)}{f'(0)^2}\right)^{-\frac{2}{4-d}}\;,
\ee
and $B(x;a,b)$ is the incomplete beta function. This result recovers the exponent derived by Mezard and Parisi \cite{mezard1991replica}. The scaling function behaves asymptotically as
\be
{\cal Q}_{\rm 1RSB}(x)\approx \begin{cases}\displaystyle \frac{\pi^{\frac{d}{2}-2}}{2(2-d)\Gamma\left(\frac{d}{2}\right)w_{\rm typ}}&\;,\;\;x\ll 1\;,\\
&\\
\displaystyle 
\frac{1}{2^d\pi^{\frac{d}{2}}\Gamma\left(\frac{2-d}{2}\right)w_{\rm typ}}x^{-\zeta_{\rm 1RSB}}&\;,\;\;x\gg 1\;.
\end{cases}
\ee
Inserting the value of $Q_{\rm typ}$ in Eq. \eqref{var_res}, the variance of ${\cal E}_{N,L}$ can be computed in the scaling limit, yielding
\be
{\rm Var}({\cal E}_{N,L})\approx NL^d{\cal V}_L(\mu)=L^{2\chi_{\rm 1RSB}}\,V_{\rm 1RSB}(\mu L^2)\;,\;\;{\rm with}\;\;\chi_{\rm 1RSB}=1-\frac{2-d}{2}\gamma=\frac{\gamma_c-\gamma}{\gamma_c}\leq 0\;,\label{var_L_1rsb}
\ee
and the scaling function
\be
V_{\rm 1RSB}(x)=\frac{g}{\gamma-1}\left({\cal Q}_{\rm 1RSB}(x)\right)^{1-\gamma}\;,
\ee
so that at large $L$ one has ${\rm Var}({\cal E}_{N,L}) \sim L^{2\chi_{\rm 1RSB}}$ at vanishingly small mass, i.e. $\mu L^2  \to 0$, while
${\rm Var}({\cal E}_{N,L}) \sim \mu^{- \tilde \chi} L^d$ with $\tilde \chi=(2-d)(1-\gamma)/2$ for fixed $\mu$. Interestingly, both the exponent $\chi$ for the ground state energy variance  and $\tilde \chi=\chi-d/2$ for the rescaled variance  are negative in the 1RSB regime. The result for the exponent $\chi$ in the 1RSB phase is, to the best of our knowledge, new, and it is particularly interesting to note that the relation $\chi=2\zeta+d-2$, which would provide $\chi=0$ is not valid in that phase.

(ii) Under the assumption of  validity of FRSB pattern the value of $Q_{\rm typ}$ satisfies $Q_{\rm typ}={f''}^{-1}\left(1/I_2(\mu)\right)$. In the limit $L\to \infty$ and $
\mu\to 0$ with a fixed dimensionless parameter $x=\mu L^2$, one obtains the scaling form
\be
Q_{\rm typ}\approx L^{2\zeta_{\rm FRSB}}{\cal Q}_{\rm FRSB}(\mu L^2)\;,\;\;\zeta_{\rm FRSB}=\frac{4-d}{2(1+\gamma)}>0\;,
\ee
where the scaling function
\be
{\cal Q}_{\rm FRSB}(x)=\left(\frac{1}{g\gamma}\frac{2\pi^{d/2}}{(2\pi)^4\Gamma(d/2)}\left[\frac{1}{2(1+\frac{x}{4\pi^2})}+\frac{d-2}{4}\left(-\frac{x}{4\pi^2}\right)^{\frac{d-4}{2}}B\left(-\frac{x}{(4\pi)^2};\frac{4-d}{2},0\right)\right]\right)^{\frac{1}{1+\gamma}}\;.
\ee
This result recovers again the scaling exponent predicted by Mezard and Parisi \cite{mezard1991replica}. The scaling function behaves asymptotically as
\be
{\cal Q}_{\rm FRSB}(x)\approx \begin{cases}\displaystyle \left(\frac{\pi^{\frac{d}{2}-4}}{8g\gamma (4-d)\Gamma\left(\frac{d}{2}\right)}\right)^{\frac{1}{1+\gamma}}&\;,\;\;x\ll 1\;,\\
&\\
\displaystyle 
 \left(-\frac{1}{g\gamma 2^d\pi^{\frac{d}{2}-1}\Gamma\left(\frac{d-2}{2}\right)\sin\left(\frac{\pi d}{2}\right)}\right)^{\frac{1}{1+\gamma}}x^{-\zeta_{\rm FRSB}}&\;,\;\;x\gg 1\;,
\end{cases}
\ee
where $B_d=-2^{d}\pi^{\frac{d}{2}-1}\Gamma\left(\frac{d}{2}-1\right)\sin\left(\frac{d\pi}{2}\right)$ appearing in the second line of the equation above takes the finite value $B_2=4\pi$ in dimension $d=2$. Inserting this scaling form in Eq. \eqref{var_res}, one can compute the scaling form for the variance as
\be
{\rm Var}({\cal E}_{N,L})\approx NL^d{\cal V}_L(\mu)=N L^{2\chi_{\rm FRSB}}\,V_{\rm FRSB}(\mu L^2)\;,\;\;\chi_{\rm FRSB}=2\zeta_{\rm FRSB}+d-2=\frac{2+\gamma(d-2)}{1+\gamma}\geq 0\;,\label{var_L_res}
\ee
where the expression of the scaling function $V_{\rm FRSB}(x)$ reads
\be
V_{\rm FRSB}(x)=\frac{g(1+\gamma)}{2\gamma(\gamma-1)}\left({\cal Q}_{\rm FRSB}(x)\right)^{1-\gamma}\;,\;\;1<\gamma\leq \gamma_c\;,\label{v0_1}
\ee
so that at large $L$ one has ${\rm Var}({\cal E}_{N,L}) \sim L^{2\chi_{\rm FRSB}}$ at vanishingly small mass, i.e. $\mu L^2  \to 0$, while
${\rm Var}({\cal E}_{N,L}) \sim \mu^{- \tilde \chi} L^d$ with $\tilde \chi=(4-d)(1-\gamma)/2/(1+\gamma)$ for fixed $\mu$. The exponent for the variance $\chi$ is positive in the FRSB regime while the exponent for the rescaled variance $\tilde \chi=\chi-d/2=2(1-\gamma)\zeta_{\rm FRSB}$ is negative. 

\begin{figure}
    \centering
    \includegraphics[width=0.7\textwidth]{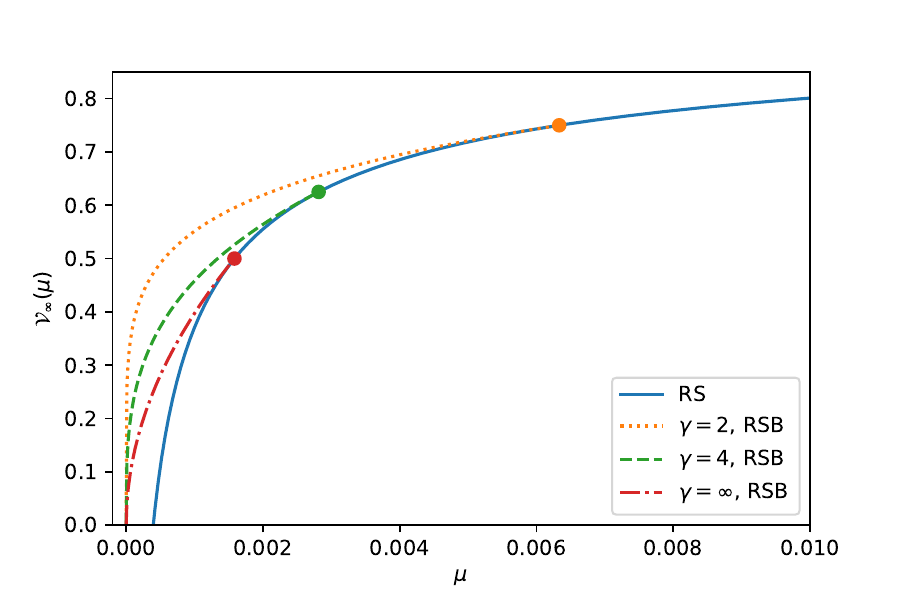}
    \caption{Plot of the rescaled variance ${\cal V}_{\infty}(\mu)$ defined in Eq. \eqref{var_res} as a function of the mass $\mu$ for the elastic manifold with internal dimension $d=3$, in the continuous limit $L\to\infty$ where $\nu({\bf k})=\mu+{\bf k}^2$ and for a short-range covariance of the disorder of the form $f_\gamma(q)=(1+q/(\gamma-1))^{1-\gamma}$ with $\gamma>1$ for $\gamma=2,4,\infty$. Below the RSB transition, i.e. for $\mu<\mu_c$ (with $\mu_c$ indicated by a dot), the variance is given by the FRSB expression (orange dotted, green dashed and red dashed dotted curves for $\gamma=2,4,\infty$ respectively) while it is given by the RS expression (solid blue curve) above the second order transition $\mu>\mu_c$.}
    \label{fig:var_GS}
\end{figure}

The exponent in Eq. \eqref{var_L_res} is in full agreement with the results of Mezard and Parisi \cite{mezard1991replica} using an approach relying on dimensional analysis and renormalisation group. Our approach relies instead on an explicit computation of the ground state energy variance and allows us to provide not only the value for the scaling exponent, but also a quantitative prediction for the associated scaling function. The criterion for the variance increasing with the system size, $\chi \geq 0$, is equivalent to the condition of validity of the FRSB ansatz (either $d \geq 2$ or $d<2$ and $\gamma < \gamma_c=2/(2-d) $).
Furthermore, using our results, one can compute the exact expression for the variance for any given correlation function $f(q)$. In Fig. \ref{fig:var_GS}, we plot the rescaled variance as a function of $\mu$ in dimension $d=3$, with elasticity term $\nu({\bf k})=\mu+{\bf k}^2$ in the continuous limit $L\to\infty$ and $a\to 0$ and for various short-range correlation functions. 

Our results show that for any value $\mu>0$, the rescaled variance ${\cal V}_L(\mu)$ in Eq. \eqref{var_res} is positive. Interestingly however as $\mu\to 0$ the value of $Q_{\rm typ}$ diverges in any internal dimension $0<d<4$ and our result for the rescaled variance in Eq. \eqref{var_res} indicates that the variance vanishes in that limit (see also Eq. \eqref{v0_1} and Fig. \ref{fig:var_GS}). 
If the value of $\chi$ is given by its FRSB expression, the exponent $\xi$ of the large deviation function reads
\be
\xi_{\rm FRSB}=\frac{d}{d-\chi_{\rm FRSB}}=\frac{d(1+\gamma)}{d+2(\gamma-1)}\;,\;d\geq 2\;\;{\rm or}\;\;0<d<2\;\;{\rm and}\;\;1<\gamma<\gamma_c=\frac{2}{2-d}\;.\label{xi}
\ee 
In any dimension $\xi_{\rm FRSB}$ is a decreasing function of $\gamma$ with $\xi_{\rm FRSB}\to 2$ as $\gamma\to 1$ and  $\xi_{\rm FRSB}\to 1$ as $\gamma\to \gamma_c$ in dimension $0<d<2$ and $\xi_{\rm FRSB}\to d/2$ as $\gamma\to \infty$ for $4>d\geq 2$.

We now provide detailed results on the large deviation function of the ground-state energy.


\subsection{Ground state energy large deviation function for short-ranged correlated disorder}

The LDF of the ground-state energy $e_{\min}$ is defined in Eq. \eqref{LD_emin} displays several branches reflecting different pattern of replica symmetry breaking.
In this section, we provide explicit results for the expression of these branches for short-ranged correlation functions $f(q)$. 
\subsubsection{Replica-symmetric branch}
The expression of the large deviation function within the replica-symmetric (RS) ansatz can be obtained either by applying the replica method or using the lower bound obtained from applying the Kac-Rice method. The RS branch can be expressed parametrically as 
\begin{align}
    \Lambda_{L,\rm RS}(e,\mu)=&\frac{s^2}{2}f\left(0\right)-\frac{1}{2}\int_{\bf k}\left[\frac{s f'(0)}{\nu({\bf k})-s f'(0)}+\ln\left(1-\frac{s f'(0)}{\nu({\bf k})}\right)\right]\;,\label{LDF_RS_res}\\
    {\rm where}\;\;e=&-s f(0)+\frac{f'(0)}{2}\int_{\bf k}\frac{1}{\nu({\bf k})-s f'(0)}\;.\label{E_RS_res}
\end{align}
This RS solution is only stable for $s\geq s_{\rm AT}=(\mu-\mu_c)/f'(0)$ or conversely $e\leq e_{\rm AT}$ obtained by inserting the value of $s_{\rm AT}$ in the expression for $\epsilon$ above. The value of $s_{\rm AT}$ is found to satisfy 
\be
f''(0)\int_{\bf k}\frac{1}{(\nu({\bf k})-s_{\rm AT}f'(0))^2}=f''(0)I_2(\mu-s_{\rm AT}f'(0))=1\;,\;\;{\rm i.e.}\;\;s_{\rm AT}=\frac{\mu-\mu_c}{f'(0)}\;,\label{s_AT_res}
\ee
where the last identity stems from $\mu_c$ being the (unique) solution to $f''(0)I_2(\mu_c)=1$. The corresponding energy value can be computed as
\be
e_{\rm AT}=(\mu_c-\mu)\frac{f(0)}{f'(0)}+\frac{f'(0)}{2}I_1(\mu_c)\;.\label{e_at}
\ee
Using that for a short range correlation function $f(0)f''(0)-f'(0)^2/2>0$, the CGF is shown to be convex in the regime of stability of the RS ansatz $s\geq s_{\rm AT}$. The RS branch of the LDF is similarly convex in the region $e\leq e_{\rm AT}$. For $\mu>\mu_c$, one has that $s_{\rm AT}<0$ and the typical ground-state energy takes its replica-symmetric expression $e_{L,\rm typ}(\mu)=f'(0)I_1(\mu)/2$. The behaviour of the LDF in the vicinity of its typical value can be investigated using Eqs. (\ref{LDF_RS_res}-\ref{E_RS_res}), yielding to the second order
\begin{align}
 \Lambda_{L,\rm RS}(e,\mu)&=\frac{\displaystyle\left(e-e_{L,\rm RS}(\mu)\right)^2}{\displaystyle 2{\cal V}_L(\mu)}+O\left(e-e_{L,\rm RS}(\mu)\right)^3\;,\\
 e_{L,\rm RS}(\mu)&=\frac{f'(0)I_1(\mu)}{2}\;,\;\;{\cal V}_L(\mu)=f(0)-I_2(\mu)\frac{f'(0)^2}{2}\;.
\end{align}
The denominator in this expression can be identified as the RS expression of the rescaled variance in Eq. \eqref{res_var}. The quadratic behaviour of the LDF is consistent with the central limit theorem describing the typical fluctuations of $e_{\min}$. The asymptotic behaviour of the large deviation function as $e\to -\infty$ is universal and reads
\be
\Lambda_{L,\rm RS}(e,\mu)\approx\frac{e^2}{2f(0)}\;,\;\;e\to -\infty\;.\label{e_large_neg}
\ee
Let us now consider the results obtained for branches with replica-symmetry breaking, starting with the case of one-step replica-symmetry breaking.


\subsubsection{One step replica symmetry broken branch}

The 1RSB branch of the large deviation function can be obtained parametrically as well using the identities
\begin{align}
 \Lambda_{L,\rm 1RSB}(e,\mu)&=\frac{s^2}{2}f\left(Q\right)-\frac{1}{2}\int_{\bf k}\left[\ln\left(1-\frac{s f'(Q)}{\nu({\bf k})}\right)+\frac{s f'(Q)}{\nu({\bf k})-s f'(Q)}\right]\;,\label{LDF_1RSB_res}\\
{\rm where}\;\;e&=-w f(0)-(s-w)\left(f(Q)-\frac{Q f'(Q)}{2}\right)+\frac{f'(0)}{2}\int_{\bf k}\frac{1}{\nu({\bf k})-s f'(Q)+w(f'(Q)-f'(0))}\label{E_1RSB_res}\;.
\end{align}
The parameters $w$ and $Q$ are obtained for fixed $e$ (or equivalently fixed $s$) as the solutions to the following equations 
\begin{align}
    Q&=\frac{1}{w}\int_{\bf k}\left[\frac{1}{\nu({\bf k})-s f'(Q)}-\frac{1}{\nu({\bf k})-s f'(Q)+w(f'(Q)-f'(0))}\right]\;,\label{Q_w_eq_1}\\
    f(0)-f(Q)+Q f'(Q)&=\frac{1}{w^2}\int_{\bf k}\left[\ln\left(1+\frac{w(f'(Q)-f'(0))}{\nu({\bf k})-s f'(Q)}\right)-\frac{w(f'(Q)-f'(0))}{\nu({\bf k})-s f'(Q)+w(f'(Q)-f'(0))}\right]\;.\label{Q_w_eq_2}
\end{align}
Note that $Q=0$ and $w=s$ is a trivial solution of the above equations and inserting these values Eqs. (\ref{LDF_1RSB_res}-\ref{E_1RSB_res}), the parametric expression for the RS branch of the large deviation function in Eqs. (\ref{LDF_RS_res}-\ref{E_RS_res}) is recovered. Thus, the 1RSB branch can only be obtained from the non-trivial solution of Eqs. (\ref{Q_w_eq_1}-\ref{Q_w_eq_2}). 

For a mass $\mu<\mu_{\rm dis}$, where the value of $\mu_{\rm dis}$ reads
\be
\mu_{\rm dis}=\mu_c-\frac{f'(0)f^{(3)}(0)}{2I_3(\mu_c)f''(0)^3}\;,
\ee
the transition from the RS to the 1RSB branch of the CGF occurs as $w\to s=s_{\rm dis}$ and the parameter $Q$ is discontinuous at the transition taking the value $0$ within the RS phase and $Q_{\rm dis}>0$ in the 1RSB phase. The parameters $s_{\rm dis},Q_{\rm dis}$ of the transition are obtained by solving the following equations
\begin{align}
    Q_{\rm dis}&=\frac{1}{s_{\rm dis}}\int_{\bf k}\left[\frac{1}{\nu({\bf k})-s_{\rm dis} f'(Q_{\rm dis})}-\frac{1}{\nu({\bf k})-s_{\rm dis} f'(0)}\right]\;,\label{Q_s_eq_1}\\
    f(0)-f(Q_{\rm dis})+Q_{\rm dis} f'(Q_{\rm dis})&=\frac{1}{s_{\rm dis}^2}\int_{\bf k}\left[\ln\frac{\nu({\bf k})-s_{\rm dis} f'(0)}{\nu({\bf k})-s_{\rm dis} f'(Q_{\rm dis})}-\frac{s_{\rm dis}(f'(Q_{\rm dis})-f'(0))}{\nu({\bf k})-s_{\rm dis} f'(0)}\right]\;.\label{Q_s_eq_2}
\end{align}
The corresponding transition in the large deviation function occurs at the energy
\be
e_{\rm dis}=-s_{\rm dis} f(0)+\frac{f'(0)}{2}\int_{\bf k}\frac{1}{\nu({\bf k})-s_{\rm dis} f'(0)}\;,
\ee
is of second order and takes the form
\be
\Lambda_{L,\rm 1RSB}(e,\mu)-\Lambda_{L,\rm RS}(e,\mu)=\frac{1}{2}\left(\frac{1}{\partial_s^2\varphi_{L,\rm 1RSB}(s_{\rm dis},\mu)}-\frac{1}{\partial_s^2\varphi_{L,\rm RS}(s_{\rm dis},\mu)}\right)\left(e-e_{\rm dis}\right)^2+O(e-e_{\rm dis})^3\;,
\ee
where the values of $\partial_s^2\varphi_{L,\rm 1RSB}(s_{\rm dis},\mu)$ and $\partial_s^2\varphi_{L,\rm RS}(s_{\rm dis},\mu)$ are given explicitly in Eq. \eqref{dis_trans}.

For a mass $\mu>\mu_{\rm dis}$, the transition for the CGF occurs as $Q\to 0$ vanishes continuously, corresponding to $s\to s_{\rm AT}$, while $w$ takes the value $w_{\rm AT}=-f^{(3)}(0)/(2(f''(0))^3 I_3(\mu_c))\geq s_{\rm AT}$ at the transition. The corresponding transition for the large deviation function occurs at the energy $e_{\rm AT}$ defined in Eq. \eqref{e_at}

\begin{figure}
    \centering
    \includegraphics[width=0.7\textwidth]{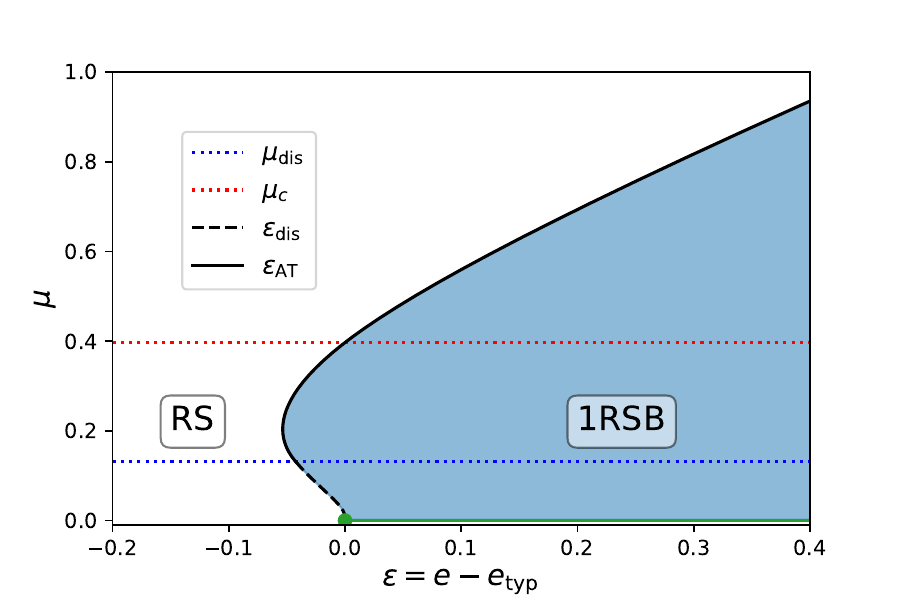}
    \caption{Phase diagram of the large deviation function ${\cal L}(\epsilon,\mu)=\lim_{L\to \infty}\Lambda_L(e_{\rm typ}+\epsilon,\mu)$ for the centred ground state energy $\epsilon_{\min}=e_{\min}-e_{\rm typ}$ of the elastic manifold $d=1$ with exponential correlation function $f_{\infty}(q)=\exp(-q)$ (see section \ref{exp-1d} for details). For any fixed value of $\mu$, the RS phase (white) and 1RSB phase (blue) are separated by a transition line. The transition occurs at $\epsilon=\epsilon_{\rm dis}$ (dashed black curve) for $\mu\leq \mu_{\rm dis}$ (indicated by the horizontal dotted blue line) and is of second order while for $\mu\geq \mu_{\rm dis}$ it is of third order and occurs at $\epsilon_{\rm AT}$ (solid black curve). The typical value of the centred energy $e_{\rm typ}$, where the LDF achieves its unique zero, corresponds to the vertical line $\epsilon=0$. For $\mu\geq \mu_{c}$ (indicated by the horizontal dotted red line), it is given by its RS expression while for $\mu<\mu_{c}$ it is given by its 1RSB expression. For $\mu=0$, one expects that the large deviation function of speed $N$ is only given by its RS expression and describes any 
    $\epsilon\leq 0$. On the other hand, the fluctuations for $\epsilon\geq 0$ (indicated by the green line) are described by a large deviation regime with higher speed.}
    \label{fig:phasediag-1rsb}
\end{figure}

The continuous transition occurring as $Q\to 0$, i.e. as the trivial and non-trivial solutions of the above equation match, occurs precisely at energy $e=e_{\rm AT}$. This continuous transition leads to a third order transition in the large deviation function 
\be
\Lambda_{L,\rm 1RSB}(e,\mu)-\Lambda_{L,\rm RS}(e,\mu)=I_3(\mu_c)\frac{f'(0)^3(\mu-\mu_c)}{6{\cal V}_L(\mu_c)^3(\mu-\mu_{\rm dis})}\left(e-e_{\rm AT}\right)^3+O(e-e_{\rm AT})^4\;,\label{third_order_1rsb}
\ee
where ${\cal V}_L(\mu_c)=f(0)-f'(0)^2/(2f''(0)$, we used the identity $\mu_c=\mu-s_{\rm AT} f'(0)$ and we remind that $I_p(x)$ is defined in Eq. \eqref{I_p}. In Fig. \ref{fig:phasediag-1rsb}, we show the phase diagram of the large deviation function, where the RS phase (in white) is separated from the 1RSB phase (in blue) by the transition line (black curve). For $\mu< \mu_{\rm dis}$, the transition is discontinuous (dashed curve), i.e. $Q_{\rm dis}>0$ at the transition while it is continuous (solid curve) for $\mu> \mu_{\rm dis}$.

In the vicinity of the typical value, the large deviation function vanishes quadratically
\begin{align}
 \Lambda_{L,\rm 1RSB}(e,\mu)&=\frac{\displaystyle\left(e-e_{L,\rm 1RSB}(\mu)\right)^2}{\displaystyle 2{\cal V}_L(\mu)}+O\left(e-e_{L,\rm 1RSB}(\mu)\right)^3\;,\\
 e_{L,\rm 1RSB}(\mu)&=\frac{f'(0)I_1(\mu)}{2}+w_{\rm typ}\left(f(Q_{\rm typ})-f(0)-\frac{Q_{\rm typ}}{2}(f'(Q_{\rm typ})+f'(0))\right)\;,\\
 {\cal V}_L(\mu)&=f(Q_{\rm typ})-I_2(\mu)\frac{f'(Q_{\rm typ})^2}{2}\;.
\end{align}
and the pair $Q_{\rm typ},w_{\rm typ}$ satisfy the integral equations
\begin{align}
    Q_{\rm typ}&=\frac{1}{w_{\rm typ}}\int_{\bf k}\left[\frac{1}{\nu({\bf k})}-\frac{1}{\nu({\bf k})+w_{\rm typ}(f'(Q_{\rm typ})-f'(0))}\right]\;,\\
    f(0)-f(Q_{\rm typ})+Q_{\rm typ} f'(Q_{\rm typ})&=\frac{1}{w_{\rm typ}^2}\int_{\bf k}\left[\ln\left(1+\frac{w_{\rm typ}(f'(Q_{\rm typ})-f'(0))}{\nu({\bf k})}\right)-\frac{w_{\rm typ}(f'(Q_{\rm typ})-f'(0))}{\nu({\bf k})+w_{\rm typ}(f'(Q_{\rm typ})-f'(0))}\right]\;.
\end{align}
This behaviour is compatible with a smooth matching between the large deviation regime and the tail of the Gaussian distribution of typical fluctuations.

Finally, the asymptotic behaviour of the large deviation function as $e\to +\infty$ can be obtained and shown to be non universal, in contrast to the behaviour as $e\to -\infty$ (see Eq. \eqref{e_large_neg}). A careful analysis shows that in this regime $Q\to \infty$ with the scaling $s=\mu/f'(Q)+O(1)$ and $e=\mu Q/2+O(1)$ while the parameter $w$ reaches an asymptotic value $w_{\infty}=O(1)$. The large deviation function thus behaves as 
\be
\Lambda_{L,\rm 1RSB}(e,\mu)\approx \frac{\displaystyle\mu^2 f\left(2e/\mu\right)}{\displaystyle 2f'\left(2e/\mu\right)^2}\;,\;\;e\to +\infty\;.\label{as_e_large}
\ee
This result concludes our results on the one-step replica symmetry broken branch. Next we describe our results for th full replica symmetry broken branch.

\subsubsection{Full replica symmetry broken branch}

\begin{figure}
    \centering
    \includegraphics[width=0.7\textwidth]{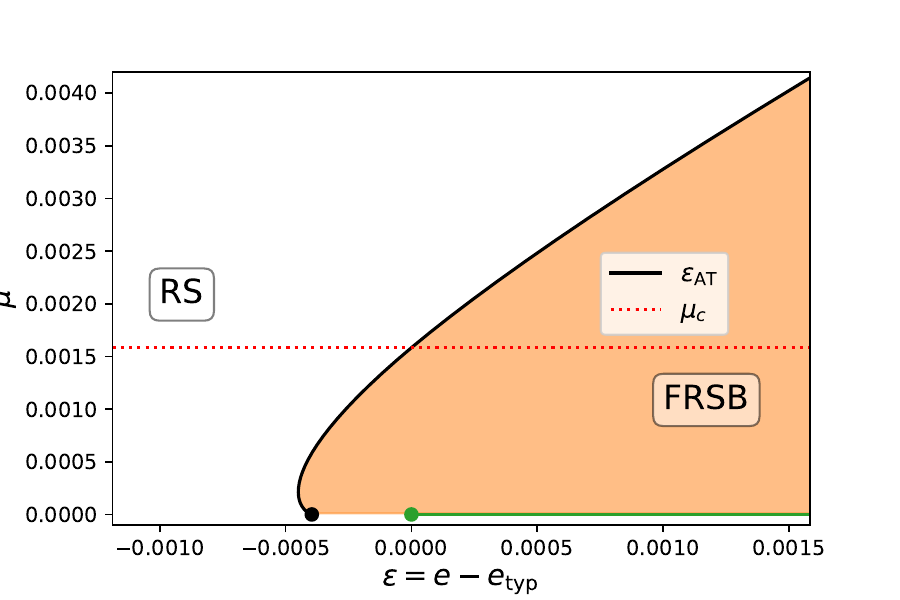}
    \caption{Phase diagram of the large deviation function ${\cal L}(\epsilon,\mu)=\lim_{L\to \infty}\Lambda_L(e_{\rm typ}+\epsilon,\mu)$ for the centred ground state energy $\epsilon_{\min}=e_{\min}-e_{\rm typ}$ of the elastic manifold of internal dimension $d=3$ with elasticity term $\nu({\bf k})=\mu+{\bf k}^2$ and exponential correlation function $f_{\infty}(q)=\exp(-q)$ (see section \ref{exp-3d} for details). For any fixed value of $\mu$, the RS phase (white) and FRSB phase (orange) are separated by a transition line. The transition occurs at $\epsilon=\epsilon_{\rm AT}$ (solid black curve) and is of third order. The typical value of the centred energy $\epsilon_{\rm typ}$, where the LDF achieves its unique zero, is represented via the solid red curve. For $\mu\geq \mu_{c}$ (indicated by the horizontal dotted red line), it is given by its RS expression while for $\mu<\mu_{c}$ it is given by its FRSB expression. For $\mu=0$, in addition to the RS phase (for energies below the black dot) and the FRSB phase (for energies between the blakc and green dot), an additional phase appears (represented by the solid green line) where the large deviation of speed $NL^d$ is infinite and a large deviation regime with higher speed describes the fluctuations.}
    \label{fig:phasediag-frsb}
\end{figure}

The FRSB branch of the large deviation function can be expressed parametrically as
\begin{align}
\Lambda_{L,\rm FRSB}(e,\mu)=&\frac{s^2}{2}f\left(Q\right)-\frac{1}{2}\int_{\bf k}\left[\ln\left(1-\frac{s f'(Q)}{\nu({\bf k})}\right)+\frac{s f'(Q)}{\nu({\bf k})-s f'(Q)}\right]\;,\;\;{\rm where}\;\;\int_{\bf k}\frac{f''(Q)}{(\nu({\bf k})-s f'(Q))^2}=1\;,\label{LDF_FRSB_res}\\
{\rm and}\;\;e=&-s f(0)-\frac{1}{2}\int_0^{Q}(z(q)-s)\left[q f''(q)-f'(q)\right]\label{E_FRSB_res}
+\frac{f'(0)}{2}\int_{\bf k}\frac{1}{\nu({\bf k})-s f'(Q)+\int_0^{Q}z(q)f''(q)dq}\;.
\end{align}
The function $z(q)$ appearing in the expression above satisfies
\be
\forall q\in [0,Q]\;,\;\;\int_{\bf k}\frac{f''(q)}{(\nu({\bf k})-s f'(Q)+\int_q^{Q}z(r)f''(r)dr)^2}=1\;.
\ee
The transition between the RS and FRSB branches occurs as $e\to e_{\rm AT}$ and is of third order
\be
\Lambda_{L,\rm FRSB}(e,\mu)-\Lambda_{L,\rm RS}(e,\mu)=I_3(\mu_c)\frac{f'(0)^3 s_{\rm AT}}{6{\cal V}_L(\mu_c)^3(s_{\rm AT}-z(0))}\left(e-e_{\rm AT}\right)^3+O(e-e_{\rm AT})^4\;.
\ee
In Fig. \ref{fig:phasediag-frsb}, we show the phase diagram of the large deviation function, where the RS phase (in white) is separated from the FRSB phase (in orange) by the transition line (black curve). The transition between a RS and FRSB phase is always continuous.

For any $\mu>0$, the large deviation function vanishes quadratically in the vicinity of the typical value with
\begin{align}
 \Lambda_{L,\rm 1RSB}(e,\mu)&=\frac{\displaystyle\left(e-e_{L,\rm FRSB}(\mu)\right)^2}{\displaystyle 2{\cal V}_L(\mu)}+O\left(e-e_{L,\rm 1RSB}(\mu)\right)^3\;,\\
 e_{L,\rm FRSB}(\mu)&=-\frac{1}{2}\int_0^{Q_{\rm typ}}dq\,z_{\rm typ}(q)\left[q f''(q)-f'(q)\right]+\frac{f'(0)}{2}\int_{\bf k}\frac{1}{\nu({\bf k})+\int_0^{Q_{\rm typ}}z_{\rm typ}(q)f''(q)dq}\;,\\
 {\cal V}_L(\mu)&=f(Q_{\rm typ})-I_2(\mu)\frac{f'(Q_{\rm typ})^2}{2}\;,\;\;{\rm where}\;\;Q_{\rm typ}={f''}^{-1}\left(\frac{1}{I_2(\mu)}\right)\;,\\
 {\rm and}\;\;\forall q\in [0,Q_{\rm typ}]&\;,\;\;\int_{\bf k}\frac{f''(q)}{(\nu({\bf k})+\int_q^{Q_{\rm typ}}z_{\rm typ}(r)f''(r)dr)^2}=1\;.
\end{align}
The expression above is fully consistent with the ground state energy $e_{\min}$ satisfying the central limit theorem.

Finally, the asymptotic behaviour of the large deviation function as $e\to +\infty$ can be explored. The asymptotic behaviour is exactly the same as for the 1RSB case in Eq. \eqref{as_e_large}. Note that this behaviour is qualitatively very different from the large deviation function of spherical spin-glasses. For the latter, as conjectured in \cite{lacroix2024replica} and subsequently proved rigorously in \cite{huang2023constructive}, the large deviation function with speed $N$ of the ground-state energy diverges at a finite value of energy indicating a transition towards a regime of large deviation with speed $N^2$. The presence of the harmonic confinement seems to prevent the occurrence of such transition for the elastic manifold. Next we will summarise the results in the limit of zero confinement, where that transition is instead expected to occur.

\subsection{Massless limit and superconcentration} \label{massless-sec}

In this section, we consider the large deviation of the centred ground-state energy $\epsilon_{\min}=e_{\min}-e_{\rm typ}$ in the massless limit $\mu\to 0$ and continuum limit $L\to \infty$. To this end, we study
\be
{\cal L}(\epsilon)=\lim_{L\to \infty}\Lambda_L(e_{\rm typ}(L,0)+\epsilon,0)\;.
\ee
To make our computation as explicit as possible, we will consider the following covariance function
\be
f_{\gamma}(q)=\left(1+\frac{q}{\gamma-1}\right)^{1-\gamma}\;,\;\;\gamma> 1\;.
\ee
Note that the asymptotic behaviour of this function is not exactly that of Eq. \eqref{asymp_MP} although quite similar but it has the advantage to have a well-defined limit $f_{\infty}(q)=e^{-q}$ as $\gamma\to \infty$.

\subsubsection{FRSB branch}

To prevent divergence of the expressions, we will work in the following with the variable $\epsilon=e-e_{\rm typ}$ which remains of order $O(1)$ in the limit $\mu\to 0$ in any dimension. Let us first consider the case $d\geq 2$ or $1\leq \gamma\leq 2/(d-2)$ where the large deviation function is given in the vicinity of its typical value by its FRSB branch. From the definition of $Q$ in that phase in Eq. \eqref{LDF_FRSB_res}, the value of $s$ can be expressed as a function of $Q$ in the limit $\mu\to 0$ as
\be
s=-\frac{1}{f'(Q)}\left(\frac{\displaystyle f''(Q)}{\displaystyle B_d}\right)^{\frac{2}{4-d}}\;,
\ee
where $B_d=-2^{d}\pi^{\frac{d}{2}-1}\Gamma\left(\frac{d}{2}-1\right)\sin\left(\frac{d\pi}{2}\right)$. Note that $B_2=4\pi$ is finite. Inserting this expression for $s$ in Eq. \eqref{LDF_FRSB_res}, the large deviation function and energy can then both be expressed parametrically in terms of the sole parameter $Q$ as
\be
{\cal L}_{\rm FRSB}(\epsilon)=\left(\frac{f''(Q)}{B_d}\right)^{\frac{4}{4-d}}\left(\frac{f(Q)}{2f'(Q)^2}-\frac{1}{d\,f''(Q)}\right)\;.
\ee
Using the explicit expression for the function $z(q)$ in Eq. \eqref{z_eq}, the energy can similarly be expressed as
\begin{align}
\epsilon
=&\left(\frac{\displaystyle f''(Q)}{\displaystyle B_d}\right)^{\frac{2}{4-d}}\left(\frac{f(Q)}{f'(Q)}-\frac{f'(Q)}{(d-2)f''(Q)}\right)-\frac{4-d}{2(d-2)}\int_{Q}^{\infty}dq\left(\frac{\displaystyle f''(q)}{\displaystyle B_d}\right)^{\frac{2}{4-d}}\;.
\end{align}
One may expect the expression above to diverge in $d=2$. However, the diverging terms cancel, yielding the finite expression
\be
\epsilon=\frac{1}{4\pi}\left[\frac{f(Q)f''(Q)}{f'(Q)}-\frac{f'(Q)}{2}-\frac{1}{2}\int_{Q}^{\infty}dq \frac{f'(q)f^{(3)}(q)}{f''(q)}\right]\;,\;\;d=2\;.
\ee

Finally, using these expressions for the explicit covariance function $f_{\gamma}(q)$, one obtains the following expression
\be
{\cal L}_{\rm FRSB}(\epsilon)=\frac{1}{\xi}\left(\left(\frac{\gamma}{B_d(\gamma-1)}\right)^{\frac{2}{4-d}}\frac{\xi(2+\gamma(d-2))}{2d\gamma}\right)^{\xi-1}(-\epsilon)^{\xi}\;,\;\;\xi=\frac{d(1+\gamma)}{d+2(\gamma-1)}\;.\label{xi_FRSB}
\ee
Note that this expression for the large deviation function is only valid in the FRSB phase, i.e. for energies $\epsilon$ larger than 
\be
\epsilon_{\rm AT}=\left(\frac{\displaystyle f''(0)}{\displaystyle B_d}\right)^{\frac{2}{4-d}}\left(\frac{f(0)}{f'(0)}-\frac{f'(0)}{(d-2)f''(0)}\right)-\frac{4-d}{2(d-2)}\int_{0}^{\infty}dq\left(\frac{\displaystyle f''(q)}{\displaystyle B_d}\right)^{\frac{2}{4-d}}\;.
\ee
For fluctuations with $\epsilon<\epsilon_{\rm AT}$, one needs to use the expression for the RS branch.

\subsubsection{1RSB branch}
The behaviour of the large deviation for $d<2$ and $\gamma\geq 2/(d-2)$ can be understood by analogy with the study of spherical spin glasses in \cite{lacroix2024replica}. It was found there that the full large deviation function is given at zero magnetic field by its RS branch while the 1RSB phase is limited to a point. One can check similarly here that the large deviation function undergoes a discontinuous transition at the value $e_{\rm dis}\equiv e_{\rm typ}$ from $Q=0$ in the RS phase to $Q\to\infty$ in the 1RSB phase. In the RS phase, the large deviation function can be obtained in the simple parametric form
\be
{\cal L}_{\rm RS}(\epsilon)=\frac{s^2}{2}f(0)-\frac{\displaystyle \left(-s f'(0)\right)^{\frac{d}{2}}}{\displaystyle d B_d}\;,
\ee
where the energy difference reads
\be
\epsilon=(w_{\rm typ}-s)f(0)-\frac{f'(0)}{(2-d)B_d}\left[\left(-w_{\rm typ}f'(0)\right)^{\frac{d}{2}-1}-\left(-sf'(0)\right)^{\frac{d}{2}-1}\right]\;.
\ee
Note that here as well, one may expect this expression to diverge in $d=2$, however the diverging terms cancel, yielding the finite expression
\be
\epsilon=(w_{\rm typ}-s)f(0)-\frac{f'(0)}{8\pi}\ln\frac{s}{w_{\rm typ}}\;,\;\;d=2\;.
\ee

Both the LDF ${\cal L}_{\rm RS}(\epsilon)$ and the expression of the energy difference $\epsilon$ must vanish as $s\to w_{\rm typ}$, which can simply be obtained from the expression of the LDF as
\be
w_{\rm typ}=\left(\frac{2 (-f'(0))^{\frac{d}{2}}}{d B_d f(0)}\right)^{\frac{2}{4-d}}\;.
\ee
Finally, one can check that the large deviation function vanishes linearly as the free energy difference vanishes, i.e.
\be
{\cal L}_{\rm RS}(\epsilon)=-w_{\rm typ}\epsilon+O(\epsilon)^2\;.\label{xi_1RSB}
\ee
As we show in appendix \ref{app_comp}, the RS expression for the LDF can be simply expressed in terms of the annealed complexity of minima $\Sigma_{\min}(e,\mu)$, i.e. 
\be
{\cal L}_{\rm RS}(\epsilon)=-\Sigma_{\min}(e_{\rm typ}+\epsilon,0)\;.
\ee
Thus, in the limit $\mu\to 0$ studied here, one can check that the typical energy $e_{\rm typ}$ coincides with the lowest argument value where the annealed complexity vanishes. This property was first obtained for the pure spherical $p$-spin in \cite{auffinger2013random} and confirmed more recently for arbitrary mixture in \cite{lacroix2024replica,huang2023constructive}.

\subsection{Long-range disorder}

In this section, we summarise the results obtained for the cumulant generating function of the ground-state energy difference $\Delta e_{\min}$. The latter reads
\be
\phi_L(s,\mu)=\lim_{NL^d\to \infty}\frac{1}{N L^d}\ln\moy{e^{-s N L^d \Delta\epsilon_{\min}}}=\begin{cases}
\displaystyle\max_{g({\bf k},z),G_0({\bf k}),v({\bf k}),w_M}{\cal S}_{1,L}(\mu)&\;,\;\;s<0\;,\\
0&\;,\;\;s=0\;,\\
\displaystyle\min_{g({\bf k},z),G_0({\bf k}),v({\bf k}),w_M}{\cal S}_{1,L}(\mu)&\;,\;\;s>0\;,
\end{cases}
\ee
where the functional ${\cal S}_{1,L}(\mu)$ reads
\begin{align}
    {\cal S}_{1,L}(\mu)=&-\frac{s}{2}\int_{\bf k}\nu({\bf k})\left[g({\bf k},s)+G_0({\bf k})\right]+\frac{s^2}{2}\left[f(0)-2f\left(\frac{1}{2}\int_{\bf k}\,\left[g({\bf k},s)+G_0({\bf k})\right]\right)\right] \label{S_action_res}\\
    &+\frac{s}{2}\left[-f'(0)\int_{\bf k}v({\bf k})+w_M\,f\left(0\right)-\int_{s}^{w_M}dz\,f\left(\int_{\bf k}g({\bf k},z)\right)\right]\nn\\
    &+\frac{1}{2}\int_{\bf k}\left[-\frac{s}{w_M}\ln v({\bf k})+\ln\left(\lambda({\bf k},s)+s G_0({\bf k})\right)-s\int_{s}^{w_M}\frac{dz}{z^2}\ln\lambda({\bf k},z)\right]\;,\nn\\
    {\rm where}&\;\;\lambda({\bf k},z)=v({\bf k})+z\,g({\bf k},z)+\int_{z}^{w_M}dw\,g({\bf k},w)\;.
\end{align}
As in the case of the short-range disorder, the (breaking point) parameter $w_M\geq s$, the functions $v({\bf k})\geq 0$ and $G_0({\bf k})\geq 0$ are positive and the function $g({\bf k},z)$ is a positive non-increasing function defined over the interval $z\in[s,w_M]$. Optimising over $w_M$ yields the explicit boundary condition $g({\bf k},z\geq w_M)=0$. The expression above is valid for any pattern of replica-symmetry breaking (replica-symmetric ($w_M=s$), one-step replica symmetry breaking, full replica-symmetry breaking).



The average/typical ground state energy value takes the same form as for short-range disorder in Eq. \eqref{e_typ_res}.
The expression for the cumulant generating function obtained above allows to derive higher-order cumulants and the rescaled variance in particular. The latter is defined as 
\begin{align}
{\cal V}_{L}^{\rm l.r.}(\mu)=&\lim_{N L^d\to \infty}N L^d {\rm Var}(\Delta e_{\min})=\partial_s^2\phi_L(0,\mu)\nn \\
=&f\left(Q_{\rm typ}\right)+f(0)-2 f\left(\frac{Q_{\rm typ}-I_2(\mu) f'\left(Q_{\rm typ}\right)}{2}\right)-\frac{I_2(\mu) f'\left(Q_{\rm typ}\right)^2}{2}\;,\;\;
{\rm where}\;\;Q_{\rm typ}=\int_{\bf k}g_{\rm typ}({\bf k},0)\;.\label{var_res_LR}
\end{align}
The integral $I_2(x)$ is defined in Eq. \eqref{I_p}. The expression in Eq. \eqref{var_res} is valid at any level of RSB while the expression of $Q_{\rm typ}$ depends instead on the level of replica symmetry breaking. In the simplest, replica-symmetric (RS) phase $Q_{\rm typ}=0$. In the full replica symmetry broken (FRSB) phase it reads instead $Q_{\rm typ}={f''}^{-1}\left(1/I_2(\mu)\right)$, i.e. it is expressed from the functional inverse of the second derivative of the covariance function. Finally, in the one-step replica symmetry broken (1RSB) phase it is obtained as the non-zero solution of the following set of self-consistent equations (eliminating the breaking-point parameter $w_{\rm typ}$ in the following equations)
\begin{align}
Q_{\rm typ}&=\int_{\bf k}\frac{1}{w_{\rm typ}}\left(\frac{1}{\nu({\bf k})}-\frac{1}{\nu({\bf k})+w_{\rm typ}(f'\left(Q_{\rm typ}\right)-f'(0))}\right)\;,\label{res_var_0}\\
f\left(Q_{\rm typ}\right)-f(0)-Q_{\rm typ} f'\left(Q_{\rm typ}\right)&=\frac{1}{w_{\rm typ}^2}\int_{\bf k}\left[\frac{w_{\rm typ}(f'\left(Q_{\rm typ}\right)-f'(0))}{\nu({\bf k})+w_{\rm typ}(f'\left(Q_{\rm typ}\right)-f'(0))}-\ln\left(1+\frac{w_{\rm typ}(f'\left(Q_{\rm typ}\right)-f'(0))}{\nu({\bf k})}\right)\right]\;.\label{res_var}
\end{align}

\section{Optimisation problem} \label{sec-opt-prob}
In this section, we derive the general expression of the cumulant generating function for the ground-state energy in terms of a functional optimisation problem.

\subsection{Integer moments}

To compute the large deviation function, we first compute exactly the integer moments of the partition function. The integer moments of the centred free-energy read \cite{fyodorov2007classical}
\begin{align}
    \moy{Z_{N,L,\eta}(\beta)^n}=&\int \prod_{a=1}^n {\cal D}{\bf u}_a({\bf x})\,\moy{e^{-\beta \sum_{a=1}^n \left({\cal H}[{\bf u}_a]-\eta{\cal H}[{\bf 0}]\right)}}\label{Z_N_1}\\
    =&\int \prod_{a=1}^n {\cal D}{\bf u}_a({\bf x})\,\exp\left[\frac{N}{2}\left(\beta^2\sum_{\bf x}\sum_{a,b}\left[f\left(\frac{({\bf u}_a({\bf x})-{\bf u}_b({\bf x}))^2}{2N}\right)-2\eta f\left(\frac{{\bf u}_a({\bf x})^2}{2N}\right)+\eta f(0)\right]\right.\right.\nn\\
    &\left.\left.-\frac{\beta}{2}\sum_{a}\sum_{{\bf x},{\bf y}}{\bf u}_a({\bf x})(\mu \delta_{{\bf x},{\bf y}}-t \Delta_{{\bf x},{\bf y}}){\bf u}_a({\bf y})\right)\right]\;,\nn
\end{align}
where the parameter $\eta=0,1$ allows to treat on the same footing short-range and long-range disorder. This expression can be rewritten in terms of the fields in momentum space as ${\bf u}_a({\bf k})$, which reads
\begin{align}
    \moy{Z_{N,L,\eta}(\beta)^n}
    &=\int \prod_{a=1}^n {\cal D}{\bf u}_a({\bf k})\,\exp\left[\frac{N}{2}\left(\beta^2\sum_{\bf x}\sum_{a,b}\left[f\left(\int_{\bf k}\int_{\bf k'}\frac{({\bf u}_a({\bf k})-{\bf u}_b({\bf k}))\cdot({\bf u}_a({\bf k}')-{\bf u}_b({\bf k}'))}{2N}e^{i{\bf x}\cdot({\bf k}+{\bf k}')}\right)\right.\right.\right.\nn\\
    &\left.\left.\left.-2\eta f\left(\int_{\bf k}\int_{\bf k'}\frac{{\bf u}_a({\bf k})\cdot{\bf u}_a({\bf k}')}{N}e^{i{\bf x}\cdot({\bf k}+{\bf k}')}\right)+\eta f(0)\right]-\frac{\beta L^d}{2}\sum_{a=1}^n\int_{\bf k}\nu({\bf k})\frac{{\bf u}_a({\bf k})\cdot{\bf u}_a(-{\bf k})}{N}\right)\right]\;.
\end{align}
In Eq. \eqref{Z_N_1}, all the terms are invariant by translation in the spatial space ${\bf x}$ such that we expect the overlaps not to depend explicitly on the position. Thus we can replace
\begin{align}
  &\sum_{\bf x}\sum_{a,b} f\left(\int_{\bf k}\int_{\bf k'}\frac{({\bf u}_a({\bf k})-{\bf u}_b({\bf k}))\cdot({\bf u}_a({\bf k}')-{\bf u}_b({\bf k}'))}{2N}e^{i{\bf x}\cdot({\bf k}+{\bf k}')}\right)\\
  &\leftrightarrow L^d\sum_{a,b} f\left(\frac{1}{L^d}\sum_{\bf x}\int_{\bf k}\int_{\bf k'}\frac{({\bf u}_a({\bf k})-{\bf u}_b({\bf k}))\cdot({\bf u}_a({\bf k}')-{\bf u}_b({\bf k}'))}{2N}e^{i{\bf x}\cdot({\bf k}+{\bf k}')}\right)\nn\\
  &=L^d\sum_{a,b} f\left(\int_{\bf k}\frac{({\bf u}_a({\bf k})-{\bf u}_b({\bf k}))\cdot({\bf u}_a(-{\bf k})-{\bf u}_b(-{\bf k}))}{2N}\right)\;.\nn
\end{align}
Using this hypothesis, one can express explicitly the integer moments of the partition function in terms of the overlap matrix ${\bf G}_{ab}({\bf k})=({\bf u}_a({\bf k})\cdot {\bf u}_b(-{\bf k}))/N$ as  
\begin{align}
    \moy{Z_{N,L,\eta}(\beta)^n}=&C_{N,n}\int \prod_{a,b=1}^n {\cal D}{\bf G}_{ab}({\bf k})\,\left(\det {\bf G}({\bf k})\right)^{\frac{n+1}{2}}\exp\left(N L^d S_n[{\bf G}({\bf k})]\right)\;,\\
    S_n=&\frac{\beta^2}{2}\sum_{a,b}\left[f\left(\int_{\bf k}\left[\frac{{\bf G}_{aa}({\bf k})+{\bf G}_{bb}({\bf k})}{2}-{\bf G}_{ab}({\bf k})\right]\right)-2\eta f\left(\int_{\bf k}\frac{{\bf G}_{aa}({\bf k})}{2}\right)+\eta f(0)\right]\\
    &+\frac{1}{2}\int_{\bf k}{\rm Tr}\ln\left[{\bf G}({\bf k})\right]-\frac{\beta}{2}\sum_{a=1}^n\int_{\bf k}\nu({\bf k}){\bf G}_{aa}({\bf k})+c_n\;,\nn\\
    {\bf G}_{ab}({\bf k})=&\frac{{\bf u}_a({\bf k})\cdot {\bf u}_b(-{\bf k})}{N}\;.
\end{align}
The expression for the moments of the partition function can be obtained at the leading (exponential) order in $NL^d$ by using a saddle-point method with respect to the matrix ${\bf G}({\bf k})$, i.e.
\be
\lim_{N\to \infty}\frac{1}{N L^d}\ln\moy{Z_{N,L,\eta}(\beta)^n}=\max_{{\bf G}({\bf k})}S_{n,\eta}[{\bf G}({\bf k})]\;.
\ee
Supposing that the limit of high-dimension $NL^d \to\infty$ and zero temperature $\beta\to \infty$ can be exchanged, the scaled cumulant generating function of $e_{\min}$ (or $\Delta e_{\min}$ for long-range disorder) can be computed by taking the limit $n\to 0$ and $\beta\to \infty$ such that the parameter $s=n\beta$ is fixed, yielding
\be
\lim_{\beta\to\infty}\lim_{N\to \infty}\frac{1}{N L^d}\ln\moy{Z_{N,L,\eta}(\beta)^{s/\beta}}=\lim_{N\to \infty}\frac{1}{N L^d}\ln\moy{e^{-s NL^d(e_{\min}-\eta e_{\rm flat})}}\;,
\ee
where $e_{\rm flat}={\cal H}[{\bf 0}]/(NL^d)$ is the intensive energy of the flat configuration ${\bf u}({\bf x})={\bf 0}$. From the expression of the cumulant generating function, we will be able to reconstruct the large deviation function by Legendre transform
\be
\Lambda_L(e,\mu)=-\min_{s\in\mathbb{R}}\left[s\epsilon+\varphi_L(s,\mu)\right]\;.
\ee
In the following, we analyse this limit explicitly.



\subsection{Parisi scheme}

Let us suppose that the matrix ${\bf G}({\bf k})$ takes the Parisi hierarchical form with $k$ replica-symmetry breaking. We introduce the $k+2$ overlap functions $G_l({\bf k})$ for $l=0,\cdots,k+1$ characterising the Parisi blocks satisfying
\be
\forall {\bf  k}\,,\,\,G_0({\bf k})\leq G_1({\bf k})\leq \cdots \leq G_k({\bf k})\leq G_{k+1}({\bf k})=G_d({\bf k})\;. 
\ee
We similarly introduce the parameters $m_l$ for $l=0,\cdots,k+1$ encoding the size of the Parisi blocks with
\be
m_0=n\geq m_1\geq \cdots \geq m_k\geq m_{k+1}=1\;.
\ee
The functional $S_{n,\eta}$ can be expressed explicitly in terms of these different parameters as
\begin{align}
    S_{n,\eta}=&\frac{n\beta^2}{2}\left[f(0)+\sum_{l=0}^k (m_l-m_{l+1})f\left(\int_{\bf k}\left[G_d({\bf k})-G_l({\bf k})\right]\right)-2n\eta f\left(\frac{1}{2}\int_{\bf k}\,G_d({\bf k})\right)+n\eta f(0)\right]\\
    &+\frac{1}{2}\int_{\bf k}\left[\ln\left(n G_0({\bf k})+\sum_{l=0}^k m_{l+1}(G_{l+1}({\bf k})-G_{l}({\bf k}))\right)+n\sum_{l=0}^{k}\left(\frac{1}{m_{l+1}}-\frac{1}{m_{l}}\right)\ln\left(\sum_{j=l}^k m_{j+1}(G_{j+1}({\bf k})-G_{j}({\bf k}))\right)\right]\nn\\
    &-\frac{n\beta}{2}\int_{\bf k}\nu({\bf k})G_{d}({\bf k})+c_n\;.\nn
\end{align}
Instead of keeping track of all the individual parameters, it will be convenient to rewrite the functional $S_{n,\eta}$ in terms of the Parisi overlap function
\be
\tilde g({\bf k},x)=G_d({\bf k})-G_0({\bf k})-\sum_{l=1}^{k+1}(G_{l}({\bf k})-G_{l-1}({\bf k}))\Theta(x-m_l)\;,
\ee
and the associated Parisi eigenvalue function
\be
\tilde \lambda({\bf k},x)=(1-m_k)\tilde g({\bf k},m_k)+x\,\tilde g({\bf k},x)+\int_{x}^{m_k}dy\,\tilde g({\bf k},y)\;.
\ee
For any fixed value of ${\bf k}$, the function $\tilde g({\bf k},x)$ is a piece-wise constant non-decreasing function of $x$ while $\tilde \lambda({\bf k},x)$ is a piece-wise constant non-increasing function. The functional $S_{n,\eta}$ can now be rewritten solely in terms of the function $\tilde g({\bf k},x)$, the parameter $m_k$ and the smallest overlap function $G_0({\bf k})$ as follows
\begin{align}
    S_{n,\eta}=&-\frac{n\beta}{2}\int_{\bf k}\nu({\bf k})\left[\tilde g({\bf k},n)+G_0({\bf k})\right]+\frac{\eta n^2\beta^2}{2}\left[ f(0)-2 f\left(\frac{1}{2}\int_{\bf k}\,\left[\tilde g({\bf k},n)+G_0({\bf k})\right]\right)\right] \\
    &+\frac{n\beta^2}{2}\left[f(0)+(m_k-1)\,f\left(\int_{\bf k}\tilde g({\bf k},m_k)\right)-\int_{n}^{m_k}dx\,f\left(\int_{\bf k}\tilde g({\bf k},x)\right)\right]\nn\\
    &+\frac{1}{2}\int_{\bf k}\left[-\frac{n}{m_k}\ln \tilde \lambda({\bf k},m_k)+\ln\left(\tilde\lambda({\bf k},n)+n G_0({\bf k})\right)-n\int_{n}^{m_k}\frac{dx}{x^2}\ln\tilde \lambda({\bf k},x)\right]+c_n\;.\nn
\end{align}
In the next section, we consider how to obtain the expression of this functional in the scaling limit $\beta\to \infty$ with $n=s/\beta$ and $s=O(1)$.

\subsection{Zero-temperature limit}

To obtain the expression of the cumulant generating function of the ground-state energy, we need to take the specific scaling limit $n\to 0$ and $\beta\to \infty$ with fixed $s=n\beta=O(1)$. In this scaling limit, we expect the other parameters to take the following scaling forms 
\begin{align}
    &v({\bf k})=\beta(G_d({\bf k})-G_k({\bf k}))=O(1)\;,\;\;w_M=\beta m_k=O(1)\;,\\
    & g({\bf k},z)=\tilde g\left({\bf k},\frac{z}{\beta}\right)\;,\;\;\lambda({\bf k},z)=\beta\tilde \lambda\left({\bf k},\frac{z}{\beta}\right)=v({\bf k})+z\,g({\bf k},z)+\int_{z}^{w_M}dw\,g({\bf k},w)\;,
\end{align}
where the function $g({\bf k},w\geq w_M)=0$. Introducing these scaling forms and taking the scaling limit in the expression of $S_n$, one obtains the following expression
\begin{align}
    {\cal S}_{\eta,L}(\mu)=&-\frac{s}{2}\int_{\bf k}\nu({\bf k})\left[g({\bf k},s)+G_0({\bf k})\right]+\frac{\eta s^2}{2}\left[f(0)-2f\left(\frac{1}{2}\int_{\bf k}\,\left[g({\bf k},s)+G_0({\bf k})\right]\right)\right] \label{S_action}\\
    &+\frac{s}{2}\left[-f'(0)\int_{\bf k}v({\bf k})+w_M\,f\left(0\right)-\int_{s}^{w_M}dz\,f\left(\int_{\bf k}g({\bf k},z)\right)\right]\nn\\
    &+\frac{1}{2}\int_{\bf k}\left[-\frac{s}{w_M}\ln v({\bf k})+\ln\left(\lambda({\bf k},s)+s G_0({\bf k})\right)-s\int_{s}^{w_M}\frac{dz}{z^2}\ln\lambda({\bf k},z)\right]\;.\nn
\end{align}
The cumulant generating function $\varphi_L(s,\mu)$ of the ground-state energy can now be expressed in terms of a functional optimisation of ${\cal S}_{0,L}(\mu)$,
\be
\varphi_L(s,\mu)=\begin{cases}
\displaystyle\max_{g({\bf k},z),v({\bf k}),w_M,G_0({\bf k})}{\cal S}_{0,L}(\mu)\;,\;\;s<0\;,\\
\\
\displaystyle\min_{g({\bf k},z),v({\bf k}),w_M,G_0({\bf k})}{\cal S}_{0,L}(\mu)\;,\;\;s\geq 0\;.
\end{cases}
\ee
The cumulant generating function $\phi_L(s,\mu)$ of the ground-state energy difference is given by a similar expression, replacing ${\cal S}_{0,L}(\mu)\to {\cal S}_{1,L}(\mu)$. The average ground-state energy (difference) can be extracted from this expression and is again obtained from a functional optimisation problem as
\be
e_{L,\rm typ}(\mu)=\lim_{N\to \infty}\moy{e_{\min}}=\lim_{N\to \infty}\moy{\Delta e_{\min}}=-\partial_s\varphi_L(0,\mu)=\max_{g({\bf k},z),v({\bf k}),w_M,G_0({\bf k})} {\cal S}_{L,\rm typ}(\mu)\;,\label{e_min}
\ee
where the functional ${\cal S}_{L,\rm typ}(\mu)$ to optimise reads
\begin{align}
    {\cal S}_{L,\rm typ}(\mu)=&\frac{1}{2}\int_{\bf k}\nu({\bf k})\left[g({\bf k},0)+G_0({\bf k})\right]+\frac{1}{2}\left[f'(0)\int_{\bf k}v({\bf k})-w_M\,f\left(0\right)+\int_{0}^{w_M}dz\,f\left(\int_{\bf k}g({\bf k},z)\right)\right]\label{S_0}\\
    &+\frac{1}{2}\int_{\bf k}\left[\frac{1}{w_M}\ln \frac{v({\bf k})}{\lambda({\bf k},0)}-\frac{G_0({\bf k})}{\lambda({\bf k},0)}+\int_{0}^{w_M}\frac{dz}{z^2}\ln\frac{\lambda({\bf k},z)}{\lambda({\bf k},0)}\right]\;.\nn
\end{align}
An explicit expression for the rescaled variance can similarly be obtained as
\begin{align}
{\cal V}_L(\mu)&=\lim_{N\to \infty}N L^d \moy{\left(e_{\min}-\eta e_{\rm flat}-\moy{e_{\min}}\right)^2}\\
&=f\left(\int_{\bf k}g_{\rm typ}({\bf k},0)\right)+\eta f(0)-2\eta f\left(\frac{1}{2}\int_{\bf k}\,\left[g_{\rm typ}({\bf k},0)+G_{\rm typ}({\bf k})\right]\right)-\frac{1}{2}\int_{\bf k}\left(\frac{G_{\rm typ}({\bf k})}{\lambda_{\rm typ}({\bf k},0)}\right)^2\;,\nn    
\end{align}
where $G_{\rm typ}({\bf k})$, $g_{\rm typ}({\bf k},z)$ and $\lambda_{\rm typ}({\bf k},z)=v_{\rm typ}({\bf k})+z\,g_{\rm typ}({\bf k},z)+\int_{z}^{w_M}dw\,g_{\rm typ}({\bf k},w)$ are the optimal functions obtained from the optimisation problem in Eq. \eqref{e_min}. Optimising explicitly the expression of ${\cal S}_{L,\rm typ}(\mu)$ in Eq. \eqref{S_0} with respect to $G_0({\bf k})$ and $g({\bf k},z)$ provides explicit expressions for $G_{\rm typ}({\bf k})$ and $\lambda_{\rm typ}({\bf k},0)$, 
\be
\lambda_{\rm typ}({\bf k},0)=\frac{1}{\nu({\bf k})}\;,\;\;G_{\rm typ}({\bf k})=-\lambda_{\rm typ}({\bf k},0)^2 f'\left(Q_{\rm typ}\right)=-\frac{f'\left(Q_{\rm typ}\right)}{\nu({\bf k})^2}\;,\;\;{\rm where}\;\;Q_{\rm typ}=\int_{\bf k} g_{\rm typ}({\bf k},0)\;.
\ee
Inserting these expressions into the rescaled variance, the rescaled variance reads
\begin{align}
{\cal V}_L(\mu)=&\lim_{N\to \infty}N L^d \moy{\left(e_{\min}-\eta e_{\rm flat}-\moy{e_{\min}}\right)^2}\nn \\
=&f\left(Q_{\rm typ}\right)+\eta f(0)-2\eta f\left(\frac{Q_{\rm typ}-I_2(\mu) f'\left(Q_{\rm typ}\right)}{2}\right)-\frac{I_2(\mu) f'\left(Q_{\rm typ}\right)^2}{2}\;,    
\end{align}
thus depending solely on $Q_{\rm typ}$ defined above and on the integrals $I_2(x)$ defined in Eq. \eqref{I_p}.
The expression for $e_{L,\rm typ}(\mu)$ within the RS ansatz is simply obtained by taking the limit $w_M\to 0$ in Eq. \eqref{S_0} where $\lambda({\bf k},0)=v({\bf k})$ and $g({\bf k},0)=0$ and optimising over the remaining parameters $v({\bf k})$ and $G_0({\bf k})$, taking respectively the values $v_{\rm typ}({\bf k})=1/\nu({\bf k})$ and $G_{\rm typ}({\bf k})=-f'(0)/\nu({\bf k})^2$. 

\subsection{Interpretation of the optimal fluctuations} \label{sec-inter}

Let us briefly discuss the probabilistic meaning of these parameters in the case where $s\geq 0$.  To this end, we define the cumulative distribution $x_{\beta}(q)$ of overlaps between replicas
\be
q_{ab}=\int_{\bf k}\frac{{\bf u}_a({\bf k})\cdot {\bf u}_b(-{\bf k})}{N}\;,
\ee
 at finite inverse temperature $\beta$. This function is a non-decreasing function of $q$ with the boundary conditions $x_{\beta}(0\leq q\leq q_0)=n$, with $q_0$ the minimal overlap and $0<n<1$, while $x_{\beta}(q\geq q_d)=1$ with $q_d=q_{aa}$ the maximal (self)-overlap. The boundary condition can be interpreted as imposing an atom at $q=0$ with probability weight $n$. If the optimal solution corresponds to a finite number $k$ of replica-symmetry breaking, the function $x_{\beta}(q)$ displays a finite number $k+1$ of steps. One may then associate an overlap function to this cumulative distribution function which reads
\be
 q_{\beta}(x)=q_0(\beta)\Theta(x-n)+\sum_{l=1}^{k+1}(q_{l}(\beta)-q_{l-1}(\beta))\Theta(x-x_l(\beta))\;,
 \ee
where $x_l(\beta)={\rm Prob}.\left[q\leq q_l(\beta)\right]$ for $l\geq 1$ and $x_0(\beta)=n$. If the solution is full replica-symmetry broken instead, the cumulative distribution function $x_{\beta}(q)$ is a continuous increasing function of $q\in[q_0(\beta),q_d(\beta)]$ and we define $q_{\beta}(x)=x_{\beta}^{-1}(x)$ as its functional inverse defined over the interval $x\in[n,1]$ while $q_{\beta}(x)=0$ for $x\in[0,n)$. In the limit of zero-temperature, where $\beta\to\infty$ and $n\to 0$ with fixed $s=n\beta=O(1)$ on which we concentrate in this manuscript, the probability distribution function of the overlap $x_{\beta}'(q)$ localises in a vicinity of order $1/\beta$ around the self-overlap $q_d(\beta)$, yielding the following scaling form for the overlap function  
\be
q_d(\beta)-q_{\beta}(x)\approx Q(\beta x)+\frac{v}{\beta}\Theta(x-1)\;,\;\;\beta\gg 1\;,
\ee
with $q_d=\displaystyle\lim_{\beta\to \infty}q_d(\beta)$ and where the parameters defined in that equation are related to the optimisation functions as
\be
Q(z)=\lim_{\beta\to \infty}\left[q_{d}(\beta)-q_{\beta}\left(\frac{z}{\beta}\right)\right]=\int_{\bf k}g({\bf k},z)\;,\;\;v=\lim_{\beta\to\infty}\beta(q_d(\beta)-q_{k}(\beta))=\int_{\bf k}v({\bf k})\;,
\ee
with the remaining optimisation parameters satisfying 
\begin{align}
w_M=\lim_{\beta\to\infty}\beta x_{k}(\beta)\;,\;\;q_0=\lim_{\beta\to\infty}q_0(\beta)=\int_{\bf k}G_0({\bf k})\;.
\end{align}
In particular, these identities yield $Q(z\geq w_M)=0$ and $Q(s)=q_d-q_0$. In the next section, we analyse the expression of the cumulant generating function and large deviation function in the different patterns of replica-symmetry.

\section{Analysis of the different phases} \label{sec-analysis}
Let us now consider in more detail the expression of the cumulant generating function, large deviation function and variance in the different phase of replica-symmetry for short-range disorder.

\subsection{Replica symmetric phase}

In the replica symmetric (RS) phase, the parameter $w_M=s$ or equivalently $g({\bf k},z\geq s)=0$. The functional ${\cal S}_{0,L}(\mu)$ to optimise simplifies drastically and takes the simpler RS form
\begin{align}
    {\cal S}_{\rm RS}[G_0,v]=&-\frac{s}{2}\int_{\bf k}\nu({\bf k})G_0({\bf k})
    +\frac{s}{2}\left[-f'(0)\int_{\bf k}v({\bf k})+s\,f\left(0\right)\right]
    +\frac{1}{2}\int_{\bf k}\ln\left(1+\frac{s G_0({\bf k})}{v({\bf k})}\right)\;,
\end{align}
which now depends only on the two functional parameters $G_0({\bf k})$ and $v({\bf k})$.  Computing the functional derivative of ${\cal S}_{\rm RS}$ with respect to $G_0({\bf k})$ and $v({\bf k})$, one obtains explicit equations satisfied by their optimal values $G_{\rm RS}({\bf k})$ and $v_{\rm RS}({\bf k})$, namely
\begin{align}
\frac{\delta {\cal S}_{\rm RS}}{\delta G_0({\bf k})}[G_{\rm RS},v_{\rm RS}]&=0=-\frac{s}{2}\nu({\bf k})+\frac{s}{2(v_{\rm RS}({\bf k})+s G_{\rm RS}({\bf k}))}\;,\\
\frac{\delta {\cal S}_{\rm RS}}{\delta v({\bf k})}[G_{\rm RS},v_{\rm RS}]&=0=-\frac{s}{2}f'(0)+\frac{1}{2}\left[\frac{1}{v_{\rm RS}({\bf k})+s G_{\rm RS}({\bf k})}-\frac{1}{v_{\rm RS}({\bf k})}\right]\;.
\end{align}
The solution to these equations reads
\be
v_{\rm RS}({\bf k})=\frac{1}{\nu({\bf k})-sf'(0)}\;,\;\;G_{\rm RS}({\bf k})=-\frac{f'(0)}{\nu({\bf k})(\nu({\bf k})-sf'(0))}\;.
\ee
Note that the parameter $v_{\rm RS}$ must be positive, which is only possible if 
\be
s>\frac{\min_{\bf k}\nu({\bf k})}{f'(0)}=\frac{\mu}{f'(0)}\;,
\ee
where we remind that $f'(0)<0$. This bond indicates that the replica-symmetric ansatz to the CGF cannot describe the function for all $s\in \mathbb{R}$.


Inserting the optimal values of $G_{\rm RS}({\bf k})$ and $v_{\rm RS}(\bf k)$ in the functional ${\cal S}_{\rm RS}$, the RS expression for the CGF reads
\begin{align}
  \varphi_{L,\rm RS}(s,\mu)=&{\cal S}_{\rm RS}[G_{\rm RS},v_{\rm RS}]\label{phi_RS}
  =\frac{s^2}{2}f(0)+\frac{1}{2}\int_{\bf k}\ln\left(1-\frac{s f'(0)}{\nu({\bf k})}\right)\;.
\end{align}
Let us define for fixed $s$ the centred energy $\epsilon_{\rm RS}(s)$ as
\begin{align}
e_{L,\rm RS}(s,\mu)=-\partial_s\varphi_{L,\rm RS}(s,\mu)=&-s f\left(0\right)+\frac{f'(0)}{2}\int_{\bf k}\frac{1}{\nu({\bf k})-sf'(0)}\;.\nn    
\end{align}
From this expression, one may easily check that 
\be
e_{L,\rm RS}(\mu)=e_{L,\rm RS}(s=0,\mu)=-\partial_s\varphi_{L,\rm RS}(s,\mu)=\frac{f'(0)}{2}\int_{\bf k}\frac{1}{\nu({\bf k})}=\frac{f'(0)}{2}I_1(\mu)\;,
\ee
where the function $I_1(x)$ is defined in Eq. \eqref{I_p}. To ensure, that each energy corresponds to a single value of $s$, one must ensure that the CGF is convex, i.e. that its second derivative 
\be
\partial_s^2\varphi_{L,\rm RS}(s,\mu)=-\partial_s e_{L,\rm RS}(s,\mu)=f(0)-\frac{1}{2}\int_{\bf k}\left(\frac{f'(0)}{\nu({\bf k})-s f'(0)}\right)^2=f(0)-\frac{f'(0)^2}{2}I_2(\mu-s f'(0))\;,\label{varphi_sec_RS}
\ee
is positive for any $s$ where the CGF is given by its RS ansatz. Thus, one must have throughout the RS phase that  $I_2(\mu-s f'(0))\leq f(0)/(2f'(0)^2)$. The rescaled variance is obtained within the RS ansatz by taking $s=0$ in the expression above
\be
{\cal V}_{\rm RS}(\mu)=\partial_s^2\varphi_{L,\rm RS}(0,\mu)=f(0)-\frac{I_2(\mu) f'(0)^2}{2}\;.
\ee
Finally, in the range where the CGF is convex, the associated large deviation function can be expressed parametrically in terms of $s$ as 
\begin{align}
    \Lambda_{L,\rm RS}(e,\mu)=&s\partial_s\varphi_{L,\rm RS}(s,\mu)-\varphi_{L,\rm RS}(s,\mu)=\frac{s^2}{2}f\left(0\right)-\frac{1}{2}\int_{\bf k}\left[\frac{s f'(0)}{\nu({\bf k})-s f'(0)}+\ln\left(1-\frac{s f'(0)}{\nu({\bf k})}\right)\right]\;,\label{LDF_RS}\\
    {\rm where}\;\;e=&-s f(0)+\frac{f'(0)}{2}\int_{\bf k}\frac{1}{\nu({\bf k})-s f'(0)}\;.\label{E_RS}
\end{align}

Both the criterion of positivity of $v_{\rm RS}({\bf k})$ and of convexity of the CGF indicate that the RS ansatz cannot describe the full range of values of $s\in\mathbb{R}$. In the next section, we will see the correct critetion ensuring the stability of the RS ansatz.

\subsection{Replica symmetry breaking}

The replica symmetric solution to the cumulant generating function can be shown to be unstable for any $s<s_{\rm AT}$ where $s_{\rm AT}$ is the solution of 
\begin{align}
 f''(0)\int_{\bf k}&\frac{1}{(\nu({\bf k})-s_{\rm AT} f'(0))^2}=f''(0) I_2\left(\mu-s_{\rm AT} f'(0)\right)=1\;.\label{s_AT_eta}
\end{align}
A similar criterion appears in the definition of the Larkin mass $\mu_c$ such that the typical energy is RS for $\mu>\mu_c$ and RSB for $\mu<\mu_c$ \cite{mezard1991replica,arous2024larkin}. In particular $\mu_c$ is defined as the solution of $f''(0) I_2\left(\mu_c\right)=1$. Comparing with the expression above, it is simple to conclude that
\be
s_{\rm AT}=\frac{\mu-\mu_c}{f'(0)}\;.
\ee
Note that $s_{\rm AT}$ only provides a criterion for continuous transitions but discontinuous transitions may occur as well. These transitions will be considered below in the section on the one-step replica symmetry breaking.
Using the definition of $s_{\rm AT}$ in Eq. \eqref{s_AT_eta} as well as the expression in Eq. \eqref{varphi_sec_RS}, and one can show that
\be
\partial_s^2\varphi_{L,\rm RS}(s_{\rm AT},\mu)=f\left(0\right)-\frac{f'(0)^2}{2}\int_{\bf k}\frac{1}{(\nu({\bf k})-s_{\rm AT} f'(0))^2}=f\left(0\right)-\frac{f'(0)^2}{2f''(0)}>0\;,\label{phi_s_AT}
\ee
indicating that the CGF is convex for any $s\geq s_{\rm AT}$. The identity in Eq. \eqref{phi_s_AT} is easily proven using the identity $f(q)=\int_0^{\infty}dk e^{-q k^2} \tilde f(k)$ with $\tilde f(k)\geq 0$ by the chain of inequalities
\begin{align}
&f''(0)f(0)-\frac{f'(0)^2}{2}=\int_0^{\infty}\int_0^{\infty} dk_1 dk_2 \tilde f(k_1)\tilde f(k_2)\left(k_1^4+k_2^4-k_1^2 k_2^2\right)\\
&>\int_0^{\infty}\int_0^{\infty} dk_1 dk_2 \tilde f(k_1)\tilde f(k_2)\left(k_1^4+k_2^4-2k_1^2 k_2^2\right)=\int_0^{\infty}\int_0^{\infty} dk_1 dk_2 \tilde f(k_1)\tilde f(k_2)\left(k_1^2-k_2^2\right)^2\geq 0\;.  \nn  
\end{align}

We have now identified one explicit condition of instability of the RS solution. Let us now analyse the expression of the cumulant generating function and large deviation function when the replica-symmetry is broken.

\subsection{One-step replica symmetry broken solution}

Let us now consider the 1RSB solution. The expression for the functional to optimise reads in that phase 
\begin{align}
   {\cal S}_{\rm 1RSB}=&-\frac{s}{2}\left[\int_{\bf k}\nu({\bf k}) (G({\bf k})+G_0({\bf k}))\right]+\frac{s}{2}\left[s f\left(Q\right)-v f'(0)+w\left(f(0)-f\left(Q\right)\right)\right]\\
    &+\frac{1}{2}\int_{\bf k}\left[\ln\left(1+\frac{s G_0({\bf k})}{v({\bf k})+w G({\bf k})}\right)-\frac{s}{w}\ln\left(\frac{v({\bf k})}{v({\bf k})+w G({\bf k})}\right)\right]\;,
\end{align}
where $v=\int_{\bf k}v({\bf k})$ and $Q=\int_{\bf k}G({\bf k})$. 

The optimisation equations of ${\cal S}_{\rm 1RSB}$ with respect to the parameters $G({\bf k}),G_0({\bf k}),v({\bf k}),w$ respectively read
\begin{align}
    \frac{\delta {\cal S}_{\rm 1RSB}}{\delta G({\bf k})}=&-\frac{s}{2}\nu({\bf k})+\frac{s}{2}(s-w)f'(Q)+\frac{1}{2}\left[\frac{s-w}{v({\bf k})+w G({\bf k})}+\frac{w}{v({\bf k})+w G({\bf k})+s G_0({\bf k})}\right]=0\;,\\
    \frac{\delta {\cal S}_{\rm 1RSB}}{\delta G_0({\bf k})}=&-\frac{s}{2}\nu({\bf k})+\frac{s}{2(v({\bf k})+w G({\bf k})+s G_0({\bf k}))}=0\;,\\
    \frac{\delta {\cal S}_{\rm 1RSB}}{\delta v({\bf k})}=&-\frac{s}{2}f'(0)-\frac{s}{2}\left[\frac{G_0({\bf k})}{(v({\bf k})+w G({\bf k}))(v({\bf k})+w G({\bf k})+s G_0({\bf k}))}+\frac{G({\bf k})}{v({\bf k})(v({\bf k})+w G({\bf k}))}\right]=0\;,\\
    \frac{\partial {\cal S}_{\rm 1RSB}}{\partial w}=&\frac{1}{2}\int_{\bf k}\left[\frac{G({\bf k})}{w}\left(\frac{s-w}{v({\bf k})+w G({\bf k})}+\frac{w}{v({\bf k})+w G({\bf k})+s G_0({\bf k})}\right)+\frac{s}{w^2}\ln\left(\frac{v({\bf k})}{v({\bf k})+w G({\bf k})}\right)\right]\\
    &+\frac{s}{2}\left(f(0)-f(Q)\right)=0\;.\nn
\end{align}
Solving the first three equations, explicit expressions for $v({\bf k})$, $G_0({\bf k})$ and $G({\bf k})$ can be found in terms of $Q$ and $w$. They read
\begin{align}
    G_0({\bf k}) &=\frac{1}{s}\left[\frac{1}{\nu({\bf k})}-\frac{1}{\nu({\bf k})-s f'(Q)}\right]\; ,\\
    G({\bf k})&=\frac{1}{w}\left[\frac{1}{\nu({\bf k})-s f'(Q)}-\frac{1}{\nu({\bf k})-s f'(Q)+\Sigma}\right]\;,\\
    v({\bf k}) &= \frac{1}{\nu({\bf k})-s f'(Q)+\Sigma}\;,\;\;{\rm where}\;\;\Sigma=w(f'(Q)-f'(0))\;.
\end{align}
The remaining equation can be re-expressed using the identities above as
\be
f(0)-f(Q)+Q f'(Q)=\frac{1}{w^2}\int_{\bf k}\left[\ln\left(1+\frac{w(f'(Q)-f'(0))}{\nu({\bf k})-s f'(Q)}\right)-\frac{w(f'(Q)-f'(0))}{\nu({\bf k})-s f'(Q)+w(f'(Q)-f'(0))}\right]\;.\label{w_eq}
\ee
Finally, using the definition of $Q$ and $q_0$, one obtains two additional self-consistent integral equations
\begin{align}
    Q(s)\equiv Q&=\frac{1}{w}\int_{\bf k}\left[\frac{1}{\nu({\bf k})-s f'(Q)}-\frac{1}{\nu({\bf k})-s f'(Q)+w(f'(Q)-f'(0))}\right]\;.\label{Q_eq}
\end{align}
The CGF can now be expressed solely in terms of the parameters $s$, $Q$ and $w$ as
\begin{align}
\varphi_{L,\rm 1RSB}(s,\mu)
=&\frac{s}{2}\left[s f\left(0\right)+(w-s)\left(f(0)-f\left(Q\right)+Q f'(Q)\right)\right]\label{phi_1rsb}\\
    &+\frac{1}{2}\int_{\bf k}\left[\ln\left(1-\frac{s f'(Q)}{\nu({\bf k})}\right)+\frac{s}{w}\ln\left(1+\frac{w(f'(Q)-f'(0))}{\nu({\bf k})-s f'(Q)}\right)\right]\;.  \nn 
\end{align}
From the expression above, one can obtain the average ground-state energy difference as
\be
e_{L,\rm 1RSB}(\mu)=-\partial_s\varphi_{L,\rm 1RSB}(0,\mu)=e_{L,\rm RS}(\mu)+w_{\rm typ}\left(f(Q_{\rm typ})-f(0)-\frac{Q_{\rm typ}}{2} \left[f'(Q_{\rm typ})+f'(0)\right]\right)\;,
\ee
where $Q_{\rm typ}=Q(s=0)$ and $w_{\rm typ}=w(s=0)$ are the optimal value of $Q$ and $w$ for $s=0$, satisfying 
\begin{align}
Q_{\rm typ}&=\int_{\bf k}\frac{1}{w_{\rm typ}}\left(\frac{1}{\nu({\bf k})}-\frac{1}{\nu({\bf k})+w_{\rm typ}(f'\left(Q_{\rm typ}\right)-f'(0))}\right)\;,\label{Q-typ-1rsb}\\
f\left(Q_{\rm typ}\right)-f(0)-Q_{\rm typ} f'\left(Q_{\rm typ}\right)&=\frac{1}{w_{\rm typ}^2}\int_{\bf k}\left[\frac{w_{\rm typ}(f'\left(Q_{\rm typ}\right)-f'(0))}{\nu({\bf k})+w_{\rm typ}(f'\left(Q_{\rm typ}\right)-f'(0))}-\ln\left(1+\frac{w_{\rm typ}(f'\left(Q_{\rm typ}\right)-f'(0))}{\nu({\bf k})}\right)\right]\;.\label{z-typ-1rsb}
\end{align}
The expression for the rescaled variance can be derived from Eq. \eqref{phi_1rsb} as
\be
{\cal V}_{L,\rm 1RSB}(\mu)=\partial_s^2\varphi_{L,\rm 1RSB}(0,\mu)=f\left(Q_{\rm typ}\right)-\frac{I_2(\mu) f'\left(Q_{\rm typ}\right)^2}{2}\;.
\ee
The transition from the RS to the 1RSB phase in the variance is always continuous and occurs at the Larkin mass $\mu_{c}$ satisfying $I_2(\mu_{c})=1/f''(0)$. Expanding Eqs. (\ref{Q-typ-1rsb}-\ref{z-typ-1rsb}) in the vicinity of $\mu_{c}$ one obtains at leading order 
$w_{\rm typ}\approx -f^{(3)}(0)/(2f''(0)^3 I_3(\mu_{c}))$ and $Q_{\rm typ}\approx -(\mu-\mu_{c})/(w_{\rm typ}f''(0))$, which yields a second order transition in the variance
\be
{\cal V}_{L,\rm 1RSB}(\mu)-{\cal V}_{L,\rm RS}(\mu)=2\frac{I_3(\mu_{c})^2 f''(0)^3}{f^{(3)}(0)}f'(0)(\mu-\mu_{c})^2+O(\mu-\mu_{c})^3\;.\label{var_trans_1rsb}
\ee

Let us investigate transitions between the RS and 1RSB phases. There are two types of transitions that may occur, i.e. continuous transitions as $Q\to 0$ and discontinuous transitions as $w\to s$. These transitions can be investigated by considering the expansion of 
\begin{align}
&\varphi_{L,\rm 1RSB}(s,\mu)-\varphi_{L,\rm RS}(s,\mu)\nn\\
&=\frac{1}{2}\int_{\bf k}\left[\ln\frac{\nu({\bf k})-s f'(Q)}{\nu({\bf k})-s f'(0)}+\frac{s}{w}\left(\frac{2w-s}{w}\ln\left(1+\frac{w(f'(Q)-f'(0))}{\nu({\bf k})-s f'(Q)}\right)+\frac{s-w}{w}\left[1-\frac{1}{1+\frac{w(f'(Q)-f'(0))}{\nu({\bf k})-s f'(Q)}}\right]\right)\right]\;,
\end{align}
in terms of these parameters. It is simple to check that this expression vanishes as the $s$ dependent parameters $Q\to 0$ or $m\to s$. In order to obtain the location of these transitions, one may expand the self-consistent equations Eqs. \eqref{Q_eq} and \eqref{w_eq} around these solutions. Expanding Eq. \eqref{Q_eq} up to first order in $Q$ or Eq. \eqref{w_eq} up to second order both yield the same equation as \eqref{s_AT_eta} for the location of the continuous transition, which occurs at $s=s_{\rm AT}$. At the transition, the parameter $w$ takes the value
\be
w_{\rm AT}\equiv w(s_{\rm AT})=-\frac{I_2(\mu-s_{\rm AT}f'(0))^3}{2I_3(\mu-s_{\rm AT}f'(0))}f^{(3)}(0)=-\frac{f^{(3)}(0)}{2(f''(0))^3 I_3(\mu_c)}\;.
\ee
Setting $w\to s$ in Eqs. \eqref{Q_eq} and \eqref{w_eq} yields the two simpler equations for the parameters $s_{\rm dis}$ and $Q_{\rm dis}=Q(s_{\rm dis})$, which read
\begin{align}
    Q_{\rm dis}&=\frac{1}{s_{\rm dis}}\int_{\bf k}\left[\frac{1}{\nu({\bf k})-s_{\rm dis} f'(Q_{\rm dis})}-\frac{1}{\nu({\bf k})-s_{\rm dis} f'(0)}\right]\;,\\
    f(0)-f(Q_{\rm dis})+Q_{\rm dis} f'(Q_{\rm dis})&=\frac{1}{s_{\rm dis}^2}\int_{\bf k}\left[\ln\frac{\nu({\bf k})-s_{\rm dis} f'(0)}{\nu({\bf k})-s_{\rm dis}  f'(Q_{\rm dis})}-\frac{s_{\rm dis}(f'(Q_{\rm dis})-f'(0))}{\nu({\bf k})-s_{\rm dis} f'(0)}\right]\;.
\end{align}
Expanding these equations as $Q_{\rm dis}\to 0$, one may again check that $s_{\rm dis}\to s_{\rm AT}$. The transition is expected to be discontinuous for sufficiently low values of $\mu<\mu_{\rm dis}$ while it is continuous for $\mu>\mu_{\rm dis}$. At the transition, one should have $s_{\rm dis}=s_{\rm AT}=w_{\rm AT}$, yielding
\be
\mu_{\rm dis}=\mu_c+f'(0)w_{\rm AT}=\mu_c-\frac{f'(0)f^{(3)}(0)}{2(f''(0))^3 I_3(\mu_c)}<\mu_c\;.
\ee
One may now investigate the nature of the transition between the two branch of the CGF. For $\mu>\mu_{\rm dis}$, the transition is continuous (in terms of $Q$) and occurs as $s\to s_{\rm AT}$. The CGF displays a third order transition
\begin{align}
 \varphi_{L,\rm 1RSB}(s,\mu)-\varphi_{L,\rm RS}(s,\mu)&=I_3\left(\mu_c\right)\frac{s_{\rm AT}f'(0)^3}{6(s_{\rm AT}-w_{\rm AT})}(s-s_{\rm AT})^3+O(s-s_{\rm AT})^4\nn\\
 &=I_3\left(\mu_c\right)\frac{\mu-\mu_c}{6(\mu-\mu_{\rm dis})}f'(0)^3(s-s_{\rm AT})^3+O(s-s_{\rm AT})^4\;.   
\end{align}
For $\mu<\mu_{\rm dis}$ instead, the transition is discontinuous (in terms of $Q$) and occurs as $s\to s_{\rm dis}$. The transition in the CGF is of second order
\begin{align}
\varphi_{L,\rm 1RSB}(s,\mu)-\varphi_{L,\rm RS}(s,\mu)=&\frac{\partial_s^2\varphi_{L,\rm 1RSB}(s_{\rm dis},\mu)-\partial_s^2\varphi_{L,\rm RS}(s_{\rm dis},\mu)}{2}(s-s_{\rm dis})^2+O(s-s_{\rm dis})^3\label{dis_trans}\\
{\rm where}\;\;\partial_s^2\varphi_{L,\rm RS}(s_{\rm dis},\mu)=&2f(0)-f'(0)^2 I_2(\mu-s_{\rm dis}f'(0))\;,\nn\\
\partial_s^2\varphi_{L,\rm 1RSB}(s_{\rm dis},\mu)=&\frac{2\left[f(Q_{\rm dis})f'(0)(f'(0)-2f'(Q_{\rm dis}))+(f(0)+Q_{\rm dis}f'(0))f'(Q_{\rm dis})^2\right]\,I_2(\mu-s_{\rm dis}f'(0))}{(f'(Q_{\rm dis})-f'(0))^2\,I_2(\mu-s_{\rm dis}f'(0))-2(f(0)-f(Q_{\rm dis})+Q_{\rm dis} f'(Q_{\rm dis}))}\nn\\
&-\frac{4f(Q_{\rm dis})[f(0)-f(Q_{\rm dis})+Q_{\rm dis}f'(Q_{\rm dis})]}{(f'(Q_{\rm dis})-f'(0))^2\,I_2(\mu-s_{\rm dis}f'(0))-2(f(0)-f(Q_{\rm dis})+Q_{\rm dis} f'(Q_{\rm dis}))}\;.\nn 
\end{align}
As $s_{\rm dis}\to s_{\rm AT}$, one has that $Q_{\rm dis}\to 0$ and the coefficients in the second order expansion above match.


The convexity of the CGF $\varphi_{L,\rm 1RSB}(s,\mu)$ is not straightforward to check. In fact even making sure that the expression for the rescaled variance is positive within the 1RSB phase is hard. Supposing this convexity, the large deviation function associated to this expression for the cumulant generating function can be obtained parametrically in terms of $s$ as
\begin{align}
\Lambda_{L,\rm 1RSB}(e,\mu)&=s\partial_s\varphi_{L,\rm 1RSB}(s,\mu)-\varphi_{L,\rm 1RSB}(s,\mu)
=\frac{s^2}{2}f\left(Q\right)-\frac{1}{2}\int_{\bf k}\left[\ln\left(1-\frac{s f'(Q)}{\nu({\bf k})}\right)+\frac{s f'(Q)}{\nu({\bf k})-s f'(Q)}\right]\;,\\
{\rm where}\;\;e&=-w f(0)-(s-w)\left(f(Q)-\frac{Q f'(Q)}{2}\right)+\frac{f'(0)}{2}\int_{\bf k}\frac{1}{\nu({\bf k})-s f'(Q)+w(f'(Q)-f'(0))}\label{e_1rsb_param}\;,
\end{align}
where $Q\equiv Q(s)$ and $w\equiv w(s)$ are the solutions of Eqs. (\ref{w_eq}-\ref{Q_eq}). The transitions of the CGF are reflected in the large deviation function with transition of the same order. In particular, the transition is of second order for $\mu<\mu_{\rm dis}$ and at the transition
\be
\Lambda_{L,\rm 1RSB}(e,\mu)-\Lambda_{L,\rm RS}(e,\mu)=\left(\frac{1}{\partial_s^2\varphi_{L,\rm 1RSB}(s_{\rm dis},\mu)}-\frac{1}{\partial_s^2\varphi_{L,\rm RS}(s_{\rm dis},\mu)}\right)\frac{(e-e_{\rm dis})^2}{2}+O(e-e_{\rm dis})^3\;,
\ee
where the value of energy at the transition reads
\be
e_{\rm dis}=-s_{\rm dis}f(0)+\frac{f'(0)}{2}\int_{\bf k}\frac{1}{\nu({\bf k})-s_{\rm dis} f'(0)}\;.
\ee


For $\mu>\mu_{\rm dis}$, the large deviation function displays a third order transition 
\begin{align}
 \Lambda_{L,\rm 1RSB}(e,\mu)-\Lambda_{L,\rm RS}(e,\mu)=I_3(\mu_c)\frac{f'(0)^3 s_{\rm AT}}{6{\cal V}_L(\mu_c)^3(s_{\rm AT}-z(0))}\left(e-e_{\rm AT}\right)^3+O(e-e_{\rm AT})^4\;,\label{cont_trans_1rsb}  
\end{align}
where the value of energy at the transition reads
\be
e_{\rm AT}=-s_{\rm AT}f(0)+\frac{f'(0)}{2}\int_{\bf k}\frac{1}{\nu({\bf k})-s_{\rm AT} f'(0)}=-(\mu-\mu_c)\frac{f(0)}{f'(0)}+\frac{f'(0)}{2}I_1(\mu_c)\;.
\ee


Finally, let us analyse the behaviour of the cumulant generating function as $s\to -\infty$. 
For a short range correlation functions, using the identity $f(q)=\int_0^{\infty}dk\,e^{-q k^2}\,\tilde f(k)$ with $\tilde f(k)\geq 0$, one can check that $f'(q)$ is a negative decreasing function of $q$ which satisfies $ f'(q)\to 0$ as $q\to \infty$. We suppose additionally that $f'(q)$ only reaches zero asymptotically as $q\to \infty$ and $q f'(q)\to 0$ in that limit. A consistent solution for $Q$ as $s\to -\infty$ is then provided by $Q\to \infty$ with fixed $s f'(Q)\to \min_{\bf k}\nu({\bf k})=\mu$. This ensures that the first term in the integral in Eq. \eqref{Q_eq} diverges. Using our assumptions, the left-hand side of Eq. \eqref{w_eq} reaches the finite value $f(0)$ as $s\to -\infty$. It is thus natural to expect that $w$ reaches a finite value $w_{\infty}$ in that limit. This asymptotic value must satisfy
\be
f(0)=\frac{1}{w_{\infty}^2}\int_{\bf k}\left[\ln\left(1-\frac{w_{\infty}f'(0)}{\nu({\bf k})-\mu}\right)+\frac{w_{\infty}f'(0)}{\nu({\bf k})-\mu-w_{\infty}f'(0)}\right]\;.
\ee
Note that the case of the toy model ($d=0$) needs to be considered separately. Using these results, the energy $e_{L,\rm 1RSB}(s,\mu)$ in Eq. \eqref{e_1rsb_param} behaves asymptotically as
\begin{align}
 e_{L,\rm 1RSB}(s,\mu)\sim s\frac{Q f'(Q)}{2}\sim \frac{\mu Q}{2}\,\;\;s\to -\infty\;.
\end{align}
The large deviation function on the other hand behaves as
\begin{align}
 \Lambda_{L,\rm 1RSB}(e_{L,\rm 1RSB}(s,\mu),\mu)\sim \frac{s^2}{2}f(Q)\sim \frac{\mu^2 f(Q)}{2f'(Q)^2}\;,\,\;\;s\to -\infty\;.
\end{align}
Thus, as $s\to -\infty$, one obtains that $\epsilon\to +\infty$ and the asymptotic behaviour of the large deviation function is independent of the dimension but depends explicitly on the correlation function with
\begin{align}
 \Lambda_{L,\rm 1RSB}(e,\mu)\approx\frac{\displaystyle\mu^2 f\left(2e/\mu\right)}{\displaystyle 2f'\left(2e/\mu\right)^2}\;,\,\;\;e\to +\infty\;.\label{asymp_1rsb}
\end{align}
Note that, in stark distinction with the large deviation for the the spherical spin model \cite{lacroix2024replica,huang2023constructive}, the LDF with speed $N$ for the elastic manifold is defined over the whole range of centred energy $e\in \mathbb{R}$ for any positive value of $\mu>0$. It is thus expected that for $\mu>0$ there is no large deviation regime with speed $N^2$. Now that we have treated in detail the case of one-step replica symmetry broken ansatz, we turn to the full replica symmetry pattern.

\subsection{Full replica symmetry broken solution}

In the full replica symmetry broken (FRSB) solution, there is no simplification to the functional ${\cal S}_{0,L}(\mu)={\cal S}_{\rm FRSB}$ in Eq. \eqref{S_action}. The optimal value of the different parameters ($g({\bf k},z),G_0({\bf k}),v({\bf k}),w_M$) is obtained by extremisation of the functional ${\cal S}_{\rm FRSB}$. We compute below the (functional) derivatives with respect to these different parameters. Taking the derivative of ${\cal S}_{\rm FRSB}$ with respect to the parameter $w_M$ and evaluating at the optimal solution yields
\begin{align}
    \frac{\partial {\cal S}_{\rm FRSB}}{\partial w_M}=
    &\frac{s}{2}\left[f\left(0\right)-f\left(\int_{\bf k}g({\bf k},w_M)\right)\right]+\frac{s}{2w_M^2}\int_{\bf k}\ln \frac{v({\bf k})}{\lambda({\bf k},w_M)}=0\;,\nn
\end{align}
which using $\lambda({\bf k},z)=v({\bf k})+z\,g({\bf k},z)+\int_{z}^{w_M}dw\,g({\bf k},w)$ is simply enforced by ensuring the boundary condition $g({\bf k},w_M)=0$. The functional derivative with respect to $g({\bf k},z)$ evaluated at the optimal solution reads
\begin{align}
    \frac{\delta {\cal S}_{\rm FRSB}}{\delta g({\bf k},z)}=&-\frac{s}{2}f'\left(\int_{\bf k}g({\bf k},z)\right)+\frac{1}{2(\lambda({\bf k},s)+s G_0({\bf k}))}-\frac{s}{2}\int_{s}^{w_M}\frac{dw\left(\Theta(z-w)+z\delta(z-w)\right)}{w^2\,\lambda({\bf k},w)}\nn\\
    &=-\frac{s}{2}f'\left(\int_{\bf k}g({\bf k},z)\right)-\frac{s}{2z\lambda({\bf k},z)}+\frac{1}{2(\lambda({\bf k},s)+s G_0({\bf k}))}-\frac{s}{2}\int_{s}^{z}\frac{dw}{w^2\,\lambda({\bf k},w)}\nn\\
    &=-\frac{s}{2}\left[f'\left(\int_{\bf k}g({\bf k},z)\right)+\frac{G_0({\bf k})}{\lambda({\bf k},s)(\lambda({\bf k},s)+s G_0({\bf k}))}-\int_{s}^{z}\frac{dw\,\partial_w g({\bf k},w)}{\lambda({\bf k},w)^2}\right]=0\;,\label{sp_FRSB}
\end{align}
where we used the identity $
\partial_w\lambda({\bf k},w)=w\,\partial_w g({\bf k},w)$. Taking the functional derivative of ${\cal S}_{\rm FRSB}$ with respect to $G_0({\bf k})$ and evaluating at the optimal solution yields
\begin{align}
    \frac{\delta {\cal S}_{\rm FRSB}}{\delta G_0({\bf k})}=&-\frac{s}{2}\nu({\bf k})+\frac{s}{2(\lambda({\bf k},s)+s G_0({\bf k}))}=0\;.
\end{align}
Using this equation as well as the limit $z\to s$ of Eq. \eqref{sp_FRSB}, one can obtain the following explicit expressions
\begin{align}
    \lambda({\bf k},s)&=\frac{1}{\nu({\bf k})-s f'\left(Q(s)\right)}\;,\\
    G_0({\bf k})&=\frac{1}{s}\left[\frac{1}{\nu({\bf k})}-\frac{1}{\nu({\bf k})-s f'\left(Q(s)\right)}\right]\;,
\end{align}
where we have defined $Q(z)=\int_{\bf k}g({\bf k},z)$.

Taking the functional derivative of ${\cal S}_{\rm FRSB}$ with respect to $v({\bf k})$ and evaluating at the optimal solution yields
\begin{align}
    \frac{\delta {\cal S}_{\rm FRSB}}{\delta v({\bf k})}=&-\frac{s}{2}f'\left(0\right)-\frac{s}{2w_M v({\bf k})}+\frac{1}{2(\lambda({\bf k},s)+s G_0({\bf k}))}-\frac{s}{2}\int_{s}^{w_M}\frac{dw}{w^2\,\lambda({\bf k},w)}\nn\\
    &=-\frac{s}{2}\left[f'\left(0\right)+\frac{G_0({\bf k})}{\lambda({\bf k},s)(\lambda({\bf k},s)+s G_0({\bf k}))}-\int_{s}^{w_M}\frac{dw\,\partial_w g({\bf k},w)}{\lambda({\bf k},w)^2}\right]=0\;.
\end{align}
Note that the equation above matches the limit $z\to w_M$ of Eq. \eqref{sp_FRSB} and thus does not provide additional information. To analyse the optimal solution further, it will be convenient to introduce the following function
\be
\sigma({\bf k},z)=\frac{1}{s(\lambda({\bf k},s)+s G_0({\bf k}))}-\frac{1}{z \lambda({\bf k},z)}-\int_{s}^{z}\frac{dw}{w^2\,\lambda({\bf k},w)}\;.
\ee
This function can be shown to coincide with the Parisi function associated to the inverse overlap matrix ${\bf G}({\bf k})^{-1}$. It can be used to re-express Eq. \eqref{sp_FRSB} as
\be
\frac{\delta {\cal S}_{\rm FRSB}}{\delta g({\bf k},z)}
    =\frac{s}{2}\left[\sigma({\bf k},z)-f'\left(\int_{\bf k}g({\bf k},z)\right)\right]=\frac{s}{2}\left[\sigma({\bf k},z)-f'\left(Q(z)\right)\right]=0\;.
\ee
Interestingly, from this expression, the optimal solution for that function is independent of the Fourier wave-vector ${\bf k}$, i.e. $\sigma({\bf k},z)\equiv \sigma(z)$. The eigenvalue function of the inverse overlap matrix ${\bf G}({\bf k})^{-1}$ is simply given by the inverse of that of ${\bf G}({\bf k})$, yielding the simple identity 

\begin{align}
\frac{1}{\lambda({\bf k},z)}&=\frac{1}{v({\bf k})}+w_M \sigma(w_M)-z\,\sigma(z)-\int_{z}^{w_M}du\,\sigma(u)=\frac{1}{\lambda({\bf k},s)}+s\sigma(s)-z\sigma(z)+\int_{s}^z du\,\sigma(u)\;. 
\end{align}
As Eq. \eqref{sp_FRSB} is valid over the whole interval $z\in[s,w_M]$, we may take an additional derivative with respect to $z$ and obtain after some simplifications
\be
\int_{\bf k}\partial_z g({\bf k},z)f''\left(\int_{\bf k}\,g({\bf k},z)\right)-\frac{\partial_z g({\bf k},z)}{\lambda({\bf k},z)^2}=Q'(z)f''\left(Q(z)\right)-\frac{\partial_z g({\bf k},z)}{\lambda({\bf k},z)^2}=0\;.
\ee
Multiplying by $\lambda({\bf k},z)^2$ and integrating this equation with respect to the Fourier wavevector ${\bf k}$ yields the simple identity
\be
\int_{\bf k}\partial_z g({\bf k},z)\left[\int_{\bf k}\lambda({\bf k},z)^2f''\left(\int_{\bf k}\,g({\bf k},z)\right)-1\right]=Q'(z)\left[\int_{\bf k}\lambda({\bf k},z)^2 f''(Q(z))-1\right]=0\;.\label{FRSB_eq}
\ee
The FRSB solution must satisfy $\partial_z g({\bf k},z)> 0$ in the interval $z\in [s,w_M]$, thus one must have
\be
f''\left(Q(z)\right)=f''\left(\int_{\bf k}\,g({\bf k},z)\right)=\frac{1}{\int_{\bf k}\lambda({\bf k},z)^2}\;.
\ee
In particular, for $z=s$, using the identity $1/\lambda({\bf k},s)=\nu({\bf k})-s f'(Q(s))$, it yields the following self-consistent equation
\be
f''\left(Q(s)\right)=\frac{1}{I_2(\mu-s f'(Q(s)))}\;.\label{Q_s_eq_f_1}
\ee

If the function $f''$ is bijective, it has a well-defined inverse ${f''}^{-1}$, allowing to express directly $Q(z)$ as well as the optimal value of $\sigma(z)$, which satisfies the self-consistent equation 
\be
\sigma(z)=f'\left[{f''}^{-1}\left(\frac{1}{\int_{\bf k}\lambda({\bf k},z)^2}\right)\right]=f'\left[{f''}^{-1}\left(\frac{1}{\int_{\bf k}\left(\lambda({\bf k},s)^{-1}+s\sigma(s)-z\sigma(z)+\int_{s}^z du\,\sigma(u)\right)^{-2}}\right)\right]\;.
\ee
Note that from the expression above, one can express explicitly
\begin{align}
\frac{1}{\lambda({\bf k},z)}&=\frac{1}{\lambda({\bf k},s)}+s f'(Q(s))-z f'(Q(z))+\int_s^z dw f'(Q(z))=\nu({\bf k})-z f'(Q(z))+\int_s^z dw f'(Q(w))\nn\\
&=\nu({\bf k})-s f'(Q(s))-\int_{Q(s)}^{Q(z)}dq\,z(q)\,f''(q)\;.    
\end{align}
It yields the identity
\begin{align}
f''(Q(z))&=\frac{1}{\int_{\bf k}\left(\nu({\bf k})-s f'(Q(s))-\int_{Q(s)}^{Q(z)}dq\,z(q)\,f''(q)\right)^{-2}}=\frac{1}{I_2(\mu(z,s))}\;,\\
\mu(z,s)&=\mu-s f'(Q(s))-\int_{Q(s)}^{Q(z)}dq\,z(q)\,f''(q)=\mu-z\sigma(z)+\int_{s}^z du\,\sigma(u)\;.\label{z_q_eq}
\end{align}
With that equation, one can simply express $Q(z)$ or its inverse function $z(q)$. Taking the form $\nu({\bf k})=\mu+{\bf k}^2$, the integral above can be computed explicitly and the following expression for the function $z(q)$ can be obtained
\be
z(q)=-\frac{2}{4-d}B_d^{-\frac{2}{4-d}}\frac{f^{(3)}(q)}{f''(q)^{\frac{2(3-d)}{4-d}}}\;,\;\;B_d=-2^{d}\pi^{\frac{d}{2}-1}\Gamma\left(\frac{d}{2}-1\right)\sin\left(\frac{d\pi}{2}\right)\;.\label{z_eq}
\ee

Coming back to the general case and taking a derivative with respect to $z$ finally yields
\be
z=-\frac{\left(\int_{\bf k}\lambda({\bf k},z)^2\right)^{3}}{2\int_{\bf k}\lambda({\bf k},z)^3}f^{(3)}\left({f''}^{-1}\left(\frac{1}{\int_{\bf k}\lambda({\bf k},z)^2}\right)\right)=-\frac{I_2(\mu(z,s))^3}{2I_3(\mu(z,s))}f^{(3)}\left({f''}^{-1}\left(\frac{1}{I_2(\mu(z,s))}\right)\right)\;.\label{z_sigma}
\ee


This equation provides an explicit expression for $z$ as a function of the combination $\mu(z,s)$, which can, after some manipulations, be used to obtain an explicit expression for $\sigma(z)$. In order for the solution to be truly FRSB, one must ensure that $Q(z)$ is a non-increasing function of $z$, i.e. $\sigma(z)=f'(Q(z))$ is a non-increasing function of $z$. Taking a derivative with respect to $z$ in Eq. \eqref{z_sigma}, one obtains
\begin{align}
-\frac{1}{z\sigma'(z)}=&\frac{3}{2}\left(2-\frac{I_4(\mu(z,s))I_2(\mu(z,s))}{I_3(\mu(z,s))^2}\right)\frac{f^{(3)}\left(Q(z)\right)}{f''\left(Q(z)\right)^2}-\frac{f^{(4)}\left(Q(z)\right)}{f^{(3)}\left(Q(z)\right)f''\left(Q(z)\right)}\label{sig_prime}\;.
\end{align}
Thus, if the right-hand-side above is positive, the solution will be full replica symmetry broken while it will have a finite level of replica symmetry breaking if the latter is negative.

The expression of $\lambda({\bf k},z)$ can be further simplified using the identity $\sigma(z)=f'(Q(z))$ as
\begin{align}
\lambda({\bf k},z)&=\frac{1}{\lambda({\bf k},s)^{-1}+s\sigma(s)-z\sigma(z)+\int_{s}^z du\,\sigma(u)}=\frac{1}{\lambda({\bf k},s)^{-1}-\int_{Q(s)}^{Q(z)}dq\,z(q)\,f''(q)}\;.
\end{align}
Using the identity $\partial_z \lambda({\bf k},z)=z \partial_z g({\bf k},z)$ as well as the boundary condition $g({\bf k},w_M)=0$, one obtains the expression
\be
g({\bf k},z)=\int_{w_M}^z dw\,\frac{Q'(w)f''(Q(w))}{\displaystyle\left(\lambda({\bf k},s)^{-1}-\int_{Q(s)}^{Q(w)}dq\,z(q)\,f''(q)\right)^2} =\int_{0}^{Q(z)}dq\,f''(q)\lambda({\bf k},q)^2\;,
\ee
where $\lambda({\bf k},q)=\lambda({\bf k},z(q))$. With these expressions, we are now able to express the CGF explicitly as
\begin{align}
    \varphi_{L,\rm FRSB}(s,\mu)=&-\frac{s}{2}\int_{\bf k}\nu({\bf k})\left[g({\bf k},s)+G_0({\bf k})\right]+\frac{s}{2}\left(-f'(0)\int_{\bf k}v({\bf k})+s\,f\left(Q(s)\right)+\int_{s}^{w_M}dz\,z\,Q'(z)\,f'\left(Q(z)\right)\right)\\
    &+\frac{1}{2}\int_{\bf k}\left[\ln\left(1+\frac{s G_0({\bf k})}{\lambda({\bf k},s)}\right)-s\int_{s}^{w_M}\frac{dz}{z}\frac{\partial_z\lambda({\bf k},z)}{\lambda({\bf k},z)}\right]\\
    =&-\frac{s}{2}\int_0^{Q(s)}dq f''(q)\int_{\bf k}\lambda({\bf k},q)^2\left(\nu({\bf k})-\frac{1}{\lambda({\bf k},q)}\right)+\frac{1}{2}\int_{\bf k}\left[\ln\left(1+\frac{s G_0({\bf k})}{\lambda({\bf k},s)}\right)-s\frac{G_0({\bf k})}{\lambda({\bf k},s)}\right]\\
    &+\frac{s^2}{2}f'(Q(s))-\frac{s}{2}\int_{0}^{Q(s)}dq\, z(q) f'(q)+\frac{s}{2}\int_{\bf k}\left[-v({\bf k})+G_0({\bf k})\left(\frac{1}{\lambda({\bf k},s)}-\nu({\bf k})\right)\right]\\
    =&-\frac{s}{2}\int_0^{Q(s)}dq \left(s f'(Q(s))+\int_{Q(s)}^q dr\,z(r) f''(r)\right)f''(q)\int_{\bf k}\lambda({\bf k},q)^2-\frac{s}{2}\int_{0}^{Q(s)}dq\, z(q) f'(q)\\
    &+\frac{1}{2}\int_{\bf k}\left[\ln\left(1+\frac{s G_0({\bf k})}{\lambda({\bf k},s)}\right)-s\frac{G_0({\bf k})}{\lambda({\bf k},s)}\right]+\frac{s^2}{2}f'(Q(s))-\frac{s}{2}\int_{\bf k}v({\bf k})-\frac{s^2}{2}f'(Q(s))\int_{\bf k}G_0({\bf k})\\
    =&\frac{s^2}{2}\left(f(Q(s))-Q(s) f'(Q(s))\right)+\frac{1}{2}\int_{\bf k}\left[\ln\left(1-\frac{s f'[Q(s)]}{\nu({\bf k})}\right)+\frac{s f'(Q(s))}{\nu({\bf k})-s f'(Q(s))}\right]\\
    &+\frac{s}{2}\int_0^{Q(s)}z(q)\left[q f''(q)-f'(q)\right]-\frac{s f'(0)}{2}\int_{\bf k}\frac{1}{\nu({\bf k})-s f'(Q(s))+\int_0^{Q(s)}z(q)f''(q)dq}\;.\nn
\end{align}

Let us now consider the expression of the variance. The latter is computed from the properties of the typical solution obtained in the limit $s\to 0$. In particular, the important parameter is $Q_{\rm typ}=Q(s=0)=\int_{\bf k}g({\bf k},0)$. In that limit, the eigenvalue function and overlap read
\begin{align}
    \lambda({\bf k},z)&=\frac{1}{\lambda({\bf k},0)^{-1}-z\sigma(z)+\int_{0}^z du\,\sigma(u)}=\frac{1}{\nu({\bf k})-z\sigma(z)+\int_{0}^z du\,\sigma(u)}\;,\\
    Q(z)&={f''}^{-1}\left(\frac{1}{\int_{\bf k}\left(\nu({\bf k})-z\sigma(z)+\int_{0}^z du\,\sigma(u)\right)^{-2}}\right)={f''}^{-1}\left(\frac{1}{I_2\left(\mu-z\sigma(z)+\int_{0}^z du\,\sigma(u)\right)}\right)\;.
\end{align}
Taking in particular its value as $z\to 0$, one obtains in the FRSB phase
\be
Q_{\rm typ}=Q(0)={f''}^{-1}\left(\frac{1}{I_2\left(\mu\right)}\right)\;.
\ee
Using that expression, the variance can be computed explicitly as a function of $\mu$ as
\begin{align}
{\cal V}_{L}(\mu)=&f(Q_{\rm typ})-\frac{f'(Q_{\rm typ})^2}{2f''(Q_{\rm typ})}
=f\left({f''}^{-1}\left(\frac{1}{I_2(\mu)}\right)\right)
-\frac{I_2(\mu)}{2}f'\left({f''}^{-1}\left(\frac{1}{I_2(\mu)}\right)\right)^2\;.
\end{align}
The variance can be shown to display a second order phase transition as $\mu\to \mu_{c}$, with 
\be
{\cal V}_{\rm FRSB}(\mu)-{\cal V}_{\rm RS}(\mu)=2\frac{I_3(\mu_{c})^2 f''(0)^3}{f^{(3)}(0)}f'(0)(\mu-\mu_{c})^2+O(\mu-\mu_{c})^3\;.\label{var_trans_full}
\ee
Comparing this result with Eq. \eqref{var_trans_1rsb}, one obtains that the leading order behaviour at the transition is independent of whether the RSB transition is from a 1RSB or FRSB phase as long as the transition is continuous. From the identity
\be
\int_{\bf k} \frac{f''(Q(s))}{\left(\nu({\bf k})-s f'(Q(s))\right)^2}=\int_{\bf k} \frac{f''(0)}{\left(\nu({\bf k})-s f'(Q(s))+\int_0^{Q(s)}z(q)f''(q)dq\right)^2}=1\;,\label{Q_s_eq_f}
\ee
one can derive the additional identity 
\be
Q'(s)=\frac{f'(Q(s))}{(z(Q(s))-s)f''(Q(s))}\;.
\ee
It can then be used to show that one can associate an energy $e_{L,\rm FRSB}(s,\mu)$ to each value of $s$, of the form 
\begin{align}
e_{L,\rm FRSB}(s,\mu)&=-\partial_s\varphi_{L,\rm FRSB}(s,\mu)\nn\\
&=-s f(0)-\frac{1}{2}\int_0^{Q}(z(q)-s)\left[q f''(q)-f'(q)\right]+\frac{f'(0)}{2}\int_{\bf k}\frac{1}{\nu({\bf k})-s f'(Q)+\int_0^{Q}z(q)f''(q)dq}\;.\label{E_FRSB_param}
\end{align}
Taking an additional derivative with respect to $s$, one can check explicitly that the CGF is convex with
\be
\partial_s^2\varphi_{L,\rm FRSB}(s,\mu)=f(Q(s))-\frac{f'(Q(s))^2}{2f''(Q(s))}> 0\;,\label{convex_FRSB}
\ee
for any $0\leq Q(s)<\infty$.
As the CGF is convex, one can compute explicitly its Legendre transform and obtain the large deviation function parametrically in terms of $s$ as
\begin{align}
    \Lambda_{L,\rm FRSB}(e=e_{L,\rm FRSB}(s,\mu),\mu)=&s\partial_s\varphi_{L,\rm FRSB}(s,\mu)-\varphi_{L,\rm FRSB}(s,\mu)\nn\\
    =&\frac{s^2}{2}f\left(Q(s)\right)-\frac{1}{2}\int_{\bf k}\left[\ln\left(1-\frac{s f'(Q(s))}{\nu({\bf k})}\right)+\frac{s f'(Q(s))}{\nu({\bf k})-s f'(Q(s))}\right]\;.\label{L_FRSB}
\end{align}
For any $\mu>0$, the large deviation function behaves quadratically in the vicinity of the typical energy
\begin{align}
\Lambda_{L,\rm FRSB}(e)&=\frac{\displaystyle(e-e_{\rm typ})^2}{\displaystyle 2{\cal V}_L(\mu)}+O(e-e_{\rm typ})^3\;,\\
{\rm with}\;\;e_{\rm typ}&=e_{L,\rm FRSB}(0,\mu)=-\frac{1}{2}\int_0^{Q_{\rm typ}}z(q)\left[q f''(q)-f'(q)\right]+\frac{f'(0)}{2}\int_{\bf k}\frac{1}{\nu({\bf k})+\int_0^{Q_{\rm typ}}z(q)f''(q)dq}\;.
\end{align}

Let us consdier the transition between the RS and FRSB branch of the cumulant generating function and large deviation function. Using the definition of $s_{\rm AT}$ in Eq. \eqref{s_AT_eta} as well as the equation for $Q(s)$ in Eq. \eqref{Q_s_eq_f_1}, one can check explicitly that $Q(s_{\rm AT})=0$. Comparing Eq. \eqref{convex_FRSB} with Eq. \eqref{phi_s_AT}, one concludes that the transition is at least of third order. Expanding the CGF difference, one obtains explicitly 
\be
\varphi_{L,\rm FRSB}(s,\mu)-\varphi_{L,\rm RS}(s,\mu)=\frac{I_3(\mu_c)^2 s_{\rm AT}(f'(0)f''(0))^3}{3\left(f^{(3)}(0)+2s_{\rm AT}f''(0)^3 I_3(\mu_c)\right)}(s-s_{\rm AT})^3+O(s-s_{\rm AT})^4\;,
\ee
confirming that the transition is of third order, except for $\mu=\mu_{c}$ where $s_{\rm AT}=0$ and the transition is of fourth order, i.e.
\be
\varphi_{L,\rm FRSB}(s,\mu)-\varphi_{L,\rm RS}(s,\mu)=\frac{I_3(\mu)^2 (f'(0)f''(0))^3}{3f^{(3)}(0)}s^4+O(s)^5\;,\;\;\mu=\mu_{c}\;.
\ee
Similarly the large deviation function difference behaves as
\be
\Lambda_{L,\rm FRSB}(e,\mu)-\Lambda_{L,\rm RS}(e,\mu)=\frac{I_3(\mu_c)^2 s_{\rm AT}(f'(0)f''(0))^3}{3{\cal V}_L(\mu_c)^3\left(f^{(3)}(0)+2s_{\rm AT}f''(0)^3 I_3(\mu_c)\right)}(e-e_{\rm AT})^3+O(e-e_{\rm AT})^4\;,
\ee
which can be obtained simply from the behaviour of the CGF above and the identity 
\be
e_{\rm FRSB}(s)\approx e_{\rm RS}(s)\approx  e_{\rm AT}+{\cal V}_L(\mu_c)\left(s-s_{\rm AT}\right)+O\left(s-s_{\rm AT}\right)^2\;.
\ee
By comparing the result above with Eq. \eqref{cont_trans_1rsb}, one concludes that the leading behaviour of the large deviation function difference at the continuous transition is independent of the nature of the RSB phase (1RSB or FRSB).

Finally, let us consider the asymptotic limit as $s\to -\infty$. In that limit, one expects that $Q(s)>0$ and thus $\int_0^{Q(s)}z(q)f''(q)dq>0$. As $s\to -\infty$, one must ensure that $f'(Q(s))\to 0$ to ensure that the left-hand-side of Eq. \eqref{Q_s_eq_f} remains finite. Thus, in the limit, one must have that $Q(s)\to +\infty$ as $s\to -\infty$. On the other hand, from Eq. \eqref{Q_s_eq_f_1}, as $f''(Q(s))\to 0$ in that limit, the integral $I_2(\mu-s f'(Q(s)))$ must diverge for their product to remain finite. Thus, one must have that at leading order $s f'(Q(s))\to \min_{\bf k}\nu({\bf k})=\mu$. The energy can be rewritten as
\begin{align}
e_{L,\rm FRSB}(s,\mu)=&-s f(Q(s))+\frac{s Q(s) f'(Q(s))}{2}-\frac{1}{2}\int_0^{Q(s)}z(q)\left[q f''(q)-f'(q)\right]\\
&+\frac{f'(0)}{2}\int_{\bf k}\frac{1}{\nu({\bf k})-s f'(Q)+\int_0^{Q}z(q)f''(q)dq}\;.\nn    
\end{align}
In the limit $s\to -\infty$, one obtains at leading order $e_{L,\rm FRSB}(s,\mu)\sim \mu Q(s)/2$. Using that behaviour as well as the expression of the large-deviation function in Eq. \eqref{L_FRSB}, one obtains that the asymptotic behaviour as $s\to -\infty$, i.e. as $\epsilon\to +\infty$ reads
\be
\Lambda_{L,\rm FRSB}(e,\mu)\approx\frac{\mu^2 f\left(2e/\mu\right)}{2f'\left(2e/\mu\right)^2}\;,\;\;e\to +\infty\;.
\ee
Thus for any positive $\mu>0$, comparing with the result obtained in the 1RSB pattern in Eq. \eqref{asymp_1rsb}, the asymptotic behaviour of the large deviation function as $e\to +\infty$ does not seem to depend on the RSB pattern or the dimension but depends explicitly on the correlation function of the disorder. For models displaying a FRSB pattern as well and $\mu>0$, as the large deviation function is well-defined for any $e\in \mathbb{R}$, one does not expect a regime of large deviation with speed $N^2$, in contrast to what has been observed for spherical spin-glasses \cite{lacroix2024replica,huang2023constructive}. In the following section, we apply our general results to a few special cases and consider in particular some cases where the unconfined limit $\mu\to 0$ can be taken.

This concludes our general analysis of the general properties of the different branch of the large deviation function. In the next section, we consider some special explicit cases.  


\section{Additional explicit results: exponential correlation function} \label{sec-special-cases}

In the following, we analyse the expression of the large deviation function in the continuous limit for an elastic term of the form $\nu({\bf k})=\mu+{\bf k}^2$. The function that we consider in particular is 
\be
{\cal L}(\epsilon,\mu)=\lim_{L\to \infty}\Lambda_L(e_{L,\rm typ}(\mu)+\epsilon,\mu)\;.
\ee
The analysis of the massless limit is done in some detail in section \ref{massless-sec}. In this section, we consider the exponential correlation function $f_{\infty}(q)=\exp(-c q)$ which corresponds to short-range disorder. The RSB pattern is of 1RSB type in dimensions $d<2$, it is marginal for $d=2$ and of FRSB type for $d>2$. In the following we detail the cases $d=1$ and $d=2$.

\subsection{Dimension one - 1RSB} \label{exp-1d}

In one dimension, the expression of the typical ground-state energy reads
\be
e_{\rm typ}(\mu)=\lim_{L\to \infty}e_{L,\rm typ}(\mu)=-\frac{c}{4\sqrt{\mu}}+w_{\rm typ}(\mu)\left(e^{-c Q_{\rm typ}(\mu)}-1+\frac{c Q_{\rm typ}(\mu)}{2} \left[1+e^{-c Q_{\rm typ}(\mu)}\right]\right)\;,
\ee
where $Q_{\rm typ}(\mu)=0$ and $w_{\rm typ}(\mu)=0$ in the RS phase, i.e. for $\mu>\mu_c=(c/2)^{4/3}$, and is given in the 1RSB phase by the non-trivial solution of
\begin{align}
Q_{\rm typ}(\mu)&=\frac{1}{2w_{\rm typ}(\mu)}\left(\frac{1}{\sqrt{\mu}}-\frac{1}{\sqrt{\mu+c w_{\rm typ}(\mu)(1-e^{-c Q_{\rm typ}(\mu)})}}\right)\;,\\
e^{-c Q_{\rm typ}(\mu)}(1+c Q_{\rm typ}(\mu))-1&=\frac{1}{w_{\rm typ}(\mu)^2}\left(\sqrt{\mu}-\frac{2\mu+c w_{\rm typ}(\mu)(1-e^{-c Q_{\rm typ}(\mu)})}{2\sqrt{\mu+c w_{\rm typ}(\mu)(1-e^{-c Q_{\rm typ}(\mu)})}}\right)\;.
\end{align}
The large deviation function displays both a RS and a 1RSB branch. The transition between the two branches is discontinuous (in terms of the order parameter $Q$) for $\mu<\mu_{\rm dis}=(c/2)^{4/3}/3=\mu_c/3$ and occurs at the energy 
\be
e_{\rm dis}(\mu)=-s_{\rm dis}(\mu)-\frac{c}{4\sqrt{\mu+c s_{\rm dis}(\mu)}}\;,
\ee
where $s_{\rm dis}(\mu)$ is the solution of the equation
\begin{align}
    Q_{\rm dis}(\mu)&=\frac{1}{2s_{\rm dis}(\mu)}\left(\frac{1}{\sqrt{\mu+c s_{\rm dis}(\mu) e^{-c Q_{\rm dis}(\mu)}}}-\frac{1}{\sqrt{\mu+c s_{\rm dis}(\mu)}}\right)\;,\\
    1-e^{-Q_{\rm dis}(\mu)}(1+c Q_{\rm dis}(\mu)) &=\frac{1}{s_{\rm dis}(\mu)^2}\left[\sqrt{\mu+c s_{\rm dis}(\mu)}-\sqrt{\mu+c s_{\rm dis}(\mu)e^{-c Q_{\rm dis}(\mu)}}-\frac{c s_{\rm dis}(\mu)(1-e^{-c Q_{\rm dis}(\mu)})}{2\sqrt{\mu+c s_{\rm dis}(\mu)}}\right]\;.
\end{align}
For $\mu>\mu_{\rm dis}=(c/2)^{4/3}/3$, the transition is continuous (in terms of the order parameter $Q$) and occurs at the energy 
\be
e_{\rm AT}(\mu)=\mu-\left(\frac{c}{2}\right)^{1/3}\;.
\ee
Using these expressions, one can plot the phase diagram in Fig. \ref{fig:phasediag-1rsb}. The transition line satisfies 
\be
\epsilon_{\rm t}(\mu)=\begin{cases}
   \displaystyle e_{\rm dis}(\mu)-e_{\rm typ}(\mu)=-s_{\rm dis}(\mu)-\frac{c}{4\sqrt{\mu+c s_{\rm dis}(\mu)}}-e_{\rm typ}(\mu)&\;,\;\;\mu\leq \mu_{\rm dis}=\frac{1}{3}\left(\frac{c}{2}\right)^{4/3}\;,\\
   \displaystyle e_{\rm AT}(\mu)-e_{\rm typ}(\mu)=\frac{\mu}{2}-\left(\frac{c}{2}\right)^{1/3}-e_{\rm typ}(\mu)&\;,\;\;\mu> \mu_{\rm dis}\;.
\end{cases}
\ee

The large deviation function can be expressed as
\begin{align}
    {\cal L}(\epsilon,\mu)&=\begin{cases}
       \displaystyle \frac{1}{2}\left(s_{\rm RS}(\epsilon)^2+\sqrt{\mu}-\sqrt{\mu+c s_{\rm RS}(\epsilon)}+\frac{c s_{\rm RS}(\epsilon)}{2\sqrt{\mu+s_{\rm RS}(\epsilon)}}\right)&\;,\;\;\epsilon\leq\epsilon_{\rm t}(\mu)\;,\\
       &\\
       \displaystyle \frac{1}{2}\left(s_{1}(\epsilon)^2 e^{-c Q_1(\epsilon)}+\sqrt{\mu}-\sqrt{\mu+c s_{1}(\epsilon)e^{-c Q_1(\epsilon)}}+\frac{c s_{1}(\epsilon)e^{-c Q_1(\epsilon)}}{2\sqrt{\mu+c s_{1}(\epsilon)e^{-c Q_1(\epsilon)}}}\right)&\;,\;\;\epsilon>\epsilon_{\rm t}(\mu)\;,
    \end{cases}\\
\epsilon&=-s_{\rm RS}(\epsilon)-\frac{c}{4\sqrt{\mu+c s_{\rm RS}(\epsilon)}}-e_{\rm typ}(\mu)\;,\\
\epsilon&=-w(\epsilon)-(s_{1}(\epsilon)-w(\epsilon))e^{-c Q_1(\epsilon)}\left(1-\frac{c Q_1(\epsilon)}{2}\right)-\frac{1}{4\sqrt{\mu+c s_{1}(\epsilon)e^{-c Q_1(\epsilon)}+c w(\epsilon)(1-e^{-c Q_1(\epsilon)})}}-e_{\rm typ}(\mu)\;,
\end{align}
where the parameters $w(\epsilon)$ and $Q_1(\epsilon)$ are solution of
\begin{align}
    Q_1(\epsilon)=&\frac{1}{2w(\epsilon)}\left(\frac{1}{\sqrt{\mu+c s_{1}(\epsilon)e^{-c Q_1(\epsilon)}}}-\frac{1}{\sqrt{\mu+c s_{1}(\epsilon)e^{-c Q_1(\epsilon)}+c w(\epsilon)(1-e^{-c Q_1(\epsilon)})}}\right)\;,\\
   1-e^{-c Q_1(\epsilon)}(1+c Q_1(\epsilon))=&\frac{1}{w(\epsilon)^2}\left[\sqrt{\mu+c s_{1}(\epsilon)e^{-c Q_1(\epsilon)}+c w(\epsilon)(1-e^{-Q_1(\epsilon)})}-\sqrt{\mu+c s_{1}(\epsilon)e^{-c Q_1(\epsilon)}}\right]\\
   &-\frac{c(1-e^{-c Q_1(\epsilon)})}{2w(\epsilon)\sqrt{\mu+c s_{1}(\epsilon)e^{-c Q_1(\epsilon)}+c w(\epsilon)(1-e^{-c Q_1(\epsilon)})}}\;.\nn
\end{align}
In the massless limit, the ground-state energy is finite and reads
\be
e_{\rm typ}(0)=-3\left(\frac{c}{4}\right)^{1/3}\;.
\ee
The large deviation function only displays a RS branch and is given parametrically as
\begin{align}
    {\cal L}(\epsilon)={\cal L}(\epsilon,0)=\frac{\sqrt{s_{\rm RS}(\epsilon)}}{2}\left(s_{\rm RS}(\epsilon)^{3/2}-\frac{\sqrt{c}}{2}\right)\;,\;\;\epsilon=-s_{\rm RS}(\epsilon)-\frac{1}{4}\sqrt{\frac{c}{s_{\rm RS}(\epsilon)}}+3\left(\frac{c}{4}\right)^{1/3}\;,\;\;\epsilon\leq 0\;.
\end{align}
The LDF vanishes linearly in that limit with ${\cal L}'(0)=-(c/2)^{1/3}$.

\subsection{Dimension two - marginal RSB}

The expression for the large deviation function is much more explicit in dimension $d=2$. In that case, the transition occurs at a value of the energy difference satisfying 
\begin{align}
  \epsilon_t(\mu)&=\epsilon_{\rm AT}(\mu)=\frac{\mu}{c}-\frac{c}{4\pi}-\frac{c}{8\pi}\ln\frac{4\pi \mu}{c^2}-\epsilon_{\rm RSB}(\mu)\Theta(\mu_c-\mu)\;,\\
  \epsilon_{\rm RSB}(\mu)&=\frac{\mu}{c}-\frac{c}{4\pi}-\frac{c^2+4\pi \mu}{8\pi c}\ln\frac{4\pi \mu}{c^2}\;,
\end{align}
where $\mu_c=\frac{c^2}{4\pi}$. The large deviation function reads
\be
{\cal L}(\epsilon,\mu)=\begin{cases}
   \displaystyle {\cal L}_{\rm RS}(\epsilon+\epsilon_{\rm RSB}(\mu)\Theta(\mu_c-\mu),\mu)\;,\;\;\epsilon\leq \epsilon_{\rm AT}(\mu)\;,\\
   \displaystyle {\cal L}_{\rm RSB}(\epsilon-\epsilon_{\rm RSB}(\mu)\Theta(\mu-\mu_c),\mu)\;,\;\;\epsilon> \epsilon_{\rm AT}(\mu)\;.
\end{cases}
\ee
where the RS branch reads 
\begin{align}
 {\cal L}_{\rm RS}(\epsilon,\mu)=&\frac{\mu}{8\pi} \left(1+\frac{4\pi \mu}{c^2}+W_{-1}\left(-\frac{8\pi \mu}{c^2}e^{\frac{8\pi(c\epsilon-\mu)}{c^2}}\right)+\ln\left(-\frac{c^2}{8\pi \mu}W_{-1}\left(-\frac{8\pi \mu}{c^2}e^{\frac{8\pi(c\epsilon-\mu)}{c^2}}\right)\right)\right) \\
 &+\frac{c^2}{128\pi^2}W_{-1}\left(-\frac{8\pi \mu}{c^2}e^{\frac{8\pi(c\epsilon-\mu)}{c^2}}\right)\left(2+W_{-1}\left(-\frac{8\pi \mu}{c^2}e^{\frac{8\pi(c\epsilon-\mu)}{c^2}}\right)\right)\;,\nn
\end{align}
with $W_{-1}(x)$ is the $-1$ branch of the Lambert W function, while the RSB branch reads
\be
{\cal L}_{\rm RSB}(\epsilon,\mu)=\frac{\mu}{8\pi}\left(e^{\frac{2c\epsilon}{\mu}}-\frac{2c\epsilon}{\mu}-1\right)\;.
\ee
Using the expressions above, one can check that the transition between the two branches is generally of third order apart for $\mu=\mu_{c}$ where the transition is of fourth order. Similarly, both branches vanish quadratically as $\epsilon\to 0$ for any finite $\mu$, which is consistent with a central limit theorem for typical fluctuation. In the limit where $\mu\to 0$, the large deviation function only displays an RS branch and reads
\be
{\cal L}(\epsilon)=\lim_{\mu\to 0}{\cal L}_{\rm RS}(\epsilon+\epsilon_{\rm RSB}(\mu),\mu)=\frac{c^2}{128\pi^2}W_{-1}\left(-2e^{\frac{8\pi}{c}\epsilon-2}\right)\left(2+W_{-1}\left(-2e^{\frac{8\pi}{c}\epsilon-2}\right)\right)\;,\;\;\epsilon\leq 0\;.
\ee
Note that in this limit instead, the large deviation function vanishes linearly with ${\cal L}'(0)=-c/(4\pi)$. 

\subsection{Dimension three - FRSB} \label{exp-3d}

In dimension three, the large deviation function displays a RS and a FRSB branch. The transition occurs at an energy difference
\begin{align}
\epsilon_t(\mu)=\epsilon_{\rm AT}(\mu)&=\frac{\mu}{c}-\frac{c\sqrt{\mu}}{8\pi}-\epsilon_{\rm FRSB}(\mu)\Theta\left(\mu_c-\mu\right)\;,\\
\epsilon_{\rm FRSB}(\mu)&=\frac{1}{4}\left(\frac{c^3}{(8\pi)^2}- \frac{c\sqrt{\mu}}{2\pi}+\frac{3\mu}{c}-\frac{\mu}{c}\ln\frac{(8\pi)^2\mu}{c^4}\right)\;,
\end{align}
with $\mu_c=\left(c^2/(8\pi)\right)^2$ and where we used that in $d=3$ one has $z(q)=-f^{(3)}(q)/(32\pi^2)=(c^3/(32\pi^2)) e^{-c q}$ and $Q_{\rm typ}(\mu)=1/(2c)\ln(\mu_c/\mu)\Theta(\mu_c-\mu)$. Using the expression above, one can obtain the phase diagram in Fig. \ref{fig:phasediag-frsb}. The large deviation function reads
\be
{\cal L}(\epsilon,\mu)=\begin{cases}
   \displaystyle {\cal L}_{\rm RS}(\epsilon+\epsilon_{\rm FRSB}(\mu)\Theta(\mu_c-\mu),\mu)&\;,\;\;\epsilon\leq \epsilon_{\rm AT}(\mu)\;,\\
   \displaystyle {\cal L}_{\rm FRSB}(\epsilon+\epsilon_{\rm FRSB}(\mu)\Theta(\mu_c-\mu),\mu)&\;,\;\;\epsilon> \epsilon_{\rm AT}(\mu)\;,
   \end{cases}
\ee
where the RS branch is expressed as
\begin{align}
{\cal L}_{\rm RS}(\epsilon)=&\frac{s_{\rm RS}(\epsilon)^2}{2}+\frac{2\mu\sqrt{\mu+c s_{\rm RS}(\epsilon)}-2\mu^{3/2}-c s_{\rm RS}(\epsilon)\sqrt{\mu+c s_{\rm RS}(\epsilon) }}{24\pi}\;,\\
{\rm with}\;\;s_{\rm RS}(\epsilon)=&-\epsilon+\frac{c}{2(8\pi)^2}\left(c^2-16\pi \sqrt{\mu}+\sqrt{(c^2-16\pi \sqrt{\mu})^2-4(8\pi)^2 c\epsilon}\right)\;,
\end{align}
while the FRSB branch is expressed as
\begin{align}
{\cal L}_{\rm FRSB}(\epsilon)&=\frac{c^6}{6(8\pi)^4}e^{-3c Q(\epsilon)}+\frac{\mu^2}{2c^2}e^{c Q(\epsilon)}-\frac{\mu^{3/2}}{12\pi}\;,\\
{\rm with}\;\;Q(\epsilon)&=\frac{2\epsilon}{\mu}-\frac{(c^2-16\pi\sqrt{\mu})^2}{2(8\pi)^2\mu c}+\frac{1}{2c} W_{-1}\left(\frac{c^4}{(8\pi)^2\mu}e^{-2c\left(\frac{2\epsilon}{\mu}-\frac{(c^2-16\pi\sqrt{\mu})^2}{2(8\pi)^2\mu c}\right)}\right)\;. 
\end{align}
In the massless limit, the transition occurs at the value
\be
\epsilon_t(0)=\epsilon_{\rm AT}(0)=-\epsilon_{\rm FRSB}(0)=-\frac{c^3}{4(8\pi)^2}\;.
\ee
The LDF still has two branches given by 
\be
{\cal L}(\epsilon)={\cal L}(\epsilon,0)=\begin{cases}
   \displaystyle \frac{1}{2}\left(-\epsilon+\frac{c^3}{4(8\pi)^2}+\frac{c}{8\pi}\sqrt{-c \epsilon}\right)^2-\frac{c^{3/2}}{24\pi}\left(-\epsilon+\frac{c^3}{4(8\pi)^2}+\frac{c}{8\pi}\sqrt{-c \epsilon}\right)^{3/2}&\;,\;\;\epsilon\leq \epsilon_{\rm AT}(0)\;,\\
   \displaystyle \frac{(-c \epsilon)^{3/2}}{6\pi}&\;,\;\;\epsilon> \epsilon_{\rm AT}(0)\;.
   \end{cases}
\ee
In particular, the LDF vanishes with a nontrivial exponent $3/2$. This concludes our analysis of the disorder with (short-range) exponential correlation function.

\section{Conclusion and discussion} \label{sec-conclusion}


We have computed explicitly the cumulant generating function and large deviation function of the rescaled ground-state energy $e_{\min}=\frac{1}{NL^d}\min_{\bf u}{\cal H}[{\bf u}]$ of the elastic manifold with internal dimension $0<d<4$ and in the limit $N\gg 1$ using replica computations. We have shown that for any positive mass $\mu>0$, the typical fluctuations of $e_{\min}$ are normal and of order $\sim (NL^d)^{-1/2}$. The associated cumulant generating function is expressed as a Parisi formula. The large deviation function $\Lambda_L(e,\mu)$ of speed $NL^d$, describing atypical fluctuations of order $O(1)$ of the energy is well defined for any $e\in \mathbb{R}$ and displays both a replica symmetric and replica symmetry broken branch for any $\mu>0$. In the massless and continuous limit $\mu=0$, the ground-state energy super-concentrates and typical fluctuations are of order $\sim N^{-1/\xi}$ where $\xi<2$ is computed explicitly. In that limit, the large deviation function ${\cal L}(\epsilon)=\lim_{L\to \infty}\Lambda_L(\epsilon+e_{L,\rm typ}(0),0)$ of speed $NL^d$ is one-sided, describing only atypical fluctuations with $\epsilon<0$ and vanishes with an exponent ${\cal L}(\epsilon)\sim (-\epsilon)^{\xi}$. We thus conjecture that the distribution of typical fluctuations displays a left tail with the same exponent $\xi$.

\begin{acknowledgments}
The research by YVF has been supported by  EPSRC grant UKRI1015 "Non-Hermitian random matrices: theory and applications". PLD acknowledges support from ANR Grant No. ANR-23-CE30-0020-01 EDIPS.

\end{acknowledgments}

\newpage

\begin{appendix}



\section{Typical ground-state energy}\label{app_typ_GS}

The average and typical ground-state energy $e_{\rm typ}\equiv e_{L,\rm typ}(\mu)$ only takes a finite value for an internal dimension $d<2$. It is obtained as
\be
e_{\rm typ}=\lim_{NL^d\to \infty}\moy{e_{\min}}=-\partial_s\varphi_L(0,\mu)=\displaystyle\max_{g({\bf k},z),G_0({\bf k}),v({\bf k}),w_M}{\cal S}_{L,\rm typ}(\mu)
\ee
where the functional ${\cal S}_{\rm typ}$ is defined as
\begin{align}
    {\cal S}_{L,\rm typ}(\mu)=&\frac{1}{2}\int_{\bf k}\nu({\bf k})g({\bf k},0)+\frac{1}{2}\left[f'(0)\int_{\bf k}v({\bf k})-w_M\,f\left(0\right)+\int_{0}^{w_M}dz\,f\left(\int_{\bf k}g({\bf k},z)\right)\right]\\
    &+\frac{1}{2}\int_{\bf k}\left[\frac{1}{w_M}\ln \frac{v({\bf k})}{\lambda({\bf k},0)}+\int_{0}^{w_M}\frac{dz}{z^2}\ln\frac{\lambda({\bf k},z)}{\lambda({\bf k},0)}\right]+\frac{1}{2}\int_{\bf k}G_0({\bf k})\left[\nu({\bf k})-\frac{1}{\lambda({\bf k},0)}\right]\;.\nn
\end{align}
where the function $\lambda({\bf k},z)$ is defined as
\be
\lambda({\bf k},z)=v({\bf k})+z\,g({\bf k},z)+\int_{z}^{w_M}dw\,g({\bf k},w)\;,
\ee
and the function $g({\bf k},z)$ satisfies the boundary condition $g({\bf k},z\geq w_M)=0$.

Let us denote with an index $_{\rm typ}$ the optimal value of the different parameters to optimise. The corresponding minimazation equations read
\begin{align}
    &\frac{\partial {\cal S}_{\rm typ}}{\partial w_M}[g_{\rm typ},v_{\rm typ},G_{\rm typ},w_{\rm typ}]=0\label{S-typ-sp-w}\\
    &=\frac{f\left(\int_{\bf k}g_{\rm typ}({\bf k},w_{\rm typ})\right)-f(0)}{2}+\int_{\bf k}\frac{g_{\rm typ}({\bf k},w_{\rm typ})}{2}\left[\int_0^{w_{\rm typ}}\frac{1}{z^2}\left(\frac{1}{\lambda_{\rm typ}({\bf k},0)}-\frac{1}{\lambda_{\rm typ}({\bf k},z)}\right)+\frac{G_{\rm typ}({\bf k})}{\lambda_{\rm typ}({\bf k},0)^2}-\frac{1}{w_{\rm typ}\lambda_{\rm typ}({\bf k},0)}\right]\;,\nn\\
    &\frac{\delta {\cal S}_{\rm typ}}{\delta G_0({\bf k})}[g_{\rm typ},v_{\rm typ},G_{\rm typ},w_{\rm typ}]=0=\frac{1}{2}\left[\nu({\bf k})-\frac{1}{\lambda_{\rm typ}({\bf k},0)}\right]\;,\label{S-typ-sp-G}\\
    &\frac{\delta {\cal S}_{\rm typ}}{\delta v({\bf k})}[g_{\rm typ},v_{\rm typ},G_{\rm typ},w_{\rm typ}]=0\label{S-typ-sp-v}\\
    &=\frac{1}{2}\left[f'(0)+\frac{G_{\rm typ}({\bf k})}{\lambda_{\rm typ}({\bf k},0)^2}+\frac{1}{w_{\rm typ}}\left(\frac{1}{v_{\rm typ}({\bf k})}-\frac{1}{\lambda_{\rm typ}({\bf k},0)}\right)+\int_0^{w_{\rm typ}}\frac{dz}{z^2}\left(\frac{1}{\lambda_{\rm typ}({\bf k},z)}-\frac{1}{\lambda_{\rm typ}({\bf k},0)}\right)\right]\;,\nn\\
    &\frac{\delta {\cal S}_{\rm typ}}{\delta g({\bf k},z_0)}[g_{\rm typ},v_{\rm typ},G_{\rm typ},w_{\rm typ}]=0\label{S-typ-sp-g}\\
    &=\frac{1}{2}\left[f'\left(\int_{\bf k}g_{\rm typ}({\bf k},z_0)\right)+\frac{G_{\rm typ}({\bf k})}{\lambda_{\rm typ}({\bf k},0)^2}+\frac{1}{z_0}\left(\frac{1}{\lambda_{\rm typ}({\bf k},z_0)}-\frac{1}{\lambda_{\rm typ}({\bf k},0)}\right)+\int_0^{z_0}\frac{dz}{z^2}\left(\frac{1}{\lambda_{\rm typ}({\bf k},z)}-\frac{1}{\lambda_{\rm typ}({\bf k},0)}\right)\right]\;.\nn
\end{align}
Note that the first equation is trivially ensured by imposing the boundary condition $g_{\rm typ}({\bf k},z\geq w_{\rm typ})=0$. Taking the limit $z_0\to 0$ in Eq. \eqref{S-typ-sp-g} and using Eq. \eqref{S-typ-sp-G} yields the simple relation
\be
G_{\rm typ}({\bf k})=-\frac{\displaystyle f'\left(\int_{\bf k}g_{\rm typ}({\bf k},0)\right)}{\displaystyle\nu({\bf k})^2}=-\frac{f'\left(Q_{\rm typ}\right)}{\nu({\bf k})^2}\;,\;\;{\rm with}\;\;Q_{\rm typ}=\int_{\bf k}g_{\rm typ}({\bf k},0)\;.
\ee
In the replica-symmetric phase, corresponding either to $w_{\rm typ}=0$ or equivalently to $g({\bf k},z)\equiv 0$, the parameters take simple expressions
\be
\lambda_{\rm typ}({\bf k},0)=v_{\rm typ}({\bf k})=\frac{1}{\nu({\bf k})}\;,\;\;G_{\rm typ}({\bf k})=-\frac{f'\left(0\right)}{\nu({\bf k})^2}\;,\;\;Q_{\rm typ}=0\;.
\ee
The expression for the average ground-state energy simplifies drastically, yielding
\be
e_{L,\rm RS}(\mu)=-\frac{f'(0)}{2}\int_{\bf k}\frac{1}{\nu({\bf k})}\;.
\ee
In the 1RSB phase instead the function $g_{\rm typ}({\bf k},z)=g_{\rm typ}({\bf k})\Theta(w_{\rm typ}-z)$, where $g_{\rm typ}({\bf k})$ and $w_{\rm typ}$ optimal values of the corresponding parameters. In addition to the equations above, which remain valid, one needs to consider the optimisation over $w_{\rm typ}$, yielding 
\be
\frac{1}{2}\left(f\left(\int_{\bf k}g_{\rm typ}({\bf k})\right)-f(0)\right)+\frac{1}{2}\int_{\bf k}\left[\frac{g_{\rm typ}({\bf k})}{\lambda_{\rm typ}({\bf k},0)}\left(\frac{G_{\rm typ}({\bf k})}{\lambda_{\rm typ}({\bf k},0)}-\frac{1}{w_{\rm typ}}\right)-\frac{1}{w_{\rm typ}^2}\ln\frac{v_{\rm typ}({\bf k})}{\lambda_{\rm typ}({\bf k},0)}\right]=0\;,
\ee
where $\lambda_{\rm typ}({\bf k},0)=v_{\rm typ}({\bf k})+w_{\rm typ}g_{\rm typ}({\bf k})$. The equations read explicitly
\begin{align}
  &\lambda_{\rm typ}({\bf k},0)=v_{\rm typ}({\bf k})+w_{\rm typ}g_{\rm typ}({\bf k})=\frac{1}{\nu({\bf k})}\;,\\
  &G_{\rm typ}({\bf k})=-\frac{f'\left(Q_{\rm typ}\right)}{\nu({\bf k})^2}\;,\;\;Q_{\rm typ}=\int_{\bf k}g_{\rm typ}({\bf k})\;,\\
  &f'(0)-f'\left(Q_{\rm typ}\right)+\frac{1}{w_{\rm typ}}\left(\left(\frac{1}{\nu({\bf k})}-w_{\rm typ}g_{\rm typ}({\bf k})\right)^{-1}-\nu({\bf k})\right)=f'(0)-f'\left(Q_{\rm typ}\right)+\frac{\nu({\bf k})^2 g_{\rm typ}({\bf k})}{1-\nu({\bf k}) w_{\rm typ}g_{\rm typ}({\bf k})}=0\;,
\end{align}
where $g_{\rm typ}({\bf k})$ can be obtained explicitly from the last equation and $Q_{\rm typ}$ can be shown to satisfy a self-consistent equation reading
\begin{align}
g_{\rm typ}({\bf k})&=\frac{1}{w_{\rm typ}}\left(\frac{1}{\nu({\bf k})}-\frac{1}{\nu({\bf k})+w_{\rm typ}(f'\left(Q_{\rm typ}\right)-f'(0))}\right)\;,\\
Q_{\rm typ}&=\int_{\bf k}\frac{1}{w_{\rm typ}}\left(\frac{1}{\nu({\bf k})}-\frac{1}{\nu({\bf k})+w_{\rm typ}(f'\left(Q_{\rm typ}\right)-f'(0))}\right)\;.\label{q_typ_eq_1_app}
\end{align}
The remaining equation can be rewritten as
\be
f\left(Q_{\rm typ}\right)-f(0)-Q_{\rm typ} f'\left(Q_{\rm typ}\right)=\frac{1}{w_{\rm typ}^2}\int_{\bf k}\left[\frac{w_{\rm typ}(f'\left(Q_{\rm typ}\right)-f'(0))}{\nu({\bf k})+w_{\rm typ}(f'\left(Q_{\rm typ}\right)-f'(0))}-\ln\left(1+\frac{w_{\rm typ}(f'\left(Q_{\rm typ}\right)-f'(0))}{\nu({\bf k})}\right)\right]\;.\label{q_typ_eq_2_app}
\ee
The expression of the ground-state energy can be simplified using these identities and reads
\be
e_{L,\rm 1RSB}(\mu)=e_{L,\rm RS}(\mu)+w_{\rm typ}\left(f(Q_{\rm typ})-f(0)-\frac{Q_{\rm typ}}{2} \left[f'(Q_{\rm typ})+f'(0)\right]\right)\;.
\ee
In particular, expanding in powers of $Q_{\rm typ}$, one observed that the continuous transition from the RS to the 1RSB phase, occurring as $Q_{\rm typ}\to 0$, is of third order. A priori, a discontinuous transition could occur if the parameter $w_{\rm typ}\to 0$ for a finite value of $Q_{\rm typ}$. Analysing Eqs. (\ref{q_typ_eq_1_app}-\ref{q_typ_eq_2_app}) in the limit $w_{\rm typ}\to 0$, one realises that the value $Q_{\rm d}$ at this transition must satisfy
\be
f(0)-f(Q_{\rm d})+\frac{Q_{\rm d}}{2} \left[f'(Q_{\rm d})+f'(0)\right]=0
\ee
for a value $Q_{\rm d}>0$. If the function $f'(q)$ is strictly concave, this equation only has the solution $Q_{\rm d}=0$. As for short-range potentials $f^{(3)}(q)=-\int dk\,k^6\,\tilde f(k)\,e^{-q k^2}<0$ with $\tilde f(k)\geq 0$ for any finite $q$, this is always the case.

Finally, in the full replica symmetry broken phase, One may rewrite Eq. \eqref{S-typ-sp-g} as
\be
f'\left(\int_{\bf k}g_{\rm typ}({\bf k},z)\right)=f'(Q_{\rm typ}(z))=\sigma_{\rm typ}(z)=\int_0^{z}dw\frac{\partial_w g_{\rm typ}({\bf k},w)}{\lambda_{\rm typ}({\bf k},w)^2}-\frac{G_{\rm typ}({\bf k})}{\lambda_{\rm typ}({\bf k},0)^2}\;,\label{eq_FRSB_typ}
\ee
which must be valid in the whole range $z_0\in[0,w_{\rm typ}]$. One may check explicitly that the function $\sigma_{\rm typ}(z)$ is the Parisi function associated to the coefficients $G_l({\bf k})^{-1}$. Thus, there exists a non-trivial relationship between the Parisi function $g_{\rm typ}({\bf k},z)$ with coefficients $G_l({\bf k})$ and the Parisi function $\sigma_{\rm typ}(z)$ which is most simply expressed as
\begin{align}
-z\sigma_{\rm typ}(z)+\int_{0}^z du\,\sigma_{\rm typ}(u)&=\frac{1}{\lambda_{\rm typ}({\bf k},z)}-\frac{1}{\lambda_{\rm typ}({\bf k},0)}\\
&=\frac{1}{v_{\rm typ}({\bf k})+z g_{\rm typ}({\bf k},z) +\int_{z}^{w_{\rm typ}} du\,g_{\rm typ}({\bf k},u)}-\frac{1}{v_{\rm typ}({\bf k})+\int_{0}^{w_{\rm typ}} du\,g_{\rm typ}({\bf k},u)}\nn\\
&=\left(\frac{1}{\nu({\bf k})}+z g_{\rm typ}({\bf k},z) +\int_{z}^{w_{\rm typ}} du\,g_{\rm typ}({\bf k},u)\right)^{-1}-\nu({\bf k})\;.\nn
\end{align}
An explicit expression for $\lambda_{\rm typ}({\bf k},z)$ in terms of $\sigma_{\rm typ}$ can thus be derived, reading
\be
\lambda_{\rm typ}({\bf k},z)=\frac{1}{\nu({\bf k})-z\sigma_{\rm typ}(z)+\int_{0}^z du\,\sigma_{\rm typ}(u)}=\frac{1}{\nu({\bf k})-\int_{0}^z du\,z\,\sigma_{\rm typ}'(u)}=\frac{1}{\nu({\bf k})+\int_{Q_{\rm typ}(z)}^{Q_{\rm typ}} du\,z_{\rm typ}(q)\,f''(q)}\;,
\ee
where we have used that $\sigma_{\rm typ}(z)=f'(Q_{\rm typ}(z))$ and defined $Q_{\rm typ}\equiv Q_{\rm typ}(0)$.

Taking a derivative with respect to $z_0$ of Eq. \eqref{eq_FRSB_typ} yields
\be
\int_{\bf k}\partial_{z}g_{\rm typ}({\bf k},z)f''\left(\int_{\bf k}g_{\rm typ}({\bf k},z)\right)-\frac{\partial_{z} g_{\rm typ}({\bf k},z)}{\lambda_{\rm typ}({\bf k},z)^2}=0\;.
\ee
After multiplication by $\lambda_{\rm typ}({\bf k},z)^2$ and integration over ${\bf k}$, one obtains the identity 
\be
\int_{\bf k}\lambda_{\rm typ}({\bf k},z)^2=\frac{1}{\displaystyle f''\left(\int_{\bf k}g_{\rm typ}({\bf k},z)\right)}\;,\;\;z\in[0,w_{\rm typ}]\;.
\ee
Introducing $Q_{\rm typ}(z)=\int_{\bf k}g_{\rm typ}({\bf k},z)$, one obtains the identity
\be
I_2\left(\mu-zf'(Q_{\rm typ}(z))+\int_{0}^z du\,f'(Q_{\rm typ}(u))\right)f''(Q_{\rm typ}(z))=1\;,\;\;z\in[0,w_{\rm typ}]\;,
\ee
where we have used that 
\begin{align}
\int_{\bf k}\lambda_{\rm typ}({\bf k},z)^p&=\int_{\bf k}\frac{1}{\left(\nu({\bf k})-z\sigma_{\rm typ}(z)+\int_{0}^z du\,\sigma_{\rm typ}(u)\right)^p}\nn\\
&=I_p\left(\mu-zf'(Q_{\rm typ}(z))+\int_{0}^z du\,f'(Q_{\rm typ}(u))\right)=I_p\left(\mu+\int_{Q_{\rm typ}(z)}^{Q_{\rm typ}} dq\,z_{\rm typ}(q)\,f''(q)\right)\;.    
\end{align}
The function $z_{\rm typ}(q)$ is the inverse function of $Q_{\rm typ}(z)$ and $Q_{\rm typ}\equiv Q_{\rm typ}(0)$. If the function $f''$ is monotonic and thus invertible, one may thus obtain the identity
\be
Q_{\rm typ}(z)=\int_{\bf k}g_{\rm typ}({\bf k},z)={f''}^{-1}\left(\frac{1}{\displaystyle \int_{\bf k}\lambda_{\rm typ}({\bf k},z)^2}\right)={f''}^{-1}\left(\frac{1}{\displaystyle \int_{\bf k}\frac{1}{\left(\nu({\bf k})+\int_{Q_{\rm typ}(z)}^{Q_{\rm typ}} dq\,z_{\rm typ}(q)\,f''(q)\right)^2}}\right)\;.
\ee
In particular, one obtains that
\be
Q_{\rm typ}\equiv Q_{\rm typ}(0)={f''}^{-1}\left(\frac{1}{\displaystyle \int_{\bf k}\frac{1}{\nu({\bf k})^2}}\right)={f''}^{-1}\left(\frac{1}{I_2(\mu)}\right)\;.
\ee
Finally, the function $\sigma_{\rm typ}(z)$ is shown to satisfy the self-consistent equation
\be
\sigma_{\rm typ}(z)=f'\left(Q_{\rm typ}(z)\right)=f'\left[{f''}^{-1}\left(\frac{1}{\int_{\bf k}\lambda_{\rm typ}({\bf k},z)^2}\right)\right]=f'\left[{f''}^{-1}\left(\frac{1}{\displaystyle \int_{\bf k}\frac{1}{\left(\nu({\bf k})-z\sigma_{\rm typ}(z)+\int_{0}^z du\,\sigma_{\rm typ}(u)\right)^2}}\right)\right]\;.
\ee
Taking an additional derivative, one obtains the expression
\be
z=-\frac{\left(\int_{\bf k}\lambda_{\rm typ}({\bf k},z)^2\right)^{3}}{2\int_{\bf k}\lambda_{\rm typ}({\bf k},z)^3}f^{(3)}\left({f''}^{-1}\left(\frac{1}{\int_{\bf k}\lambda_{\rm typ}({\bf k},z)^2}\right)\right)=-\frac{f^{(3)}\left(Q_{\rm typ}(z)\right)}{2f''\left(Q_{\rm typ}(z)\right)^3 I_3\left(\mu-zf'(Q_{\rm typ}(z))+\int_{0}^z du\,f'(Q_{\rm typ}(u))\right)}\;.
\ee
Taking an additional derivative with respect to $z$, one obtains
\begin{align}
-\frac{1}{z\sigma_{\rm typ}'(z)}=&\frac{3}{2}\left(2-\frac{\int_{\bf k}\lambda_{\rm typ}({\bf k},z)^4\int_{\bf k}\lambda_{\rm typ}({\bf k},z)^2}{\left(\int_{\bf k}\lambda_{\rm typ}({\bf k},z)^3\right)^2}\right)\frac{f^{(3)}\left(Q_{\rm typ}(z)\right)}{f''\left(Q_{\rm typ}(z)\right)^2}-\frac{f^{(4)}\left(Q_{\rm typ}(z)\right)}{f^{(3)}\left(Q_{\rm typ}(z)\right)f''\left(Q_{\rm typ}(z)\right)}\label{sig_prime_typ}\;.
\end{align}

Finally, let us use the identity $\partial_z \lambda ({\bf k},z)=z\partial_z g({\bf k},z)$ to express 
\be
g({\bf k},z)=\int_{w_M}^{z}dz_0\frac{Q_{\rm typ}'(z_0)f''[Q_{\rm typ}(z_0)]}{\left(\nu({\bf k})+\int_{Q_{\rm typ}(z_0)}^{Q_{\rm typ}} du\,z_{\rm typ}(q)\,f''(q)\right)^2}=\int_{0}^{Q_{\rm typ}(z)}\frac{f''(q)dq}{\left(\nu({\bf k})+\int_{q}^{Q_{\rm typ}} dr\,z_{\rm typ}(r)\,f''(r)\right)^2}\;.
\ee

To simplify the expression of the typical energy, we re-express
\begin{align}
    {\cal S}_{\rm typ}+\frac{f'(0)}{2}\int_{\bf k}\frac{1}{\nu({\bf k})}=&\frac{1}{2}\int_{\bf k}\nu({\bf k})\left[g({\bf k},0)+G_0({\bf k})\right]+\frac{1}{2}\left[f'(0)\int_{\bf k}v({\bf k})-w_M\,f\left(0\right)+\int_{0}^{w_M}dz\,f\left(Q(z)\right)\right]\\
    &+\frac{1}{2}\int_{\bf k}\left[\frac{1}{w_M}\ln \frac{v({\bf k})}{\lambda({\bf k},0)}-\frac{G_0({\bf k})}{\lambda({\bf k},0)}+\int_{0}^{w_M}\frac{dz}{z^2}\ln\frac{\lambda({\bf k},z)}{\lambda({\bf k},0)}\right]\;,\nn\\
    =&\frac{f'(0)}{2}\int_{\bf k}\frac{1}{\nu({\bf k})+\int_{0}^{Q_{\rm typ}} dq\,z_{\rm typ}(q)\,f''(q)}+\frac{1}{2}\int_0^{w_M} dz\,z\,Q'(z)\,f'(Q(z))\nn\\
    &+\frac{1}{2}\int_{\bf k}\int_{0}^{w_M}\partial_z g({\bf k},z)\left[\frac{1}{\lambda({\bf k},z)}-\nu({\bf k})\right]\nn\\
    =&\frac{f'(0)}{2}\int_{\bf k}\frac{1}{\nu({\bf k})+\int_{0}^{Q_{\rm typ}} dq\,z_{\rm typ}(q)\,f''(q)}+\frac{1}{2}\int_0^{w_M} dz\,Q'(z)\,\left[z f'(Q(z))+\int_{Q(z)}^{Q_{\rm typ}}dq\,z_{\rm typ}(q)\,f''(q)\right]\nn\\
    =&\frac{f'(0)}{2}\int_{\bf k}\frac{1}{\nu({\bf k})+\int_{0}^{Q_{\rm typ}} dq\,z_{\rm typ}(q)\,f''(q)}-\frac{1}{2}\int_0^{Q_{\rm typ}} dq\,\left[z_{\rm typ}(q) f'(q)+\int_{q}^{Q_{\rm typ}}dr\,z_{\rm typ}(r)\,f''(r)\right]\nn
\end{align}
In the FRSB phase, the typical energy can thus be expressed as 
\begin{align}
e_{L,\rm FRSB}(\mu)=-\frac{1}{2}\int_0^{Q_{\rm typ}}z_{\rm typ}(q)\left[q f''(q)-f'(q)\right]+\frac{f'(0)}{2}\int_{\bf k}\frac{1}{\nu({\bf k})+\int_0^{Q_{\rm typ}}z_{\rm typ}(q)f''(q)dq}\;.  
\end{align}

\section{Complexity}\label{app_comp}

Let us compute the density of minima of the Hamiltonian at fixed energy density $e$, which reads
\begin{align}
    \rho_{\min}(e)=&\int {\cal D}{\bf u}({\bf x})\,\moy{\delta^N\left(\frac{\delta {\cal H}[{\bf u}]}{\delta {\bf u}({\bf x})}\right)\det {\cal K}[{\bf u}({\bf x})]\Theta({\cal K}[{\bf u}({\bf x})])\delta\left(NL^d e-{\cal H}[{\bf u}]\right)}\\
    =&\sum_{\alpha:{\rm minima}}\moy{\delta\left(\frac{{\cal H}[{\bf u}_{\alpha}]}{NL^d}-e\right)}\geq P_{\min}(e)=\moy{\delta\left(e_{\min}-e\right)}\;,\label{bound_1}
\end{align}
where ${\cal K}_{i {\bf x},j {\bf y}}=\frac{\delta^2 {\cal H}[{\bf u}]}{\delta u_i({\bf x})\delta u_j({\bf y})}$. In fact, one can compute exactly the double-sided Laplace transform of this quantity, i.e.
\begin{align}
\tilde \rho_{\min}(s)&=NL^d\int_{-\infty}^{\infty}\,de\,\rho_{\min}(e)\,e^{-NL^d s e}=\int {\cal D}{\bf u}({\bf x})\,\moy{\delta^N\left(\frac{\delta {\cal H}[{\bf u}]}{\delta {\bf u}({\bf x})}\right)\det {\cal K}[{\bf u}({\bf x})]\Theta({\cal K}[{\bf u}({\bf x})])e^{-s{\cal H}[{\bf u}]}}\\
&=\sum_{\alpha:{\rm minima}}\moy{e^{-s {\cal H}[u_{\alpha}]}}\geq \moy{e^{-s N L^d e_{\min}}}\;.
\end{align}
Note that for $s=0$, $\tilde \rho_{\min}(s=0)=\moy{{\cal N}_{\min}(\mu)}$ is simply the average number of minima. To perform this computation, one should compute the statistics of the Hessian conditioned on (i) a vanishing gradient and (ii) a fixed value of the disordered potential. One can simply realise using the translationally invariant form of the covariance, similarly to the computation of $\moy{{\cal N}_{\min}(\mu)}$ \cite{fyodorov2020manifolds}, that the gradient $\frac{\delta {\cal H}[{\bf u}]}{\delta {\bf u}({\bf x})}$ is uncorrelated with both the Hessian ${\cal K}[{\bf u}({\bf x})]$ and the disordered potential $V({\bf u},{\bf x})$. One can compute explicitly the conditional expectation of the Hessian as
\be
\moy{{\cal K}_{i{\bf x},j{\bf y}}|V({\bf u},{\bf x})=Nv({\bf x})}=-t\Delta_{{\bf x},{\bf y}}\delta_{ij}+\left(\mu+\frac{f'(0)}{f(0)}v({\bf x})\right)\delta_{ij}\delta_{{\bf x}{\bf y}}
\ee
while its conditional covariance reads
\be
{\rm Cov}\left[\frac{\partial^2}{\partial u_i\partial u_j}V({\bf u},{\bf x}),\frac{\partial^2}{\partial u_k\partial u_l}V({\bf u},{\bf x})|V({\bf u},{\bf x})=Nv({\bf x})\right]=\frac{f''(0)}{N}(\delta_{ik}\delta_{jl}+\delta_{il}\delta_{jk})+\frac{1}{N}\left(f''(0)-\frac{f'(0)^2}{f(0)}\right)\delta_{ij}\delta_{kl}\;.
\ee
Computing explicitly the Gaussian integration over the field ${\bf u}({\bf x})$, one obtains
\begin{align}
&\int {\cal D}{\bf u}({\bf x})\,\moy{\delta^N\left(\frac{\delta {\cal H}[{\bf u}]}{\delta {\bf u}({\bf x})}\right)e^{-s{\cal H}[{\bf u}]}}\\
&=\left(-2\pi f'(0)\right)^{-\frac{NL^d}{2}}\int {\cal D}{\bf u}({\bf x})\,e^{\frac{1}{2f'(0)}\sum_{{\bf x},{\bf y}}{\bf u}({\bf x})\left[(\mu\mathbb{I}-t\Delta)((\mu-s f'(0))\mathbb{I}-t\Delta)\right]_{{\bf x}{\bf y}}{\bf u}({\bf y})}\moy{e^{-s\sum_{\bf x}V({\bf u},{\bf x})}}\\
&=\prod_{\bf x}\int_{-\infty}^{\infty}\frac{dv({\bf x})}{\sqrt{2\pi f(0)/N}}e^{-N\left[s v({\bf x})+\frac{v({\bf x})^2}{2f(0)}\right]}\det\left(K(\mu)K(\mu-s f'(0))\right)^{-1/2}\;,
\end{align}
where $K_{i{\bf x},j{\bf y}}(\mu)=\delta_{ij}\left(\mu\delta_{{\bf x}{\bf y}}-t\Delta_{{\bf x}{\bf y}}\right)$. The function $\tilde \rho_{\min}(s)$ can be computed exactly as
\begin{align}
    \tilde \rho_{\min}(s)=&\prod_{\bf x}\int_{-\infty}^{\infty}\frac{N dv({\bf x})d\xi({\bf x})}{2\pi\sqrt{f''(0)f(0)-f'(0)^2}}e^{-N\left[s v({\bf x})+\frac{v({\bf x})^2}{2f(0)}+\frac{\left(\xi({\bf x})-\frac{f'(0)}{f(0)}v({\bf x})\right)^2}{2\left(f''(0)-\frac{f'(0)^2}{f(0)}\right)}-s e_{\rm RS}\right]}\\
&\times\frac{\det(K(\mu)+X+\sqrt{f''(0)}H)\Theta\left(K(\mu)+X+\sqrt{f''(0)}H\right)}{\det\left(K(\mu)^{1/2}K(\mu-s f'(0))^{1/2}\right)}\\
    &=\prod_{\bf x}\int_{-\infty}^{\infty}\sqrt{\frac{N}{2\pi f''(0)}}d\xi({\bf x})e^{-N\left[\frac{\left(\xi({\bf x})+s f'(0)\right)^2}{2 f''(0)}-\frac{s^2}{2}f(0)\right]}\frac{\det(K(\mu)+X+\sqrt{f''(0)}H)\Theta\left(K(\mu)+X+\sqrt{f''(0)}H\right)}{\det\left(K(\mu)^{1/2}K(\mu-s f'(0))^{1/2}\right)}\\
    &=e^{NL^d\frac{s^2 f(0)}{2}}\left(\frac{\det(K(\mu-s f'(0))) }{\det(K(\mu))}\right)^{1/2}\moy{{\cal N}_{\min}(\mu-s f'(0))}\;,
\end{align}
where $X_{i{\bf x},j{\bf y}}=\delta_{ij}\delta_{{\bf x}{\bf y}}\xi({\bf x})$, $H_{i{\bf x},j{\bf y}}=\delta_{{\bf x}{\bf y}}h_{ij}({\bf x})$ where the $h({\bf x})$'s are iid standard GOE matrices. Note that the computation above is exact. One can then use the results from \cite{fyodorov2020manifolds} to express the annealed complexity of minima as
\be
\Sigma_{\min}(\mu)=\lim_{NL^d\to \infty}\frac{1}{NL^d}\ln \moy{{\cal N}_{\min}(\mu)}=\begin{cases}
    \displaystyle 0\;,\;\;\mu\geq \mu_{c}\;,\\
    &\\
    \displaystyle -\frac{(\mu-\mu_{c})^2}{2}+\int_{\mu}^{\mu_{c}}dx\int_{\bf k}\left[\frac{1}{\nu({\bf k})-\mu+x}-\frac{1}{\nu({\bf k})-\mu+\mu_{c}}\right]\;,\;\;\mu\leq \mu_{c}\;,
\end{cases}
\ee
where $\mu_{c}$ satisfies $f''(0)I_2(\mu_c)=1$. Thus, as a function of $s$, 
\begin{align}
\Sigma_{\min}(\mu-s f'(0))&=\lim_{NL^d\to \infty}\frac{1}{NL^d}\ln \moy{{\cal N}_{\min}(\mu-s f'(0))}\\
&=\begin{cases}
    \displaystyle 0\;,\;\;s\geq s_{\rm AT}\;,\\
    &\\
    \displaystyle -\frac{f'(0)^2(s-s_{\rm AT})^2}{2}+\int_{\bf k}\left[\ln\frac{\nu({\bf k})-s f'(0)}{\nu({\bf k})-s_{\rm AT} f'(0)}+\frac{(s-s_{\rm AT})f'(0)}{\nu({\bf k})-s_{\rm AT}f'(0)}\right]\;,\;\;s\leq s_{\rm AT}\;,
\end{cases}    
\end{align}
where the parameter $s_{\rm AT}$ satisfies
\be
s_{\rm AT}=\frac{\mu-\mu_c}{f'(0)}\;.
\ee
For general value of $s$, one obtains that 
\begin{align}
\tilde\Sigma_{\min}(s,\mu)=\lim_{NL^d\to \infty}\frac{1}{NL^d}\ln \tilde \rho_{\min}(s)&=\frac{s^2 f(0)}{2}+\frac{1}{2}\int_{\bf k}\ln\left(1-\frac{s f'(0)}{\nu({\bf k})}\right)+\Sigma_{\min}(\mu-s f'(0))\label{bound}\\
&\geq \lim_{NL^d\to \infty}\frac{1}{NL^d}\ln \moy{e^{-N L^d s e_{\min}}}=\varphi_L(s,\mu)\;.  \nn  
\end{align}
In the regime $s>s_{\rm AT}$ where $\Sigma_{\min}(\mu-s f'(0))=0$ one obtains in particular that 
\begin{align}
\tilde\Sigma_{\min}(s,\mu)&=\frac{s^2 f(0)}{2}+\frac{1}{2}\int_{\bf k}\ln\left(1-\frac{s f'(0)}{\nu({\bf k})}\right)=\varphi_{L,\rm RS}(s,\mu)\;.    
\end{align}
Thus, if the transition is continuous, i.e. it occurs at $s=s_{\rm AT}$, the bound in Eq. \eqref{bound} is tight for any $s>s_{\rm AT}$. If the transition is  is a discontinuous transition, i.e. the transition occurs at $s_{\rm dis}>s_{\rm AT}$, this bound is still tight for any $s>s_{\rm dis}$.


By Legendre transform, the expression for the annealed complexity of minima at fixed centred energy reads
\be
\Sigma_{\min}(e,\mu)=\min_{s\in \mathbb{R}}\left[s e+\tilde\Sigma_{\min}(s,\mu)\right]\geq -\Lambda_L(e,\mu)\;,
\ee
where the bound is a simple consequence of Eq. \eqref{bound_1}. One recovers similarly that the bound is tight in the regime of validity of the RS ansatz. If the transition is continuous, the bound is tight for any $e\leq e_{\rm AT}$ while if the transition is discontinuous, it is tight only for $e\leq e_{\rm dis}<e_{\rm AT}$.

\section{Some integrals}

In the expressions below, we use the following integrals:
\begin{align}
&\frac{S_{d-1}}{(2\pi)^d} L^{d-
2}\int_{2\pi/L}^{\infty}dk\,\frac{k^{d-1}}{\mu+k^{2}}={\cal D}_{1,d}(\mu L^2)\\
&{\cal D}_{1,d}(x)=\frac{2\pi^{d/2}}{2(2\pi)^2\Gamma(d/2)}\left(-\frac{x}{4\pi^2}\right)^{\frac{d-2}{2}}B\left(-\frac{x}{(4\pi)^2};\frac{2-d}{2},0\right)\;,\\
&\frac{S_{d-1}}{(2\pi)^d} L^{d-
4}\int_{2\pi/L}^{\infty}dk\,\frac{k^{d-1}}{(\mu+k^{2})^2}={\cal D}_{2,d}(\mu L^2)\\
&{\cal D}_{2,d}(x)=\frac{2\pi^{d/2}}{(2\pi)^4\Gamma(d/2)}\left[\frac{1}{2(1+\frac{x}{4\pi^2})}+\frac{d-2}{4}\left(-\frac{x}{4\pi^2}\right)^{\frac{d-4}{2}}B\left(-\frac{x}{(4\pi)^2};\frac{4-d}{2},0\right)\right]\;.
\end{align}

\end{appendix}

\bibliography{ref}    

\end{document}